\documentclass[%
 reprint,
 amsmath,amssymb,
 aps,
prb,
]{revtex4-2}

\usepackage{graphicx}
\usepackage{dcolumn}
\usepackage{bm}

\usepackage{caption}
\usepackage{subcaption}
\usepackage{ragged2e}

\begin{document}

\preprint{APS/123-QED}

\title{Unifying principle for Hall coefficient in systems near magnetic instability}

\author{Jalaja Pandya}%
\email{jalajapandya@gmail.com}
\author{Navinder Singh}
\email{navinder.phy@gmail.com}
\affiliation{Physical Research Laboratory, Ahmedabad, India, PIN: 380054.}%

\date{\today}

\begin{abstract}
Typically, Hall coefficient of materials near magnetic instabilities exhibits pronounced temperature dependence. To explore the reasons involved, we studied the temperature dependence of Hall coefficient in $Cr_{1-x}V_x$, $V_{2-y}O_3$ and some high-$T_c$ superconducting cuprates. We argue that it can be rationalized using the following unifying principle:\textit{ When a system is near a magnetic instability and temperature is reduced towards the instability, there is a progressive ``loss" of carriers (progressive ``tying down" of electrons) as they participate in long-lived and long-ranged magnetic correlations.} In other words, magnetic correlations grow in space and are longer-lived as temperature is reduced towards the magnetic instability. This is the mechanism behind reduced carrier density with reducing temperature and leads to an enhancement of the Hall coefficient. This unifying principle is implemented and quantitative analysis is done using the Gor'kov Teitel'baum Thermal Activation (GTTA) model. We also show that the Hall angle data can be understood using one relaxation time (in contrast to the ``two-relaxation" times idea of Anderson) by taking into consideration of temperature dependence of carrier density. This unifying principle is shown to be working in above studied systems, but authors believe that it is of much more general validity.

\end{abstract}

\maketitle

\section{\label{sec:level1}Introduction to the origin of the temperature dependence of Hall coefficient in ``magnetic metals"}
The Hall coefficient $R_H$ carries valuable information regarding the nature of carriers in a material. In the approximation of free electron model $R_H$ is given by, $R_H=\frac{1}{ne}$ where $n$ is the number density of carriers and $e$ is the charge of each carrier (positive for holes and negative for electrons). The alkali metals (Na, K, Rb, etc.) having simple closed Fermi surface obey this Drude theory result for Hall coefficient to a very good approximation. In 1969, Alderson and Farrell studied the temperature dependence of $R_H$ for Li, Na and K \cite{alderson1969hall}. They analyzed the effective charge carriers per atom (n*) for these metals. The value of n* is close to unity which is in accordance with the free electron model. The relative change in n* from unity for Li is displayed in Fig. 1(a). The temperature dependence of $R_H$ for alkali metals is also reported by Fletcher \cite{fletcher1977righi,fletcher1973righi} at low temperatures. $R_H$ is roughly constant with temperature and shows very less deviation from the free electron model. The group 1B metals show a temperature dependence of $R_H$ at low temperatures (less than 100K), it is constant between $\sim100 K-350 K$ after which it is linear in temperature \cite{hurd2012hall}. This temperature dependence of $R_H$ is due to their complicated Fermi surface topology \cite{hurd2012hall}. 
\

On the other hand, magnetic metals (Ni, Co, Fe, etc.) show drastic temperature dependence of $R_H$ which is in sharp contrast to the $R_H(T)$ observed in group 1B metals. In the ferromagnetic state (below $T_C$), the Hall resistivity is $\rho_H(B)=R_oB+4\pi M_sR_s$, where $R_o$ and $R_s$ are ordinary and anomalous Hall coefficients \cite{kaul1979anomalous,karplus1954hall,maranzana1967contributions}. The anomalous Hall coefficient $R_s$ shows stronger temperature dependence compared to the ordinary Hall coefficient $R_o$ \cite{allison1956temperature}. These materials also show strong temperature dependence of $R_H$ in paramagnetic phase as well \cite{doi:10.1143/JPSJ.17.717}. The question is: why does Hall coefficient shows strong temperature dependence in magnetic metals in paramagnetic phase? Consider for example the case of pure Ni especially in the paramagnetic state ($T>T_N$). As temperature increases, $R_H$ decreases \cite{doi:10.1143/JPSJ.17.717} i.e. the effective charge carrier density increases. The relative change in $R_H$ with temperature (with respect to $R_H$ (at 972K)) for $T>T_c$ is displayed in Fig. 1(b). A stronger temperature dependence of $R_H$ is observed in Ni compared to Li. A remarkably similar temperature dependence of $R_H$ is observed in high temperature superconducting cuprates.   
\begin{figure}[h!]
 \centering
  \subfloat[$Li$]{\includegraphics[width=0.5\linewidth]{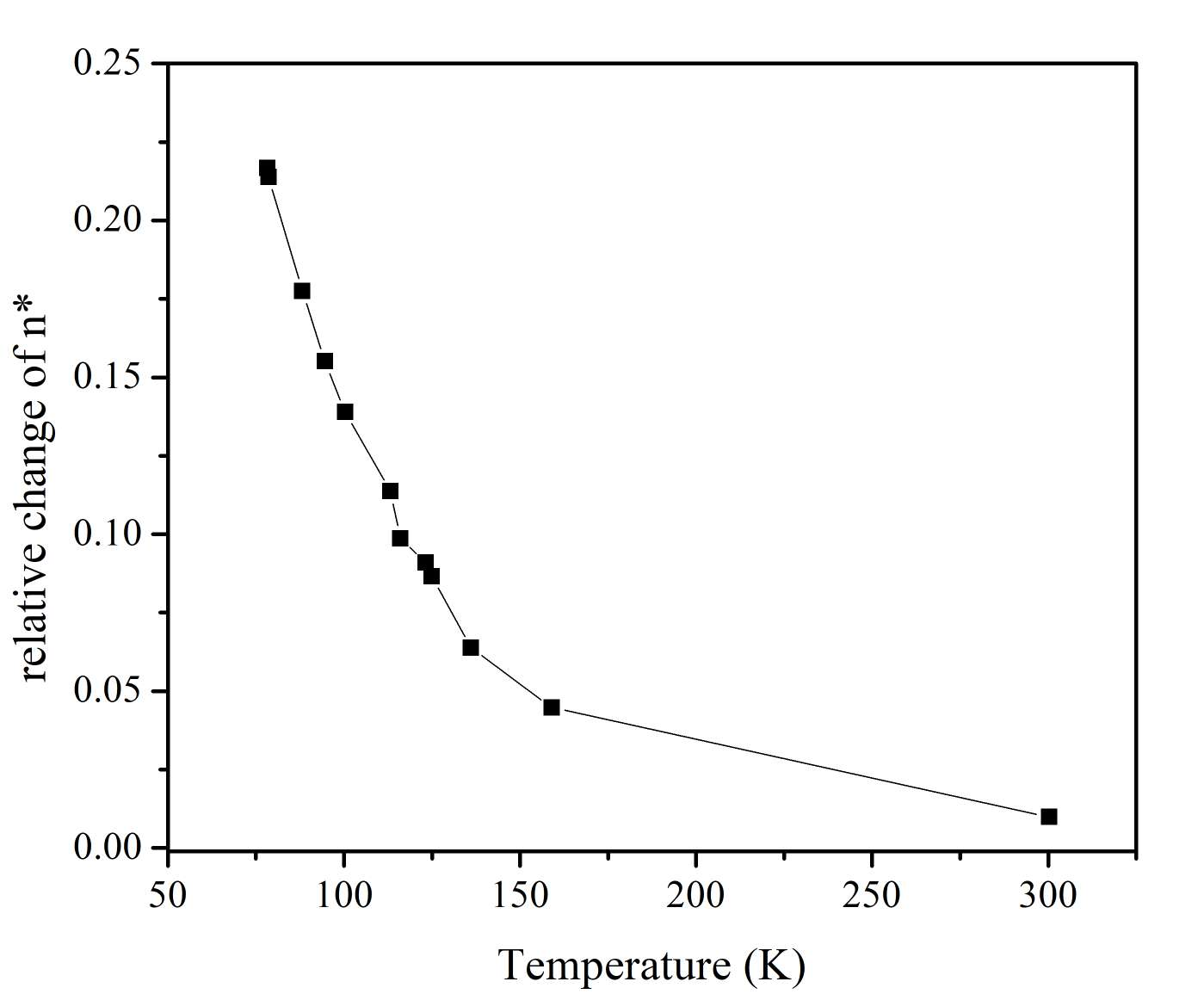}}
  \subfloat[$Ni$]{\includegraphics[width=0.5\linewidth]{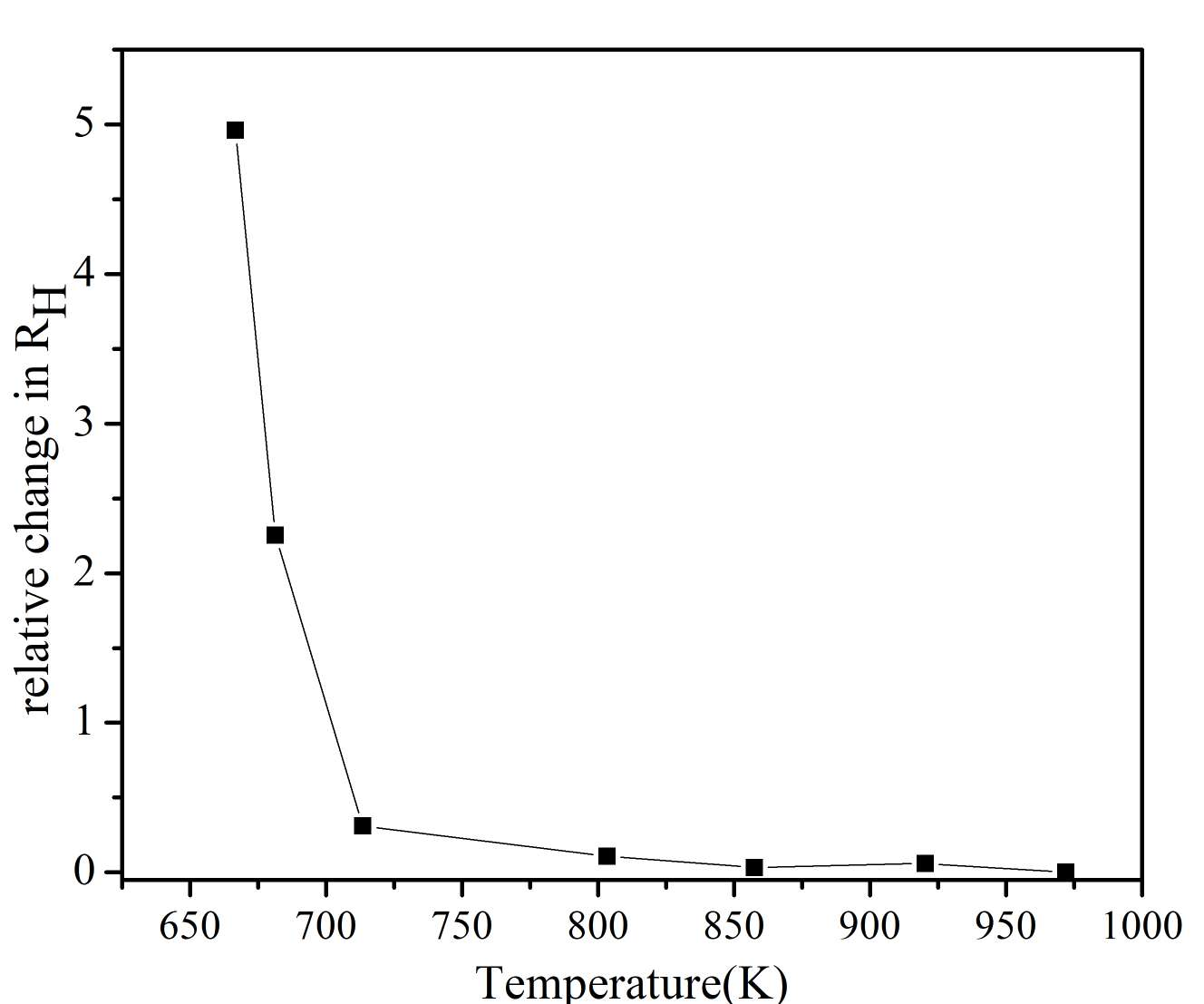}}
  \caption{\justifying (a) The relative change in the carrier concentration n* of Li extracted from Fig. 1 of \cite{alderson1969hall}, and (b) The relative change in the Hall coefficient $R_H$ of pure Ni extracted from Fig. 1 of \cite{doi:10.1143/JPSJ.17.717}. In Ni (b), the relative change in $R_H$ is an order of magnitude greater than that in Li (a) which is clear from the scale of the graphs where in (a) the relative change in y-axis is displayed till 0.25 value while in (b) it is till 5.}
  \label{fig:1}
\end{figure}

\ It is easy to understand the temperature dependence of $R_H$ in semiconductors in which thermal excitation of carriers above the energy gap increase the number of carriers available for conduction, thus $R_H$ reduces with increase in temperature \cite{turner1961physical}. More accurate treatment involves the consideration of both electron and hole mobilities and their concentration \cite{hackmann1981determination}. \textit{However, in paramagnetic phase of magnetic metals, no such energy-gaps are present.}\textit{ Then why is temperature dependent $R_H$ observed for such materials? To answer this question we put forward a unifying principle first hinted in} \cite{yeh2002quantum}.\

``When a system is towards magnetic instability and temperature is reduced, there is a progressive ``loss" of carriers as they exhibit longer-lived and longer-ranged magnetic correlations. In other words, it is the progressive ``tying down" of electrons as temperature is reduced which means lower carrier density and larger Hall coefficient with reducing temperature." We will observe that within this paradigm it is easy to understand the temperature dependent $R_H$ in magnetic systems from a unified point of view. This work is devoted to understand temperature dependence of $R_H$ using this unifying principle. In this light we could rationalize the temperature dependence of $R_H$ in $Cr_{1-x}V_x$, $V_{2-y}O_3$ and some high-$T_c$ superconducting cuprates. However, the unifying principle must be tested for a greater variety of systems. In addition, mathematical formulation of the principle in terms of Gor'kov Teitel'baum Thermal Activation (GTTA) model is only ad-hoc (or provisional). More accurate mathematical expression (from a microscopic calculation) for the principle is needed. \

The paper is organized in the following way; In section II we consider the case of $Cr_{1-x}V_x$ wherein the temperature dependence of $R_H$ is studied using the GTTA model and the behaviour Hall angle is also understood. The transport properties of $V_{2-y}O_3$ is analyzed in Section III. Section IV includes the Hall angle analysis of $La_{2-x}Sr_xCuO_4$ using the GTTA model. The temperature dependence of $R_H$ and Hall angle (cot$\theta_H$) of other high-$T_C$ superconducting cuprates viz. $YBa_2Cu_3O_{6+\delta}$, $TlBa_{1+x}La_{1-x}CuO_{5}$ and $Bi_2Sr_{2-x}La_xCuO_{6}$ is studied from the point of view of the proposed unifying principle in Section V to Section VII. An updated phase diagram is drawn for all these materials. We discuss and conclude all our results in Section VIII. 


\section{\label{sec:level1}Case of $Cr_{1-x}V_x$}
Chromium being an antiferromagnetic metal, has strong temperature dependence on $R_H$ \cite{de1957hall}. Alloying of Chromium with other transition metals brings a shift in transition temperature and its electronic properties \cite{de1959transition}. Vanadium doping has a strong effect on the physical properties of the metal near the Neel temperature ($T_N$) in the paramagnetic phase \cite{fawcett1994spin}. In 2002, in a seminal paper, Yeh et al. performed the Hall effect analysis on $Cr_{1-x}V_x$ near its antiferromagnetic quantum phase transition \cite{yeh2002quantum}. The ordered magnetic moment vanished completely at the critical doping $x_c=0.035$ (refer to phase diagram in Fig. 2). The longitudinal resistivity $\rho(T)$ for different vanadium doping concentrations is reported in  Fig 2. of \cite{yeh2002quantum}. $\rho(T)$ increases with the increase in doping concentration $x$ implying stronger scattering of charge carriers due to vanadium doping. At low temperatures $\rho(T)$ follows a $T^3$ power law. The inverse of Hall coefficient ($\frac{1}{R_H}$) shows strong temperature dependence beyond $T_N$ i.e. in paramagnetic phase \cite{yeh2002quantum} which indicates loss in charge carries as the temperature decreases which can be rationalized using the unifying principle discussed in Section 1. This unifying principle was first intimated in \cite{yeh2002quantum} wherein authors state that: \\

 \textit{ ``There are strong antiferromagnetic fluctuations extending to very high energy ($0.4 eV$) for paramagnetic $Cr_{1-x}V_x$...such fluctuations will tie down electrons from within the Fermi surface in the same way that magnetic order ties down electrons, implying an apparent loss of carriers on cooling as the fluctuations slow down...the T-dependent Hall effect is reminiscent of that for much more exotic metals, and, together with susceptibility data, implies a pseudogap in the spectrum of electronic excitations" \cite{yeh2002quantum}.} \\

\begin{figure}[h!]
    \includegraphics[width=1.0\linewidth]{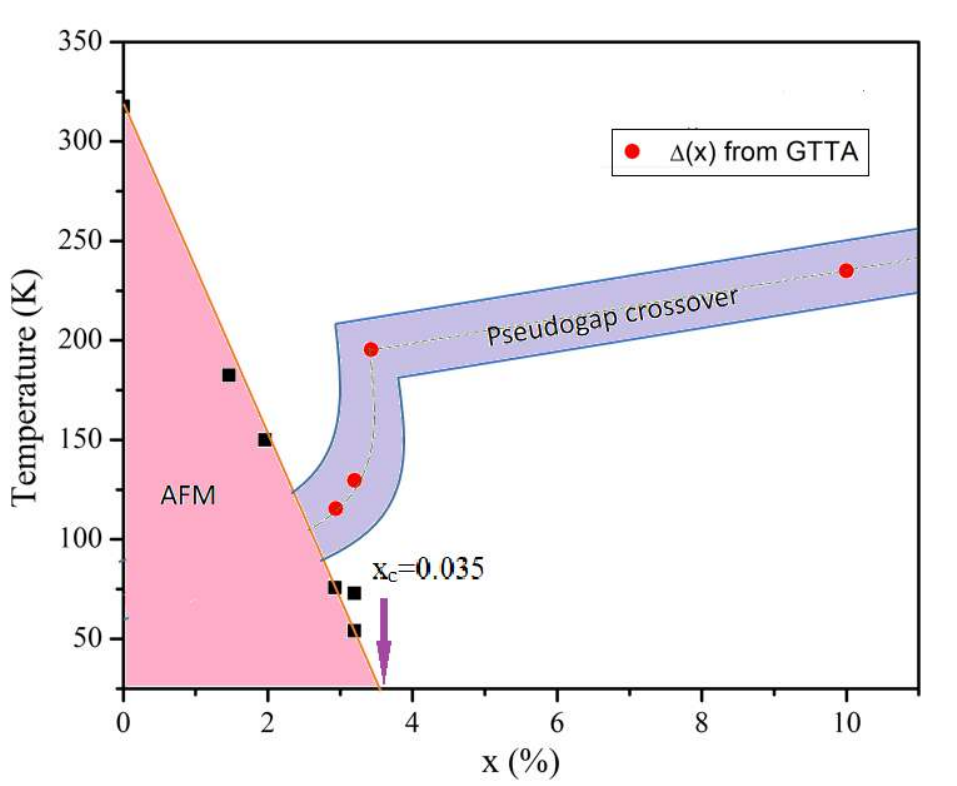}
    \caption{\justifying Updated phase diagram of $Cr_{1-x}V_x$. The Neel temperatures are extracted from Fig. 1(a) of \cite{yeh2002quantum} and the PG crossover is obtained from the value of $\Delta(x)$ using the GTTA model (present work). The pink shaded region marks the Antiferromagnetic phase of $Cr_{1-x}V_x$.}
\label{CrV-phase}
\end{figure}

 Thus, this prompts us to formulate this idea as a unifying principle and to estimate the value of a possible Pseudogap (PG) crossover in this material proposed by the authors of \cite{yeh2002quantum}. Here we use the very successful Gor'kov Teitel'baum Thermal Activation (GTTA) model discussed in Appendix A to determine quantitatively the PG crossover for $Cr_{1-x}V_x$ and draw an updated phase diagram (Fig. \ref{CrV-phase}). We would like to stress here that an effective gap brought about by the mechanism of ``tying down" of electrons by magnetic correlations (as temperature reduces) is not a hard gap (like forbidden gaps in semiconductors) but is an effective gap of thermal activation nature to which authors of \cite{yeh2002quantum} attributed as PG crossover. And its boundary is only a very gradual crossover, definitely not a phase transition with some emergent ordered parameter. Our aim is to quantify this crossover.\ 

Fig 2. of \cite{yeh2002quantum} shows the temperature dependence of $\rho(T)$ and $(eR_H)^{-1}$, the data points of which are extracted and analyzed using the GTTA model (equation (1) of Appendix A). The $n_0(x)$, $n_1$ and $\Delta(x)$ are extracted for $Cr_{1-x}V_x$. The doping dependence of $n_0(x)$ and $\Delta(x)$ of $Cr_{1-x}V_x$ is displayed in Fig \ref{CV-GTTA}. $n_1$ is roughly constant ($\sim0.8$) for all the doping concentrations. As discussed in Appendix A, $n_0(x)$ is the temperature independent term which corresponds to the effective charge carries due to the external doping. In $Cr_{1-x}V_x$, $n_0(x)$ increases rapidly with $x$ and tends to saturate after the critical point $x_c=0.035$. This doping dependence of $n_0(x)$ is in good agreement with doping dependence of carrier concentration at 0K reported in Fig. 1(c) of \cite{yeh2002quantum}. The $\Delta(x)$, which is interpreted as the pseudogap crossover, at low doping is small since pure Cr is a metal. On adding vanadium $\Delta(x)$ increases rapidly till the quantum critical point $x_c=0.035$ after which $\Delta(x)$ seems to increase roughly linearly. \ 

The next question is that whether the conjectured and estimate PG has real physical meaning in the phase diagram (Fig. \ref{CrV-phase})? To this end, we propose experiments (such as NMR $\frac{1}{T_1T}$  relaxation rate) to determine the proposed PG crossover. The deviation in $\frac{1}{T_1T}$ would show some anomaly in the range of 200K for doping range $4\%$ to $10\%$. However, the signature may be weak and one may need to look at the derivatives of the relaxation curve.

\begin{figure}[h!]
 \centering
  \subfloat[$n_0(x)$]{\includegraphics[width=1.0\linewidth]{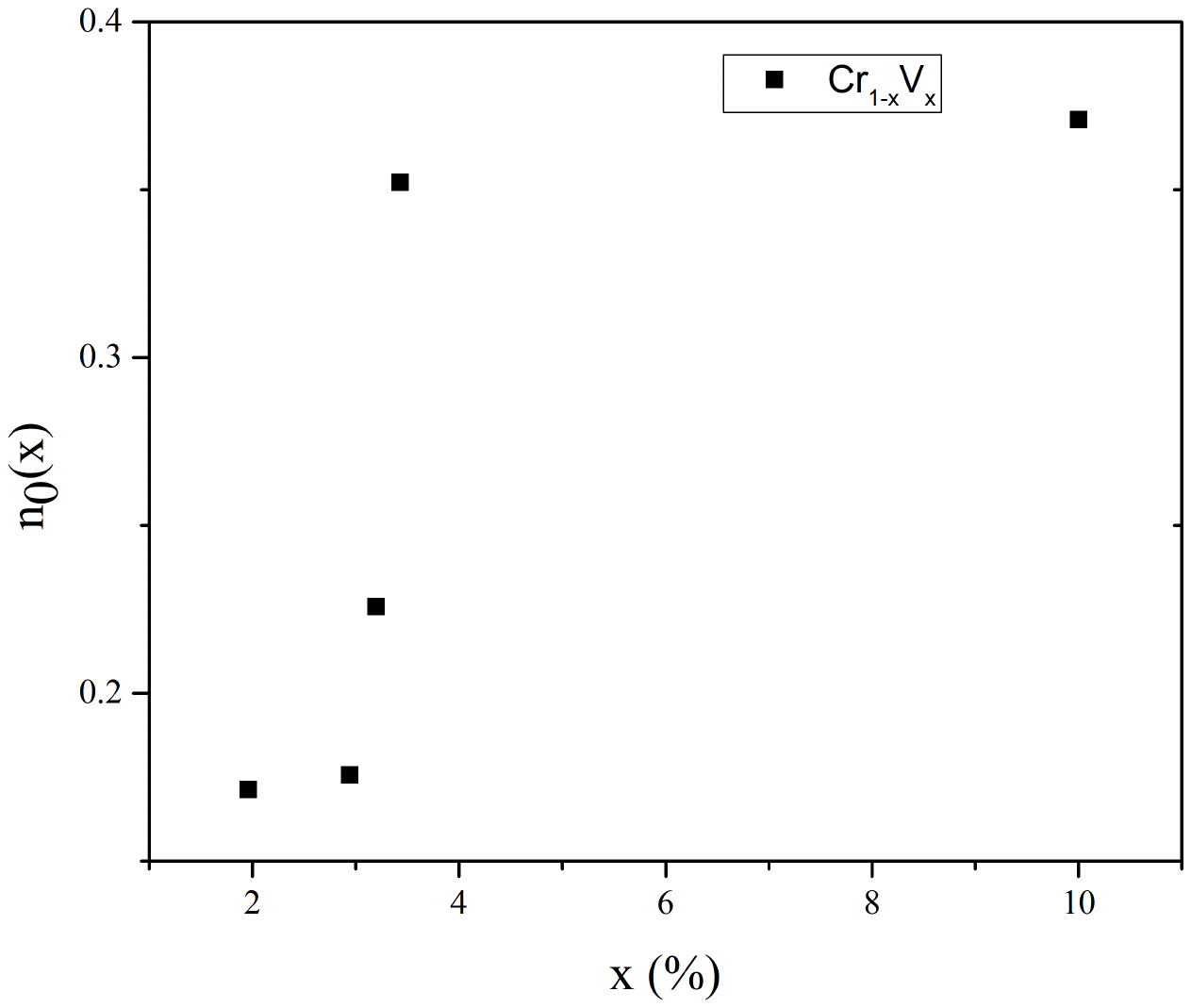}} \\
  \subfloat[$\Delta(x)$]{\includegraphics[width=1.0\linewidth]{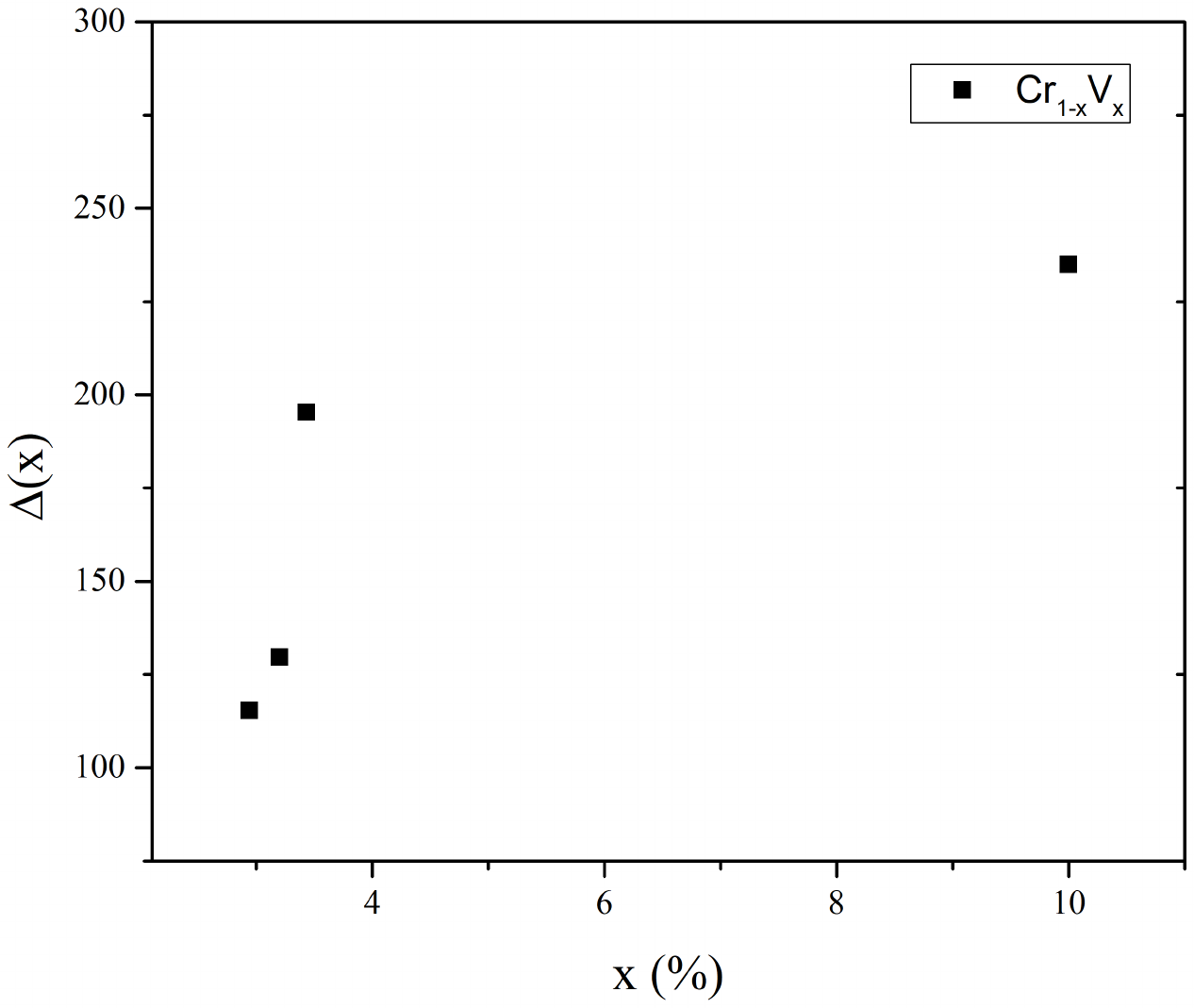}}
  \caption{\justifying The doping dependence of (a) $n_0(x)$ and (b) $\Delta(x)$ of $Cr_{1-x}V_x$ from the Hall coefficients extracted from Fig. 2(b) of \cite{yeh2002quantum} using the GTTA model.}
\label{CV-GTTA}
\end{figure}
\

Since this extraordinary behaviour of $R_H$ in $Cr_{1-x}V_x$ is similar to the high-$T_c$ superconductors the temperature dependence of Hall angle (cot$\theta_H$) is also reported by authors of \cite{yeh2002quantum} which showed an approximate $T^2$ dependence unlike the temperature dependence seen for $\rho(T)$. The D.C. resistivity $\rho(T)$ shows $T^3$ behaviour in the low temperature regime ($T<<T_N$) and roughly T-linear behaviour above $T_N$. From these observations one would like to argue that there are two or more than two relaxation rates in $Cr_{1-x}V_x$: one relaxation rate which scales as $T^2$ governs cot$\theta_H$ behaviour and the other relaxation rate which scales as T governs the behaviour of D.C. resistivity (in high temperature limit $T>>T_N$) and different relaxation rate in low temperature limit. In case of cuprates, this sort of discrepancy in the temperature dependence of the transverse (cot$\theta_H$) and longitudinal ($\rho(T)$) relaxation rates was rationalized by Anderson using ``two-relaxation times" idea (discussed in Appendix B) \cite{singh2024review}. In contrast to this ``two-relaxation times" idea, Gor'kov and Teitel'baum (in case of cuprates) proposed a new way of looking at this problem wherein the charge carrier density $n$ is doping and temperature dependent (As discussed in Appendix A) \cite{gor2006interplay,gor2008mobility}. It turns out that the underlying relaxation rate (both for Hall angle and resistivity) scales as $T^2$. But the resistivity scales differently because in $\rho(x,T)=\frac{m*}{n(x,T)e^2}\frac{1}{\tau(T)}$, $n(x,T)$ is temperature and doping dependent. Similar ideas are advocated in \cite{barivsic2022high}. In fact the underlying relaxation rate can be determined directly from the experimental data if we consider $\frac{1}{\tau(T)}=\frac{e^{*2}}{m*}\rho(x,T)(x,T)$. For $La_{2-x}Sr_xCuO_4$, Gor'kov and Teitel'baum showed that $\rho(x,T)(x,T)$ for $x=0.08, 0.12$ and $0.15$ actually scales as $T^2$ in agreement with experiments \cite{gor2008mobility}. \
\begin{figure}[h!]
    \begin{subfigure}[b]{0.5\columnwidth}
    \includegraphics[width=\textwidth]{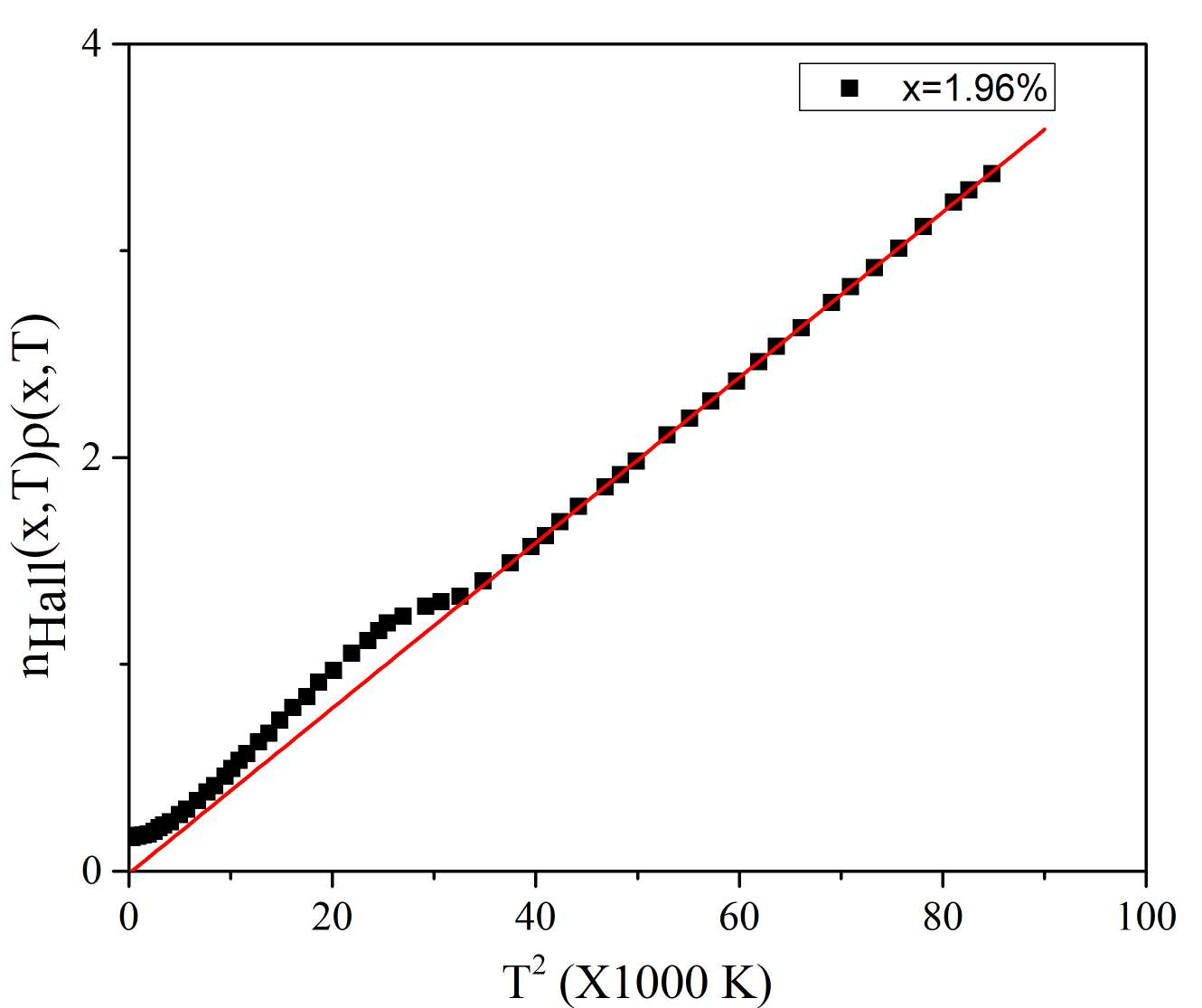}
    \caption{$x=1.96\%$}
  \end{subfigure}%
  \hfill
  \begin{subfigure}[b]{0.5\columnwidth}
    \includegraphics[width=\textwidth]{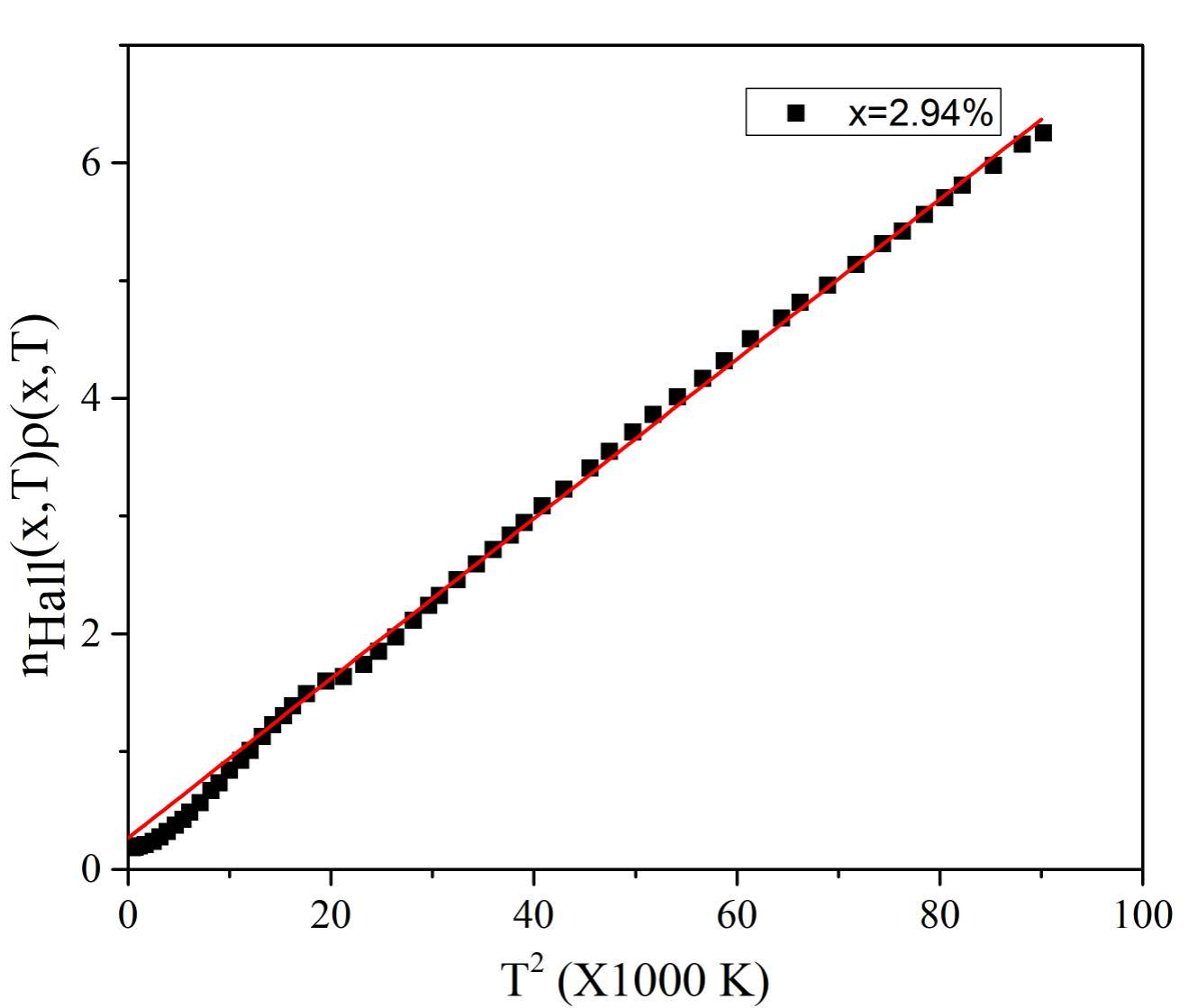}
    \caption{$x=2.94\%$}
  \end{subfigure}%

\medskip
\begin{subfigure}[b]{0.5\columnwidth}
    \includegraphics[width=\textwidth]{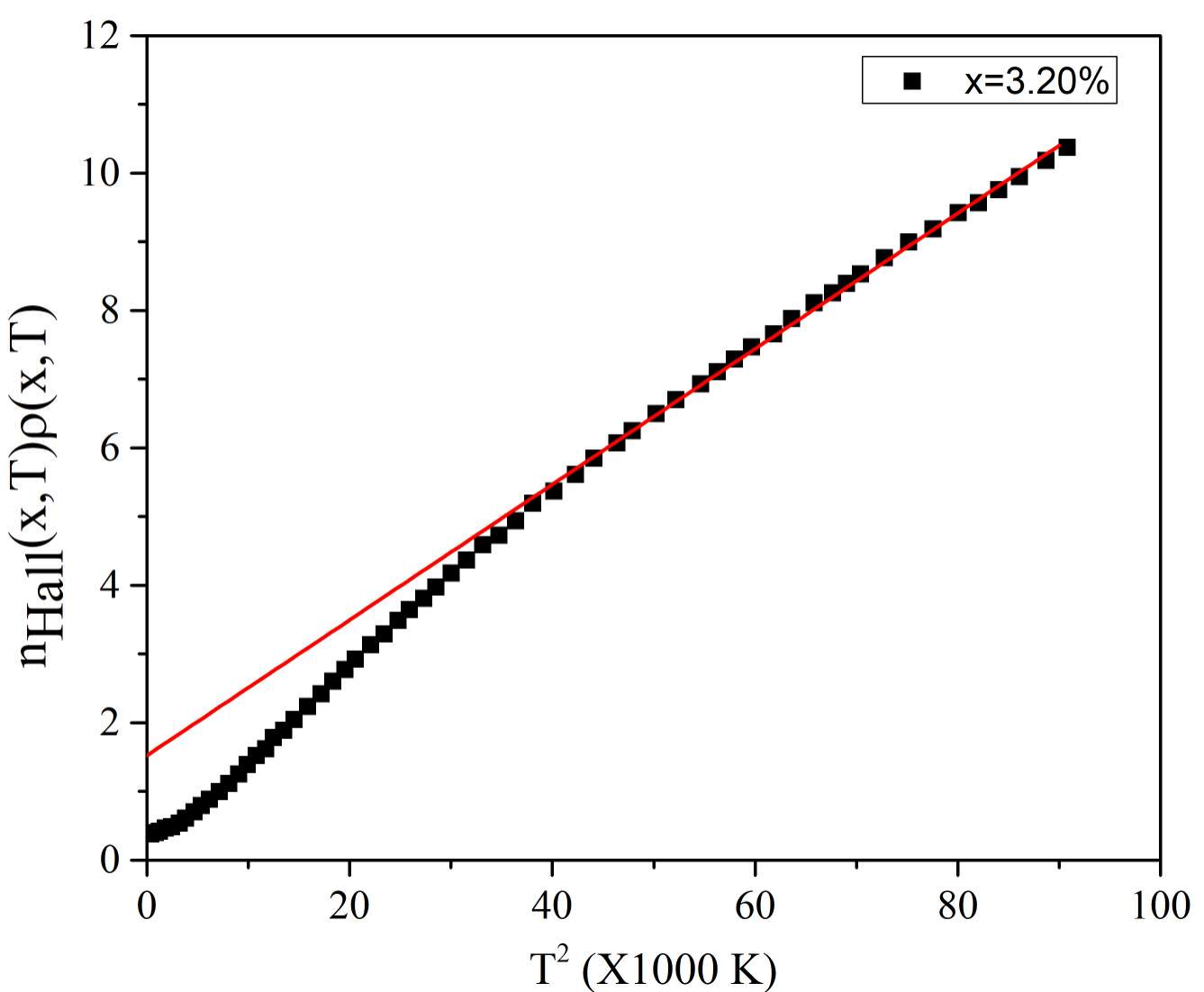}
    \caption{$x=3.20\%$}
  \end{subfigure}%
 \hfill
\begin{subfigure}[b]{0.5\columnwidth}
    \includegraphics[width=\textwidth]{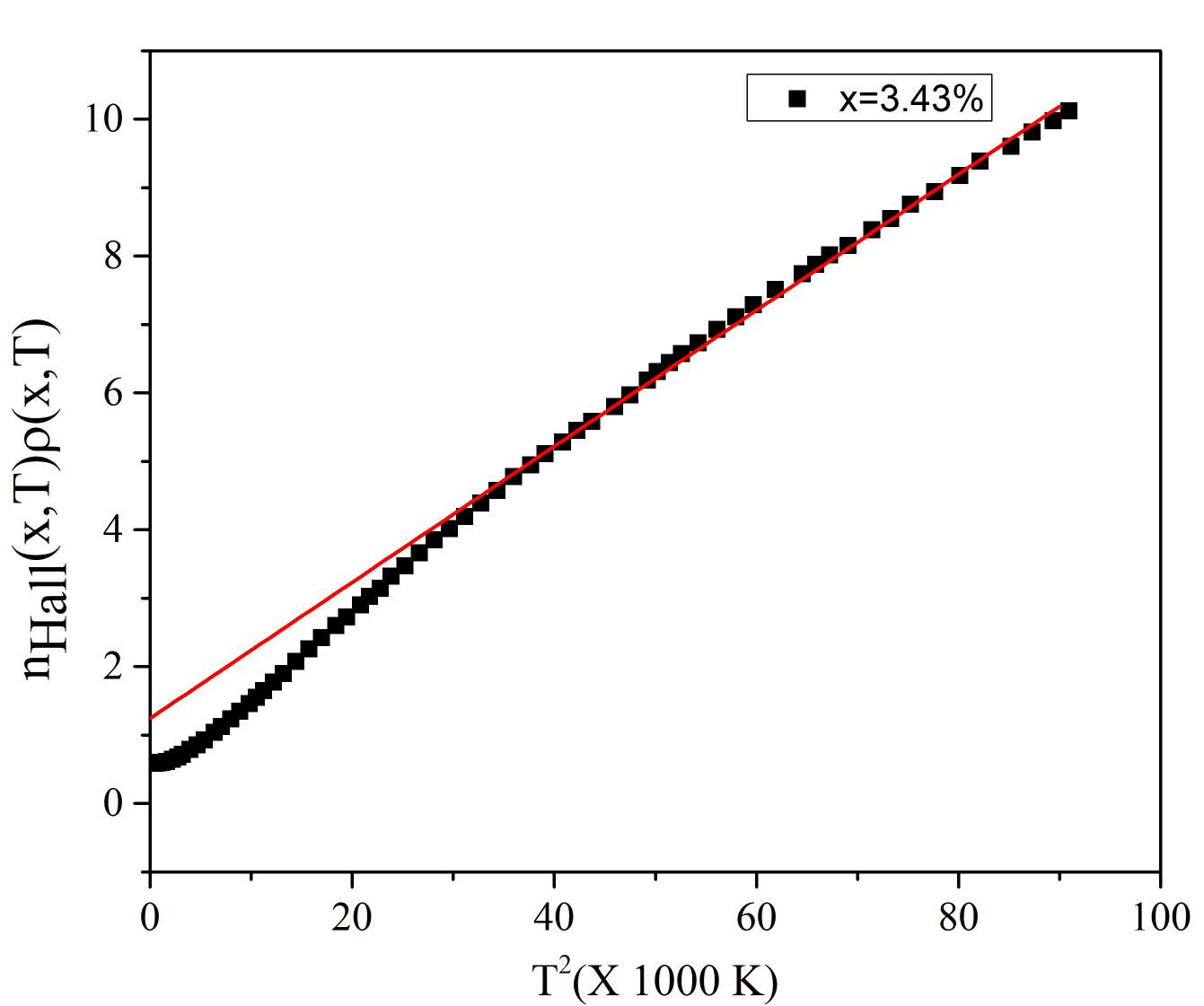}
    \caption{$x=3.43\%$}
  \end{subfigure}%
  \hfill
\begin{subfigure}[b]{0.5\columnwidth}
    \includegraphics[width=\textwidth]{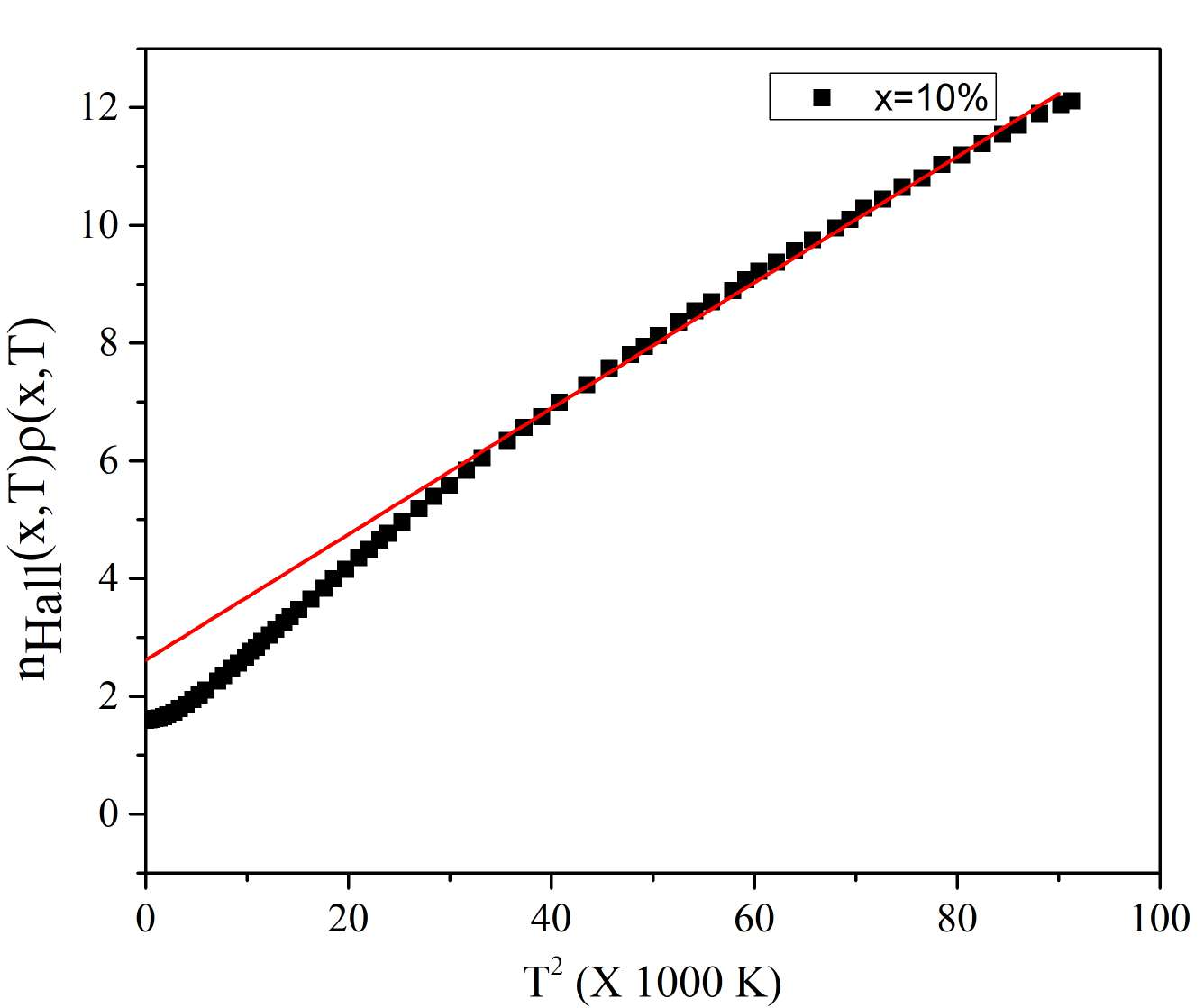}
    \caption{$x=10\%$}
    \end{subfigure}%
    \centering

  \caption{\justifying The $\rho(x,T)n_{Hall}(x,T)$ ($\propto$cot$\theta_H$) vs $T^2$ of $Cr_{1-x}V_x$. Longitudinal resistivity $\rho(x,T)$ is extracted from Fig. 2(a) of \cite{yeh2002quantum}. $n_{Hall}(x,T)$ is the carrier concentration calculated from the $R_H(T)$ data reported in Fig. 2(b) of \cite{yeh2002quantum} using the GTTA model. We can notice the $T^2$ behaviour of relaxation rate begins at roughly $\sim175K$.}
\label{CV-cot}
\end{figure}
In case of $Cr_{1-x}V_x$, we obtained $n_{Hall}(x,T)$ from GTTA analysis and plotted $\rho(x,T)n_{Hall}(x,T)$ ($\propto$cot$\theta_H$) with respect to $T^2$ in Fig. \ref{CV-cot}. The cot$\theta_H$ shows a roughly quadratic temperature dependence for all the doping concentrations which is in agreement with the experimental cot$\theta_H$ data (Fig 3(a) of \cite{yeh2002quantum}). On close analysis of Fig. \ref{CV-cot}, a deviation in $T^2$ behaviour of $\rho(x,T)n_{Hall}(x,T)$ is seen. For $x=0.0196$ (at $\sim170K$) and $x=0.0294$ (at $\sim140K$), a kink is observed at low temperatures. A similar behaviour of cot$\theta_H$ is noted for $x=0.0196$ at $\sim 145K$. In case of larger doping a downturn is seen at low temperatures. This downturn starts at $\sim187 K$ for $x=0.032$ and $\sim180K$ for $x=0.1$ doping which is also seen in Fig 3(a) of \cite{yeh2002quantum}. This further validates the temperature dependent $n_{Hall}(x,T)$ obtained using the GTTA model. Therefore for $T>T_N$ in $Cr_{1-x}V_x$ there is only one relaxation rate which scales as $T^2$!

\section{Case of $V_{2-y}O_{3}$}
\label{sec:level1}
Vanadium sesquioxide $V_{2-y}O_3$ shows an antiferromagnetic-insulator to metallic sharp transition at around 170 K \cite{mcwhan1973metal}. The transition temperature changes with external pressure, doping and Vanadium deficiency \cite{klimm2001new}. In 1988, Husmann et al. studied the temperature dependence of $R_H$ and longitudinal resistivity $\rho(T)$ of $V_{2-y}O_3$  \cite{rosenbaum1998temperature}. The temperature dependence of $R_H$ observed for this material is similar to $Cr_{1-x}V_x$, wherein $R_H$ continuously decreases with the increase in temperature. In $V_{2-y}O_3$, $R_H$ peaks at very low temperature ($\sim10K$). As $y$ (V deficiency) is increased the temperature dependence of $R_H$ is quenched. This behaviour of $R_H$ at low temperatures is quite similar to temperature dependence of $R_H$ seen in many cuprates and $Cr_{1-x}V_x$. Thus, our proposed unifying principle comes into play in $V_{2-y}O_3$ as well which means that the ``tying down" of electrons with decrease in temperature is responsible for the increase in $R_H$. Therefore an effective PG crossover similar to the one observed in $Cr_{1-x}V_x$ may be seen in $V_{2-y}O_3$ too. An evidence of pseudogap formation in Cr doped $V_{2-y}O_3$ from infrared spectroscopy is reported by Baldassarre et al \cite{baldassarre2008quasiparticle}.

\begin{figure}[h!]
    \includegraphics[width=1.0\linewidth]{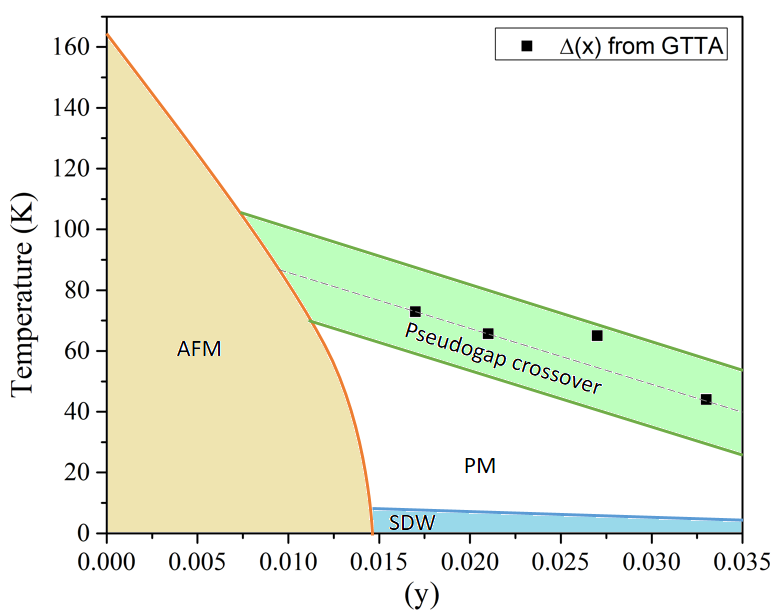}
    \caption{\justifying Updated phase diagram of $V_{2-y}O_3$. The orange shaded region shows the antiferromagnetic phase. The blue shaded is the spin density wave phase of the material. The boundary of these phases is reproduced from \cite{klimm2001new}. The PG crossover is shaded green obtained from the $\Delta(x)$ extracted from the GTTA model (present work).}
\label{VO-phase}
\end{figure}

To quantitatively analyse this possible effective PG crossover in $V_{2-y}O_3$, we use the GTTA model to the temperature dependent $R_H$ reported in Fig. 1 of \cite{rosenbaum1998temperature} and extract the value of this PG crossover in $V_{2-y}O_3$ (updated phase diagram in Fig. \ref{VO-phase}). The $n_0(y)$ and $\Delta(y)$ values of $V_{2-y}O_3$ are displayed in Fig. \ref{VO-GTTA} which are obtained using equation (1) of Appendix A. $n_0(x)$ increases very rapidly with vanadium deficiency which indicates an increase in charge carrier concentration with increase in vanadium deficiency. The value of $\Delta(x)$ decreases roughly linearly with $y$. These results are consistent with the fact that $V_{2-y}O_3$ undergoes a insulator to metallic transition on V deficiency.

\begin{figure}[h!]
 \centering
  \subfloat[$n_0(y)$]{\includegraphics[width=0.8\linewidth]{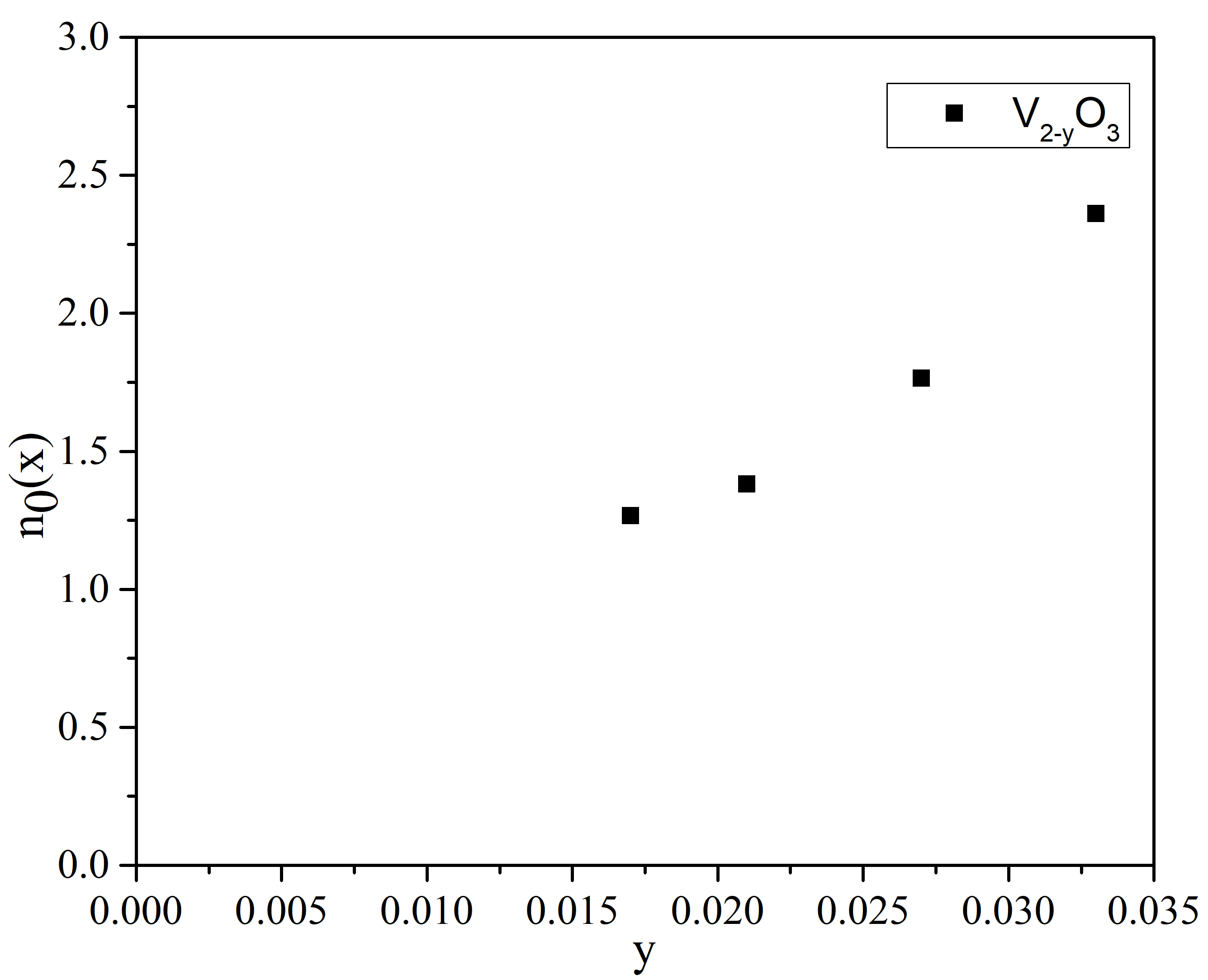}}\\
  \subfloat[$\Delta(y)$]{\includegraphics[width=0.8\linewidth]{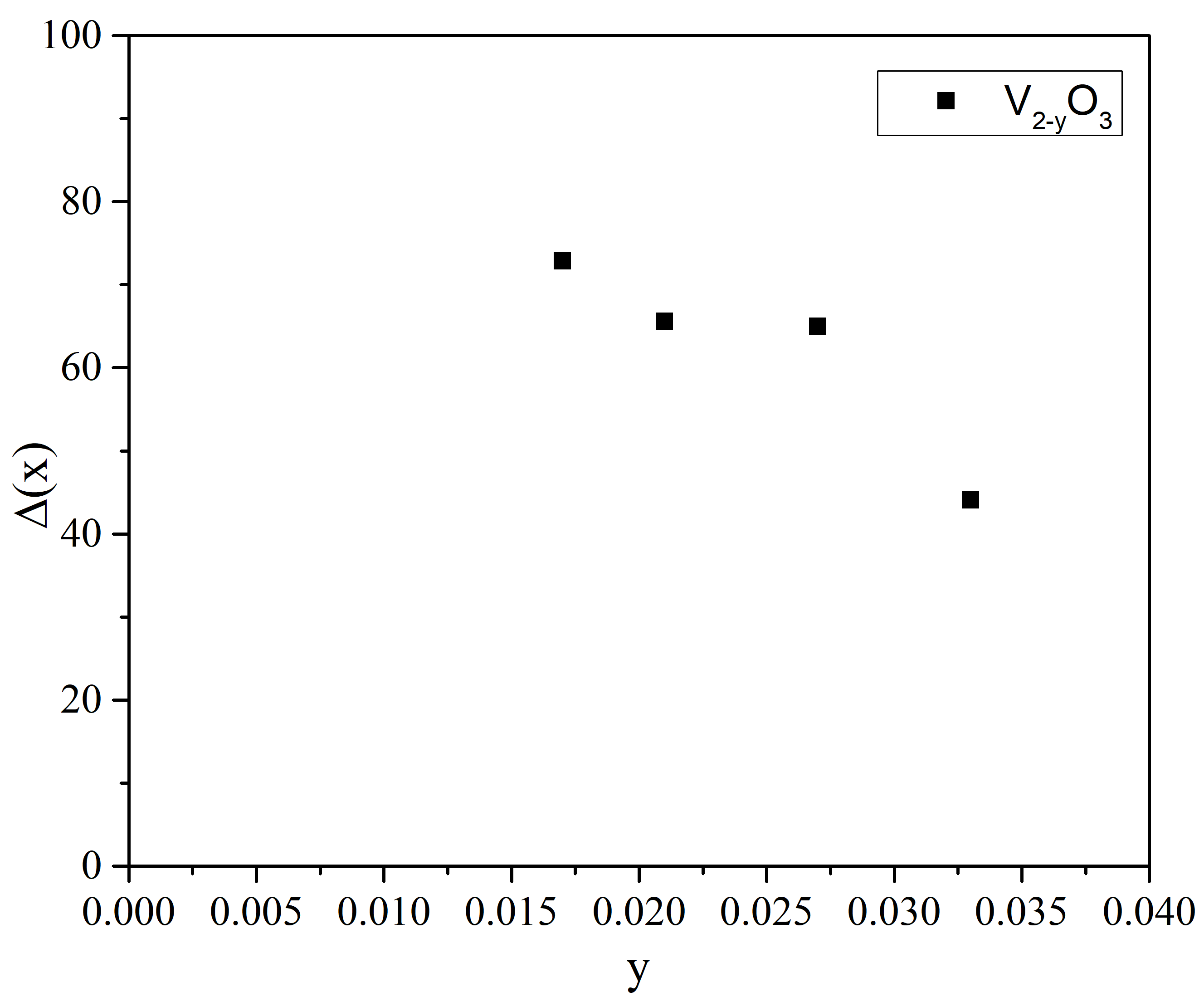}}
  \caption{ \justifying The doping dependence of (a) $n_0(y)$ and (b) $\Delta(y)$ of $V_{2-y}O_3$ obtained using the GTTA model from the $R_H(T)$ extracted from Fig. 1 of \cite{rosenbaum1998temperature}.}
  \label{VO-GTTA}
\end{figure}
\
The longitudinal resistivity $\rho_{xx}$ is reported in Fig. 1 of \cite{rosenbaum1998temperature}. They are motivated by the self-consistent renormalization theory which explained the neutron diffraction data and plotted $\rho_{xx}$ with respect to $T^\frac{3}{2}$ (typical of spin fluctuations mediated resistivity). However, the authors note a deviation of $\rho_{xx}$ from the model tending towards the temperature powers to be less than 1.5. Thus we re-analyzed the temperature dependence of $\rho_{xx}$ (without considering it to be $T^\frac{3}{2}$) and it is observed that $\rho_{xx}$ deviates considerably from $T^\frac{3}{2}$ for $y=0.017,0.021$ and $0.027$ (it roughly shows a T-linear behaviour) whereas for $y=0.033$ it seems to obey $T^\frac{3}{2}$ 
The Hall angle cot$\theta_H$ follows the $T^2$ law (Fig.2 of \cite{rosenbaum1998temperature}). The authors rationalize the temperature dependence of $\rho_{xx}$ and cot$\theta_H$ by suggesting two different relaxation rates (scattering mechanisms) for the carrier motion in parallel and transverse direction. Again, using the same set of arguments discussed in Section II (and in Appendix B), we stress that the carrier density $n$ is temperature dependent in this system also which influences the overall temperature dependence of $\rho_{xx}$.\ 

\begin{figure}[h!]
    \begin{subfigure}[b]{0.5\columnwidth}
    \includegraphics[width=\textwidth]{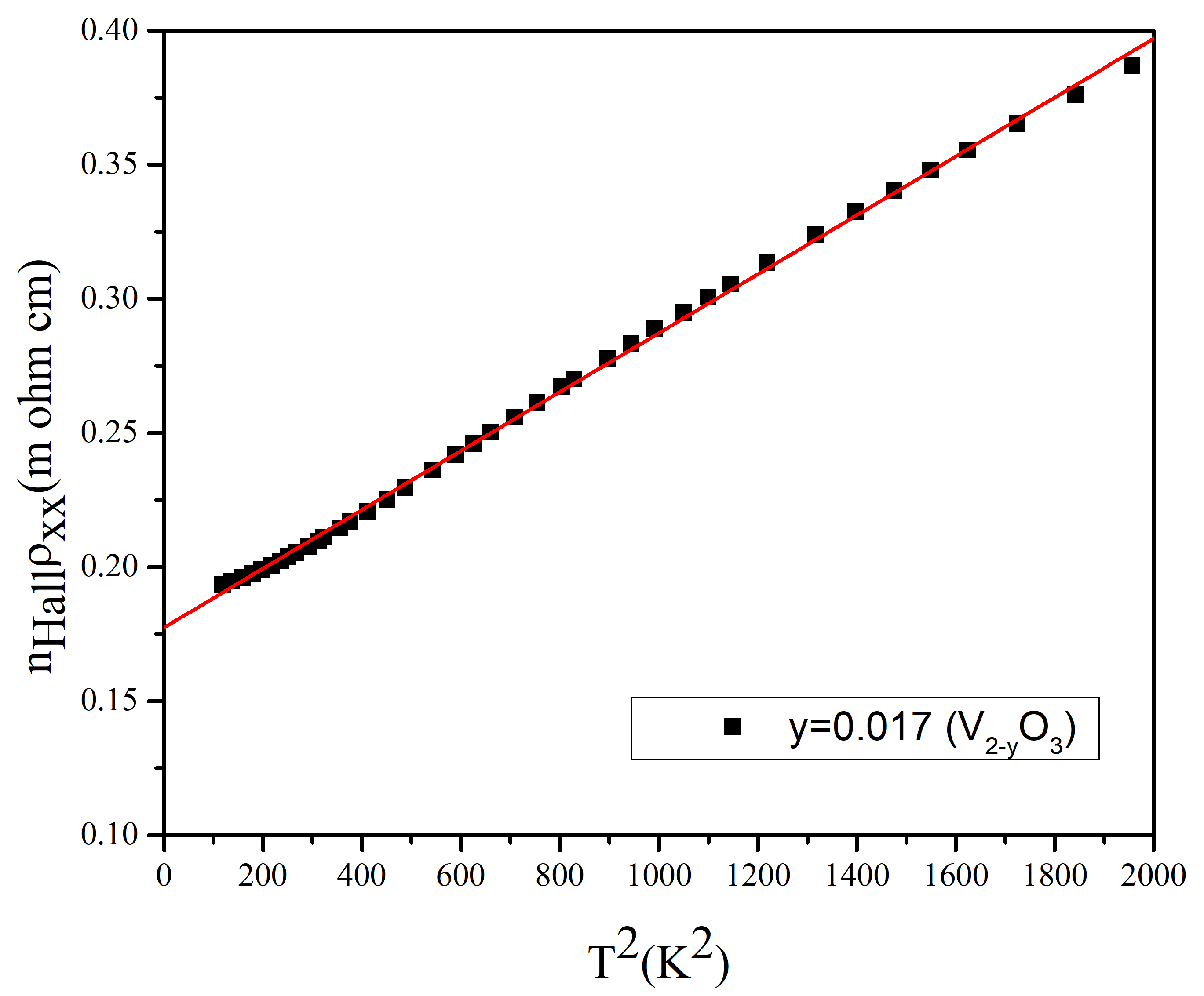}
    \caption{$y=0.017$}
  \end{subfigure}%
  \hfill
  \begin{subfigure}[b]{0.5\columnwidth}
    \includegraphics[width=\textwidth]{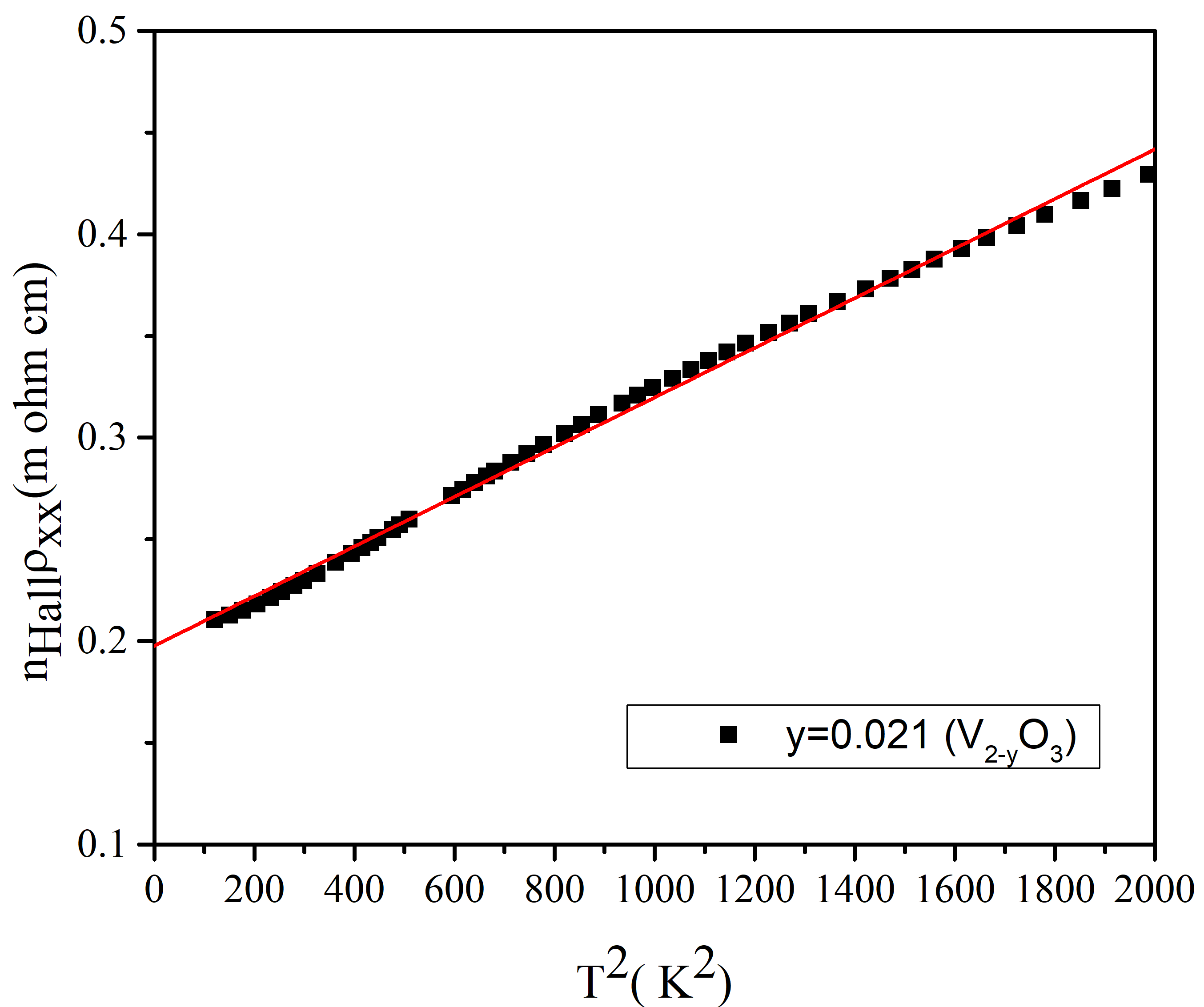}
    \caption{$y=0.021$}
  \end{subfigure}%
\\
\begin{subfigure}[b]{0.5\columnwidth}
    \includegraphics[width=\textwidth]{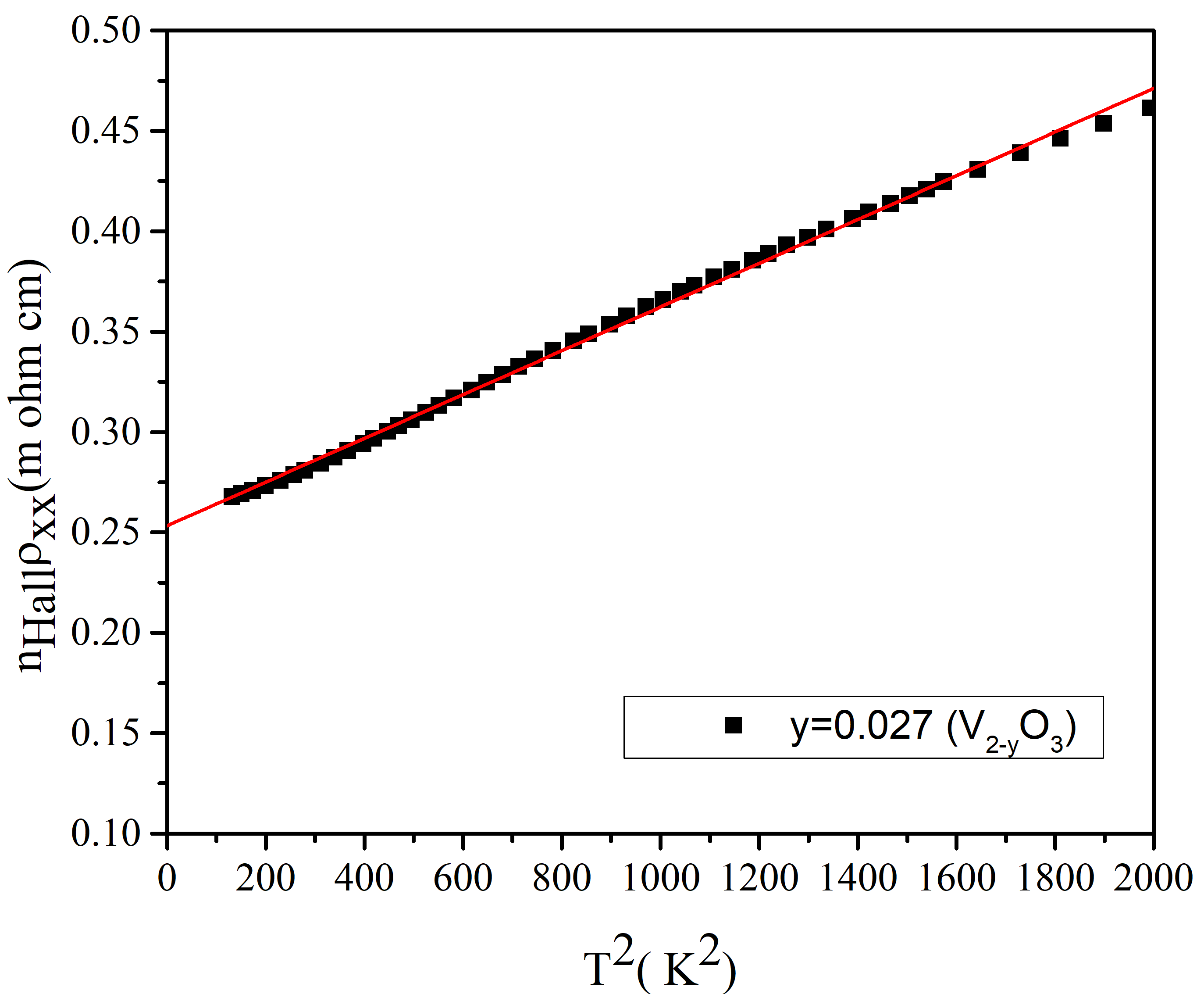}
    \caption{$y=0.027$}
  \end{subfigure}%
 \hfill
\begin{subfigure}[b]{0.5\columnwidth}
    \includegraphics[width=\textwidth]{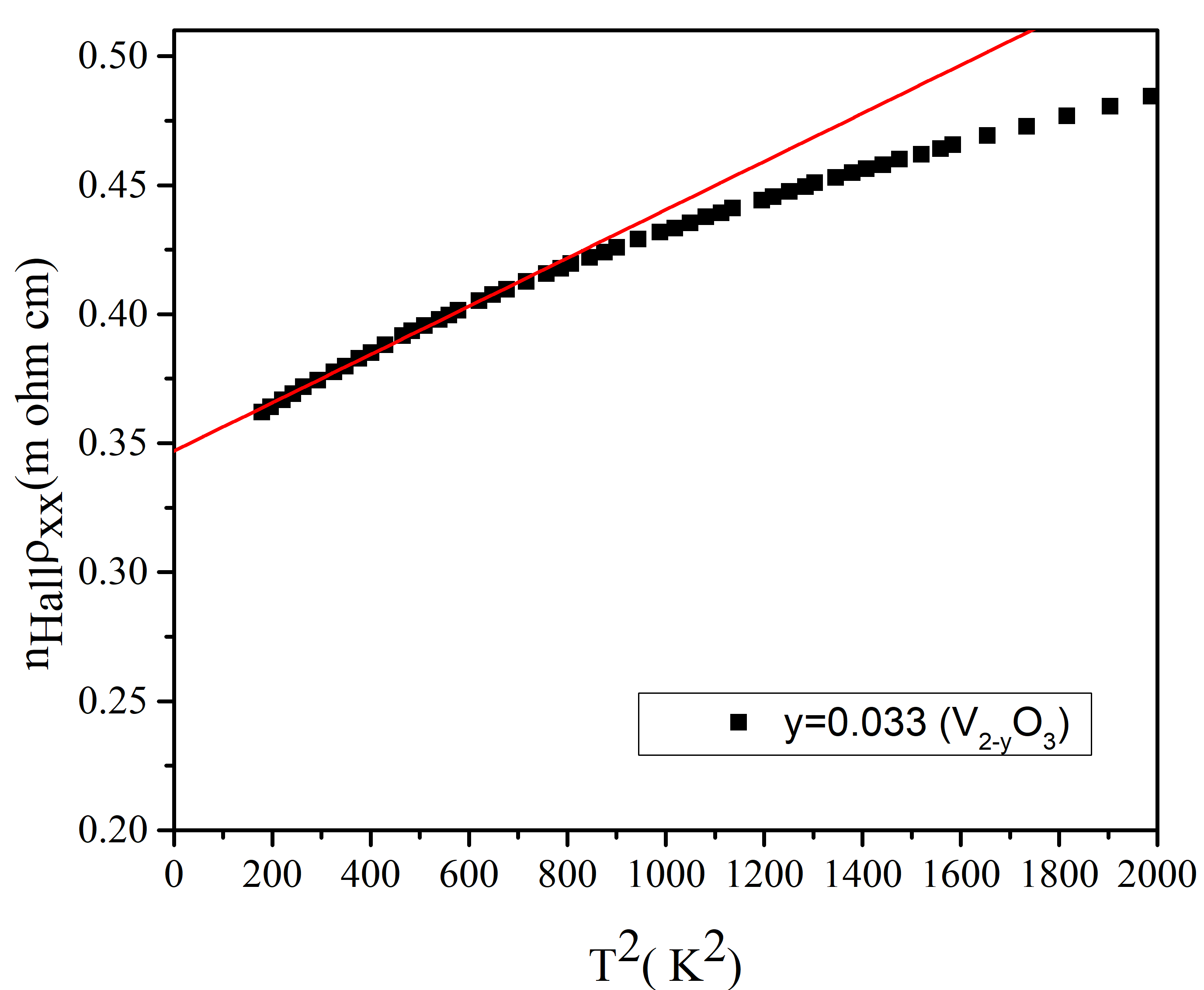}
    \caption{$y=0.033$}
  \end{subfigure}%
  \caption{\justifying The $\rho(y,T)n_{Hall}(y,T)$ (cot$\theta_H$) vs $T^2$ of $V_{2-y}O_3$. The $n_{Hall}(y,T)$ are obtained using the GTTA model by extracted $R_H$ data from Fig. 1 of \cite{rosenbaum1998temperature}. The resistivity $\rho(y,T)$ is extracted from Fig. 1 of \cite{rosenbaum1998temperature}.}
\label{VO-cot}
\end{figure}

We calculate temperature dependence cot$\theta_H$ $\propto$ $\rho(y,T)n_{Hall}(y,T)$ using the GTTA model. Fig. \ref{VO-cot} displays the plot of $\rho(y,T)n_{Hall}(y,T)$ vs $T^2$ for various concentration of vanadium deficiency $y$. The calculated cot$\theta_H$ obeys the $T^2$ law for $y=0.017,0.021$ and $0.027$. Similar dependency of cot$\theta_H$ on temperature is observed from the experimental measurements (Fig. 2 of \cite{rosenbaum1998temperature}). Thus, the argument that there is only one relaxation rate for $V_{2-y}O_3$ is true. As seen in Fig. \ref{VO-cot}(d), for $y=0.033$ at which $R_H$ is roughly constant with temperature \cite{rosenbaum1998temperature}, deviates considerably from $T^2$ at roughly $\sim30K$ which is also observed in Fig. 2 of \cite{rosenbaum1998temperature}. This behaviour of cot$\theta_H$ for $y=0.033$ can be justified from its PG crossover value ($\Delta(y)\sim44K$). Below this temperature, the phenomena of ``tying down" of electrons takes place, which progressively increases (i.e. charge carriers decrease) as the temperature is further reduced. Above the PG crossover temperature the magnetic correlations rapidly vanishes and electrons are no longer ``tied down". Thus, $V_{1.67}O_3$ behaves like a metal (beyond $\Delta(y)\sim44K$) which is also reflected in the Hall coefficient plot (Fig. 1 of \cite{rosenbaum1998temperature}) wherein $R_H(T)$ is roughly constant. Hence cot$\theta_H$ at $y=0.033$ does not show $T^2$ dependence beyond $\sim30K$. The downward deviation of $\rho(y,T)n_{Hall}(y,T)$ for $y=0.033$ indicates that the temperature dependence of cot$\theta_H$ moves towards the lower power of T.\

Thus, the temperature dependence of resistivity $\rho_{xx}$ and Hall angle cot$\theta_H$ can be rationalized by only one relaxation rate which scales as $T^2$. It is the temperature dependent carrier density $n(T)$ which makes the resistivity $\rho_{xx}$ scale differently.

\section{\label{sec:level1}Case of $La_{2-x}Sr_xCuO_4$}
The T-linear dependence of in-plane resistivity $\rho$ in $La_{2-x}Sr_xCuO_4$ was reported by Nakano et al \cite{nakano1994magnetic}. In 2004, Ando et al reported a detailed $T^2$ behaviour of cot$\theta_H$ for $YBa_2Cu_3O_{6.30}$ and $La_{2-x}Sr_xCuO_4$. The linear cot$\theta_H vs T^2$ behaviour was observed for $0.2\leq x\leq0.16$ in $La_{2-x}Sr_xCuO_4$ from $\sim 75K$ till room temperature ($T\simeq 300$) \cite{ando2004evolution}. In 2007, Ando et al extended the temperature range for cot$\theta_H$ measurements to 1000K for $La_{2-x}Sr_xCuO_4$ which indicated a breakdown of $T^2$ law at higher temperatures ($\sim450K$) \cite{ono2007strong}. The quadratic temperature dependence of the inverse Hall mobility of $La_{2-x}Sr_xCuO_4$ for low dopings ($x\leq0.15$) was reported in 1994 by Beschoten et al \cite{beschoten1994fermion}. This was advocated as an evidence of Fermion-Fermion scattering in the material. For larger doping the hall mobility inverse did not obey the $T^2$ law \cite{beschoten1994fermion}. \ 

We use the Hall coefficient and resistivity data reported in \cite{ono2007strong} for our analysis. $R_H$ values for different $x$ is extracted from Fig 1. of \cite{ono2007strong} and is fitted to the GTTA model. This $R_H(T)$ was also analyzed by Gor'kov and Teitel'baum \cite{gor2006interplay}. The extracted $\Delta(x)$ were in good agreement with the experimental PG signatures obtained by ARPES measurements \cite{yoshida2003metallic,yoshida2006systematic}. Our extracted parameter values are also in agreement with the values reported in \cite{gor2006interplay}. The doping dependence of $n_0(x)$, $n_1(x)$ and $\Delta(x)$ for $La_{2-x}Sr_xCuO_4$ calculated are reported in \cite{pandya2024pseudogap}. \

The idea of ``two-relaxation times" was formulated by Anderson to essentially address the anomalous transport properties of cuprates (discussed in Appendix B). However, as argued in section II and section III, it is the temperature dependent $n$ that makes the in-plane resistivity $\rho(x,T)$ T-linear ($\tau \propto T^2$). Thus, we calculate $\rho(x,T)n_{Hall}(x,T)$ for $La_{2-x}Sr_xCuO_4$ to understand the temperature dependence of cot$\theta_H$ using the GTTA model. $\rho(x,T)$  is extracted from Fig. 7 of \cite{ono2007strong} which depicts the evolution of in-plane resistivity with temperature.
\
\begin{figure}[h!]
    \begin{subfigure}[b]{0.5\columnwidth}
    \includegraphics[width=\textwidth]{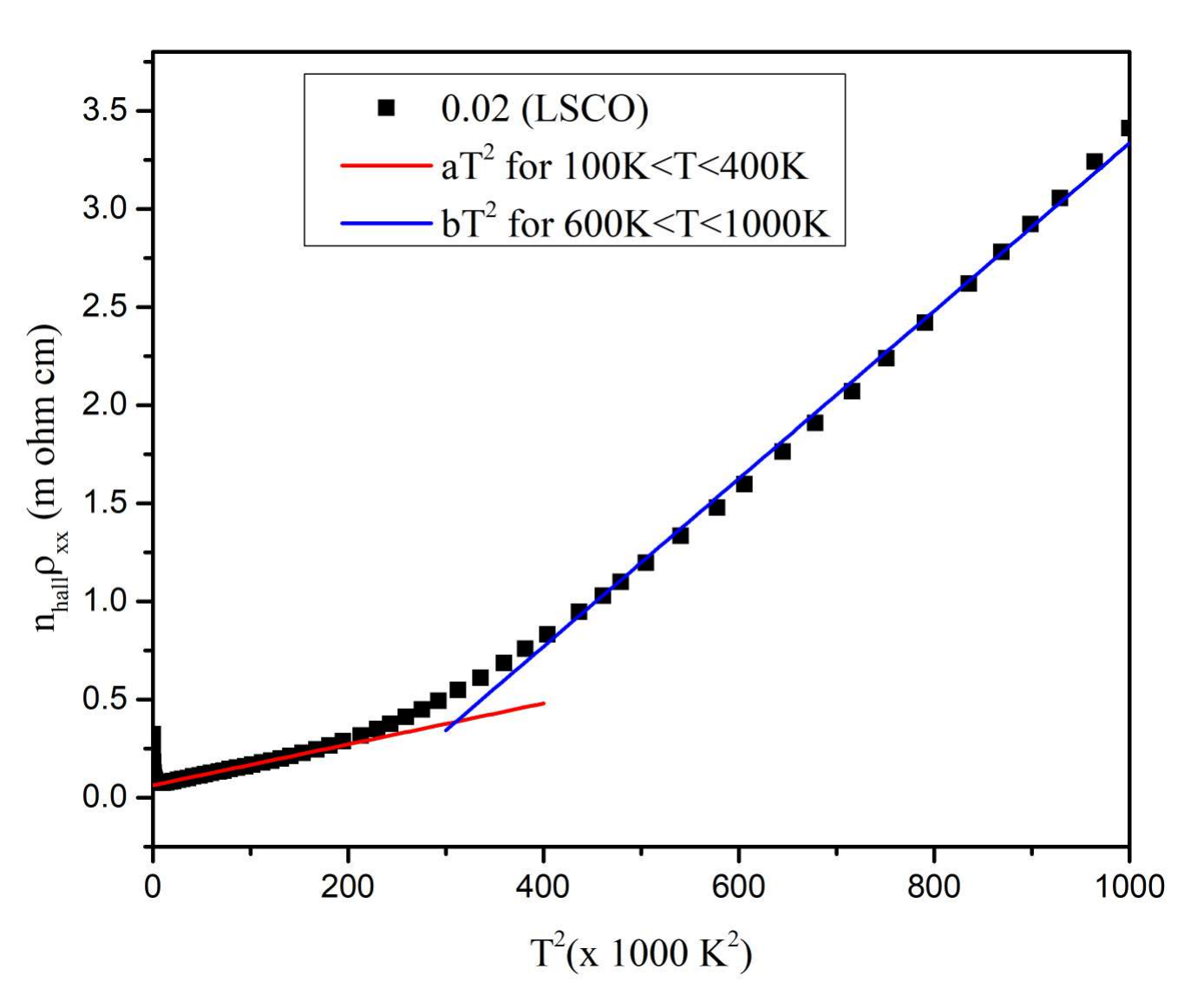}
    \caption{$x=0.02$}
  \end{subfigure}%
  \hfill
  \begin{subfigure}[b]{0.5\columnwidth}
    \includegraphics[width=\textwidth]{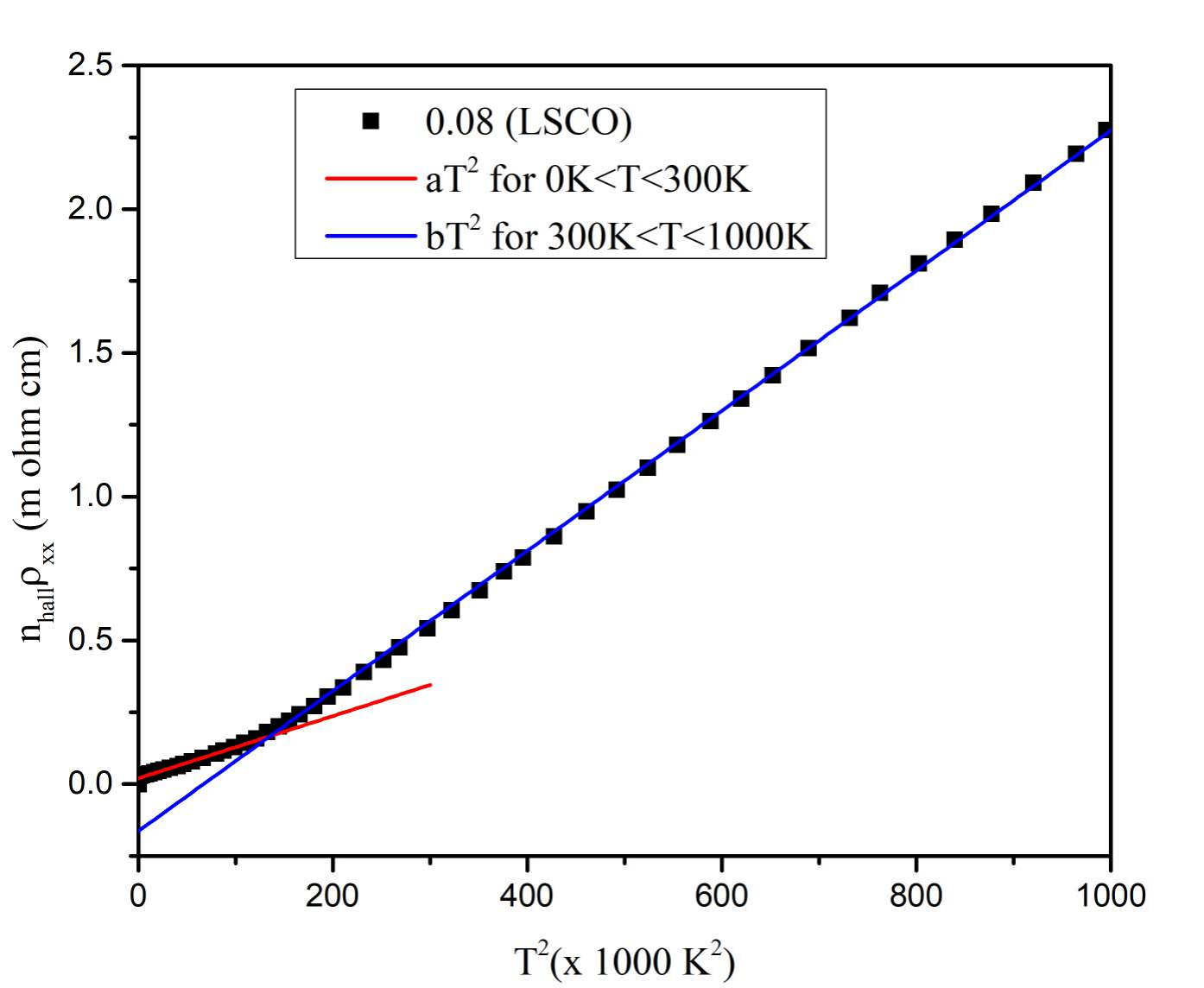}
    \caption{$x=0.08$}
  \end{subfigure}%

\medskip
\begin{subfigure}[b]{0.5\columnwidth}
    \includegraphics[width=\textwidth]{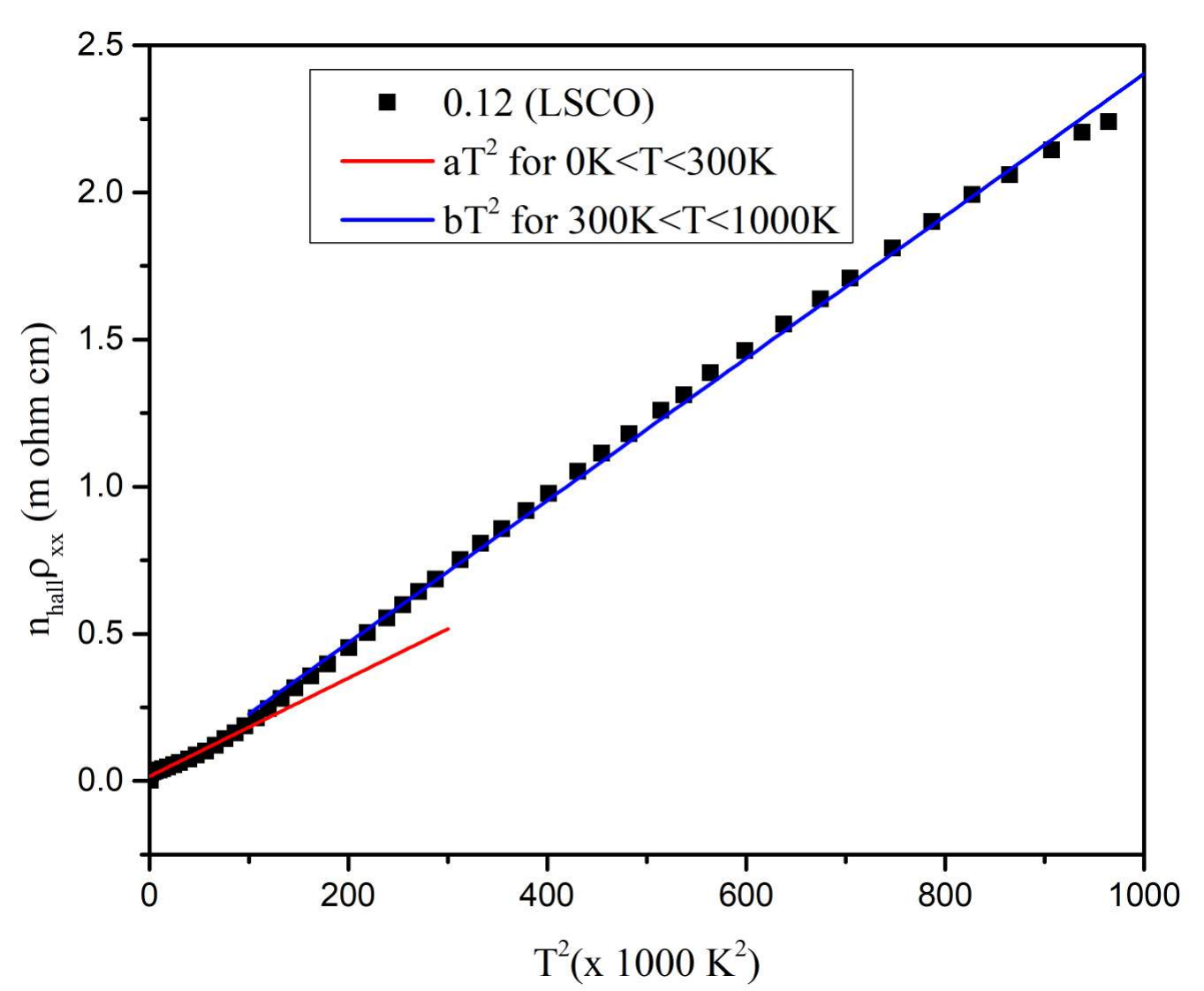}
    \caption{$x=0.12$}
  \end{subfigure}%
 \hfill
\begin{subfigure}[b]{0.5\columnwidth}
    \includegraphics[width=\textwidth]{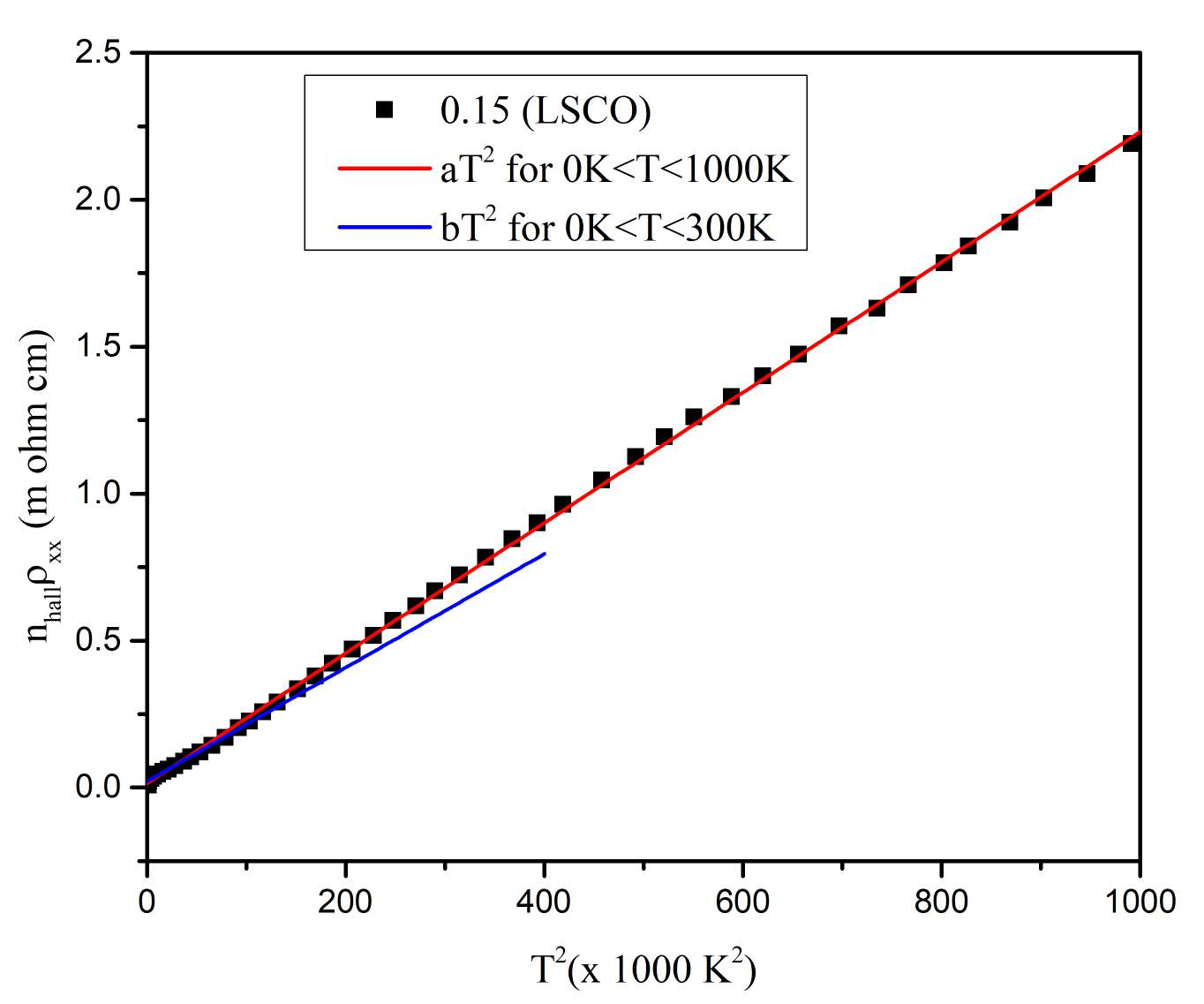}
    \caption{$x=0.15$}
  \end{subfigure}%
  \hfill
\begin{subfigure}[b]{0.5\columnwidth}
    \includegraphics[width=\textwidth]{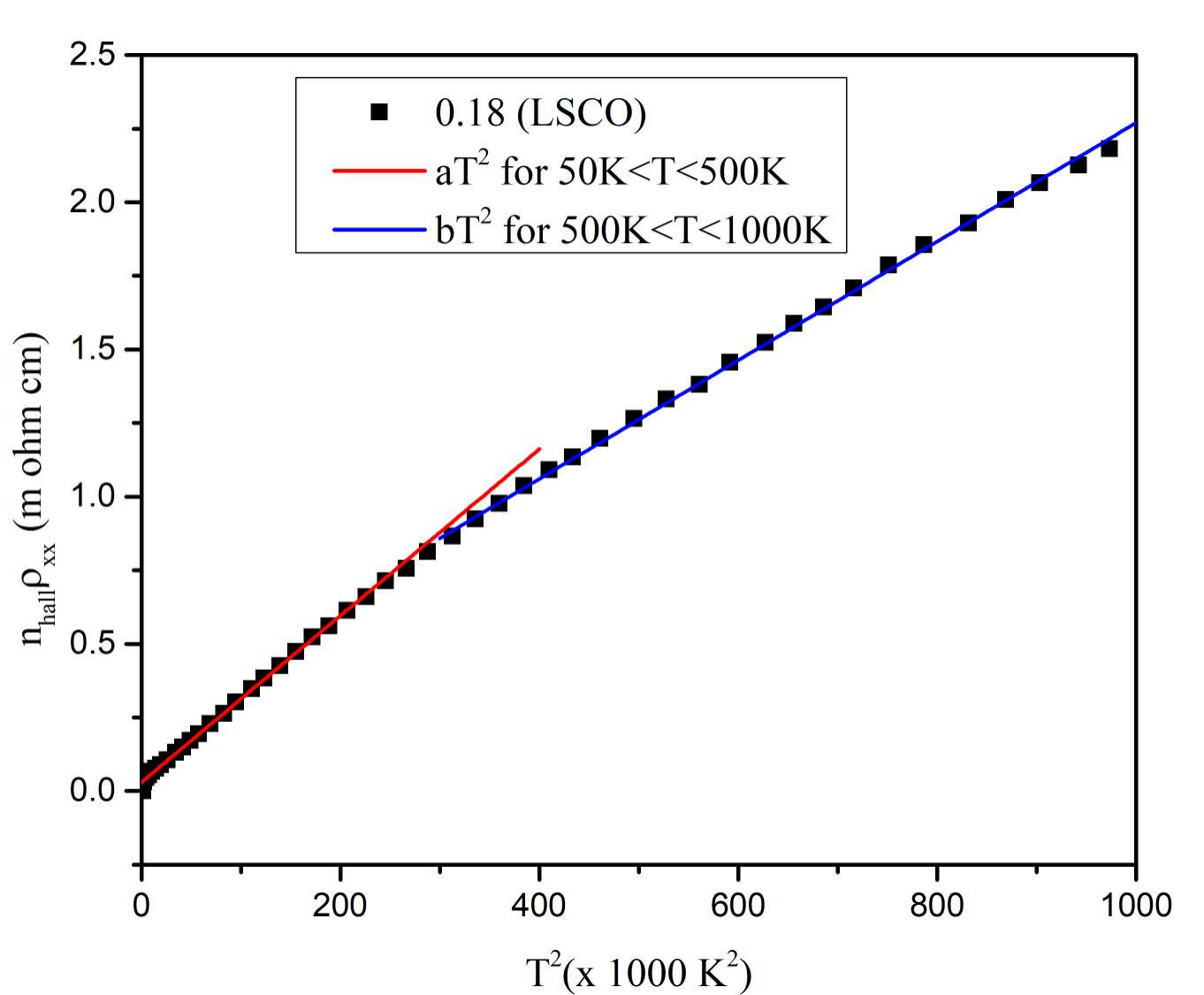}
    \caption{$x=0.18$}
    \end{subfigure}%
    \centering
  \caption{\justifying The $\rho(x,T)n_{Hall}(x,T)$ (cot$\theta_H$) vs $T^2$ of $La_{2-x}Sr_xCuO_4$. The $n_{Hall}(y,T)$ are obtained using the GTTA model by extracted $R_H$ data from Fig. 1 of \cite{ono2007strong}. The resistivity $\rho(y,T)$ is extracted from Fig. 7 of \cite{ono2007strong}. Notice the break in the slopes for underdoped samples which corresponds to a structural phase change.}
\label{LSCO-cot}
\end{figure}

The plots of $\rho(x,T)n_{Hall}(x,T)$ vs $T^2$ for various doping concentrations are displayed in Fig. \ref{LSCO-cot}. For $x=0.02$, as observed in Fig. \ref{LSCO-cot}(a), $\rho(x,T)n_{Hall}(x,T)$ is not $T^2$ dependent in the entire temperature range ($0K\leq T\leq1000K$). However, we see two separate regions where the plot is linear. At low temperatures (below $\sim400K$), $aT^2$ fits reasonably well with the data points. Similarly, a good fit with $bT^2$ function is seen for $T\geq600K$. This behaviour of $\rho(x,T)n_{Hall}(x,T)$ with respect to temperature is similar to the one observed in the experimentally measured cot$\theta_H$ (refer to Fig. 8a of \cite{ono2007strong}). The temperature where $\rho(x,T)n_{Hall}(x,T)$ breaks the $T^2$ law indicates some sort of phase change taking place in the system. For x=0.02, this phase change is observed at $\sim558K$ (temperature at which the fitted lines, $aT^2$ and $bT^2$, intersect) which is in very good agreement with the temperature at which $La_{2-x}Sr_xCuO_4$ undergoes a structural transition from orthorhombic to tetragonal ($473K$ for $x=0.02$) \cite{takagi1992disappearance}. \ 

At doping concentration $x=0.08$, $\rho(x,T)n_{Hall}(x,T)$ vs $T^2$ plot (Fig. \ref{LSCO-cot}(b)) shows behaviour similar to the one observed for $x=0.02$. The temperature range where $\rho(x,T)n_{Hall}(x,T)$ is not $T^2$ dependent is considerably less for this doping compared to $x=0.02$. For $x=0.08$, the structural phase change temperature is  $\sim372K$, which again matches considerably well with the experimental value ($385K$) \cite{takagi1992disappearance}. The $\rho(x,T)n_{Hall}(x,T)$ vs $T^2$ plot for $x=0.08$ obtained using the GTTA model is also similar to the experimentally obtained cot$\theta_H$ (refer to Fig. 8b of \cite{ono2007strong}). \

In case of $x=0.12$, a moderate $T^2$ dependence is seen in the entire temperature range. On careful analysis, a phase transformation is observed at $\sim207K$ which is again in good agreement with the structural phase change temperature reported in \cite{takagi1992disappearance}. At optimal doping i.e. $x=0.15$, an excellent $T^2$ fit can be seen in the entire temperature range. However, a small deviation from $T^2$ is noticed at $\sim137K$. This deviation from $T^2$ dependence is also observed in the experimentally obtained cot$\theta_H$ reported in Fig. 8c of \cite{ono2007strong}. As we move to the overdoped region , the $T^2$ dependency of $\rho(x,T)n_{Hall}(x,T)$ vanishes. Fig. \ref{LSCO-cot}(e) displays $\rho(x,T)n_{Hall}(x,T)$ vs $T^2$ plot at $x=0.18$. A complete breakdown of $T^2$ is clearly observed. \

\begin{figure}[h!]
    \includegraphics[width=1.0\linewidth]{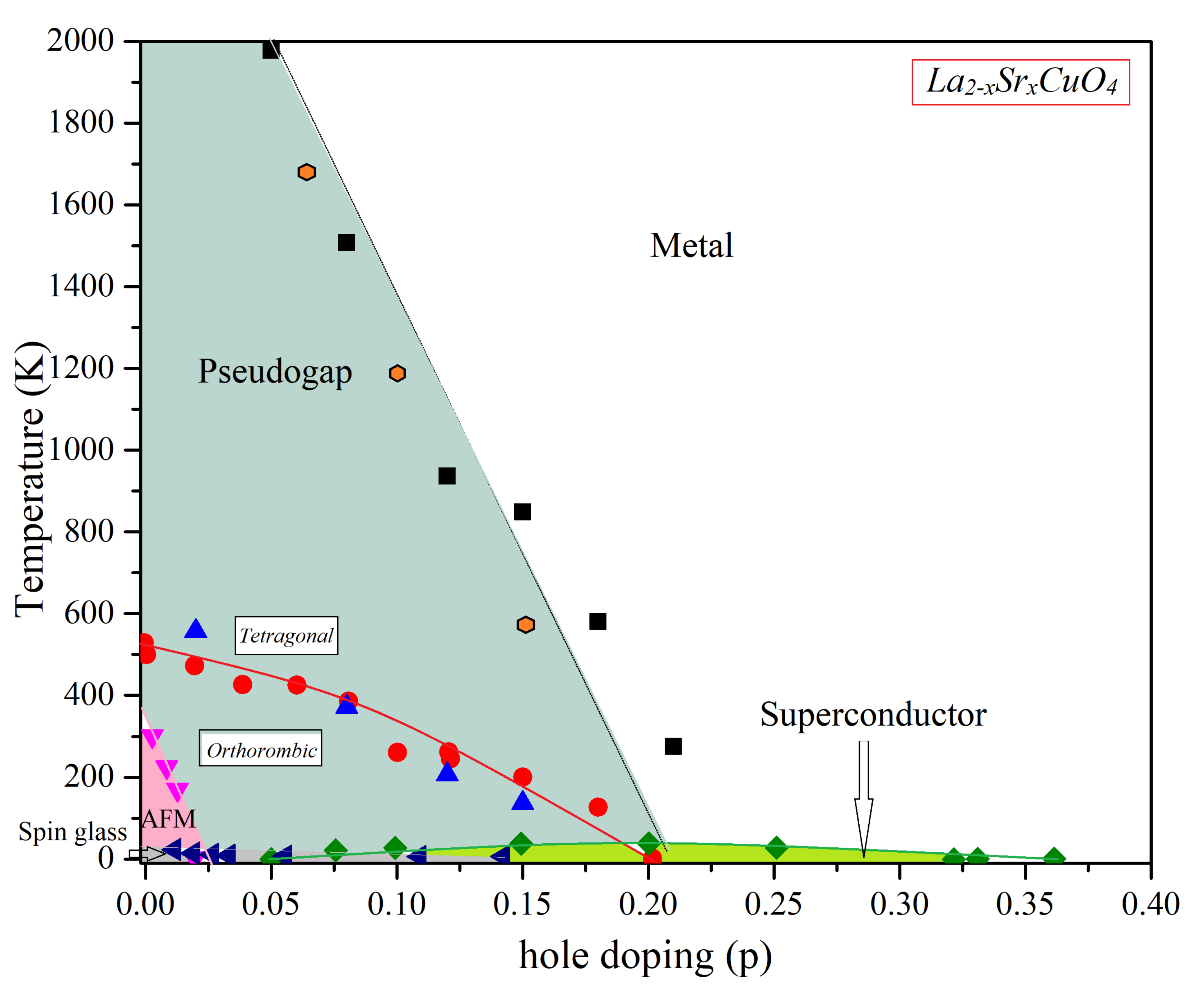}
    \caption{\justifying Phase diagram for $La_{2-x}Sr_xCuO_4$. The black squares represents the extracted $\Delta(x)$ from GTTA model. The orange hexagons represents the PG values obtained by ARPES measurements \cite{yoshida2003metallic,yoshida2006systematic}. We notice an excellent agreement between the two. The phase change obtained from the $\rho(x,T)n_{Hall}(x,T)$ vs $T^2$ plots are marked as blue triangles. The red solid circle shows the structural phase change crossover extracted from Fig. 4 of \cite{takagi1992disappearance}. Again, an excellent agreement between the two is observed. The other data points are re-plotted from Fig. 4 of \cite{birgeneau1990richtmyer}.}
\label{LSCO-phase}
\end{figure}

As discussed, a structural phase change is observed in the $\rho(x,T)n_{Hall}(x,T)$ (cot$\theta_H$) vs $T^2$ plots wherein the structure $La_{1-x}Sr_xCuO_4$ changes from orthogonal to tetragonal. The temperatures at which this phase change takes place decreases as the doping concentration increases and are in excellent agreement with the experiments \cite{takagi1992disappearance}. From this analysis of $La_{1-x}Sr_xCuO_4$ (using the GTTA model) an updated phase diagram of the material is drawn and displayed in Fig. \ref{LSCO-phase}. The structural phase boundary obtained by GTTA is in agreement with the one reported in \cite{takagi1992disappearance}. The PG crossover obtained by extracting $\Delta(x)$ using the GTTA model is also marked in Fig. \ref{LSCO-phase} which are in agreement with ARPES measurements \cite{yoshida2003metallic,yoshida2006systematic}.

\section{\label{sec:level1}Case of $YBa_2Cu_3O_{6+\delta}$}
$YBa_2Cu_3O_{6+\delta}$ is another high-$T_c$ superconductor whose transport property has been extensively studied \cite{segawa2004intrinsic,ito1993systematic,alexandrov2004hall}. Chein et al reported the in-plane resistivity, Hall coefficient and Hall angle measurements on Zn doped $YBa_2Cu_3O_7$ \cite{chien1991effect}. The Hall angle data (cot$\theta_H$) when plotted with respect to $T^2$ fell on a straight line. This $T^2$ dependence of cot$\theta_H$ was also observed in Pr doped $YBa_2Cu_3O_7$ by Jiang et al \cite{jiang1992hall}. In $YBa_2Cu_3O_{6+\delta}$, the resistivity deviates considerably from its T linear behaviour for small $\delta$ \cite{ito1993systematic}. The Hall response of $YBa_2Cu_3O_{6+\delta}$ was also measured by Segawa and Ando in 2004 \cite{segawa2004intrinsic}. The cot$\theta_H$ for $YBa_2Cu_3O_{6+\delta}$ fits well with $AT^2$ function for $\delta\geq0.85$ whereas for $\delta\leq0.75$, it fits as ($B+CT^\alpha$) for $\alpha > 2$ \cite{segawa2004intrinsic}.  \

 The Hall effect of $YBa_2Cu_3O_{6+\delta}$ is also studied using the GTTA model in this work. The Hall coefficient data reported in Fig. 3 of \cite{segawa2004intrinsic} is extracted and $n_{Hall}(x,T)$ is calculated from equation (1) of Appendix A. The $n_0(x)$, $n_1(x)$ and $\Delta(x)$ is calculated for different oxygen concentrations. $n_1(x)$ is roughly constant ($\sim1.5$) till $\delta=0.95$ ($p=0.16$, the number of hole doped per Cu ($p$) is calculated from the law $\frac{T_c}{T_{cmax}}=1-82.6(p-0.16)^2$). \ 
 
\begin{figure}[h!]
 \centering
  \subfloat[$n_0(p)$]{\includegraphics[width=0.8\linewidth]{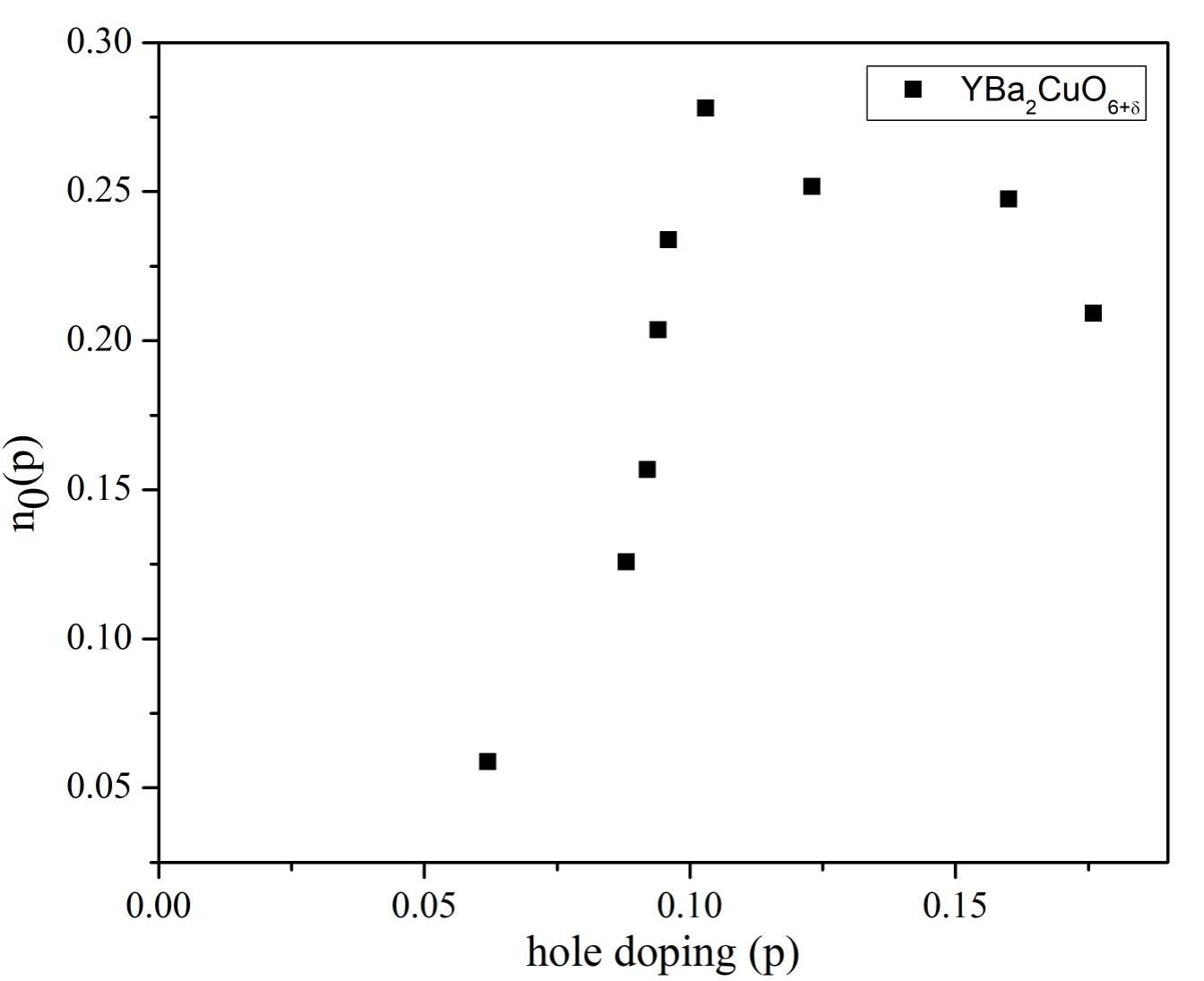}}\\
  \subfloat[$\Delta(p)$]{\includegraphics[width=0.8\linewidth]{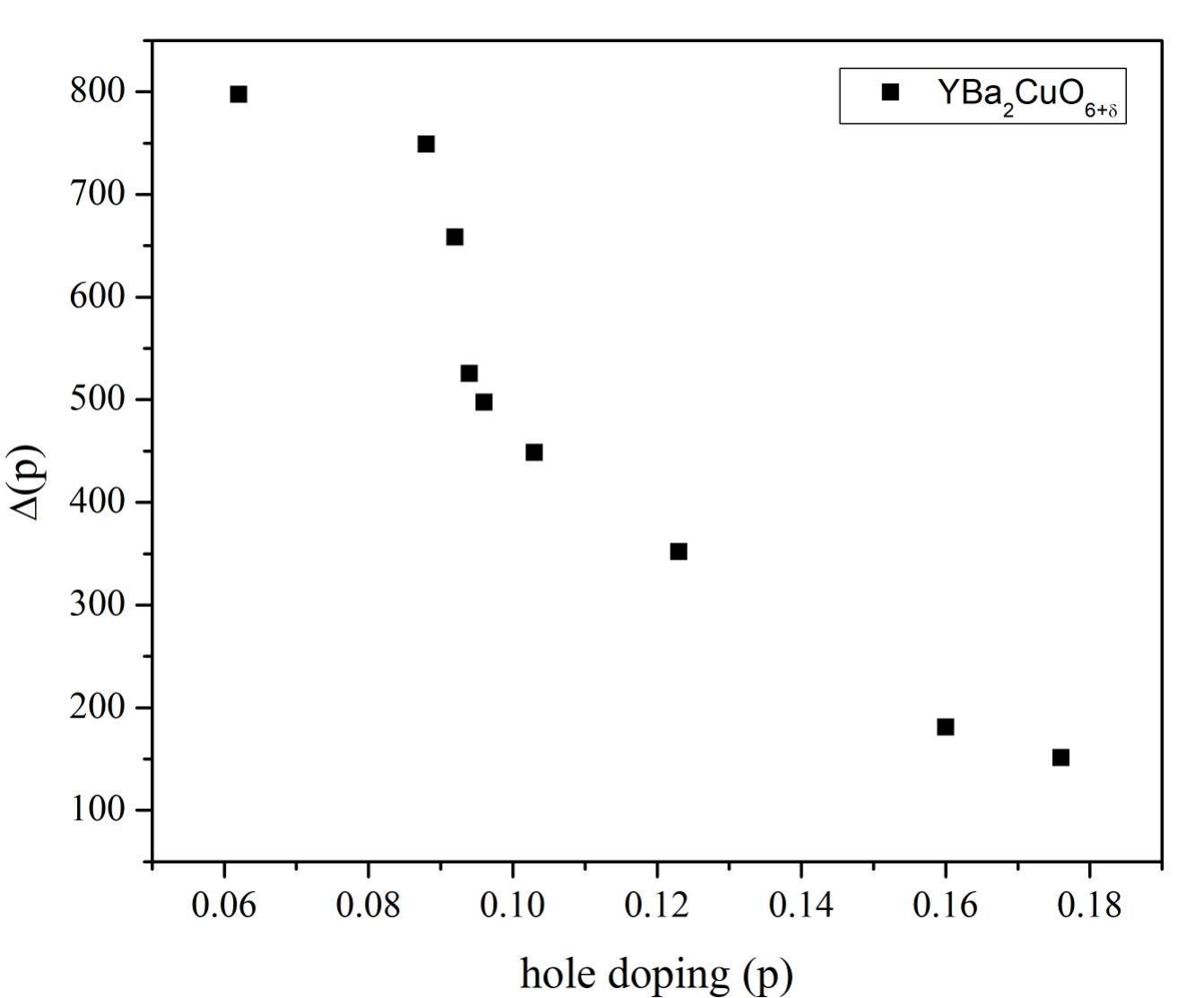}}
  \caption{\justifying The doping dependence of (a) $n_0(p)$ and (b) $\Delta(p)$ of $YBa_2Cu_3O_y$ obtained using the GTTA model from the $R_H(T)$ extracted from Fig. 3 of \cite{segawa2004intrinsic}.}
\label{YBCO-GTTA}
\end{figure}

 Moving on to the analysis of $n_0(p)$ and $\Delta(p)$, Fig. \ref{YBCO-GTTA} displays $n_0(p)$ and $\Delta(x)$ as a function of $p$. $n_0(p)$ increases rapidly as hole doping increases till $p\leq0.10$. For $p>0.10$ a sudden decrease in $n_0(p)$ is observed.  This decrease in $n_0(p)$ may be due to charge density waves present in this system. Charge density wave order of roughly $\sim20$ unit cell domain size is observed from x-ray scattering measurements in $YBa_2Cu_3O_{6+\delta}$ \cite{PhysRevLett.109.167001,PhysRevLett.110.187001}. A pronounced charge density wave is observed for doping concentration $p\sim0.12$ and it is not seen till $p\sim0.085$ \cite{blanco2014resonant}. \ 

\begin{figure}[h!]
    \includegraphics[width=1.0\linewidth]{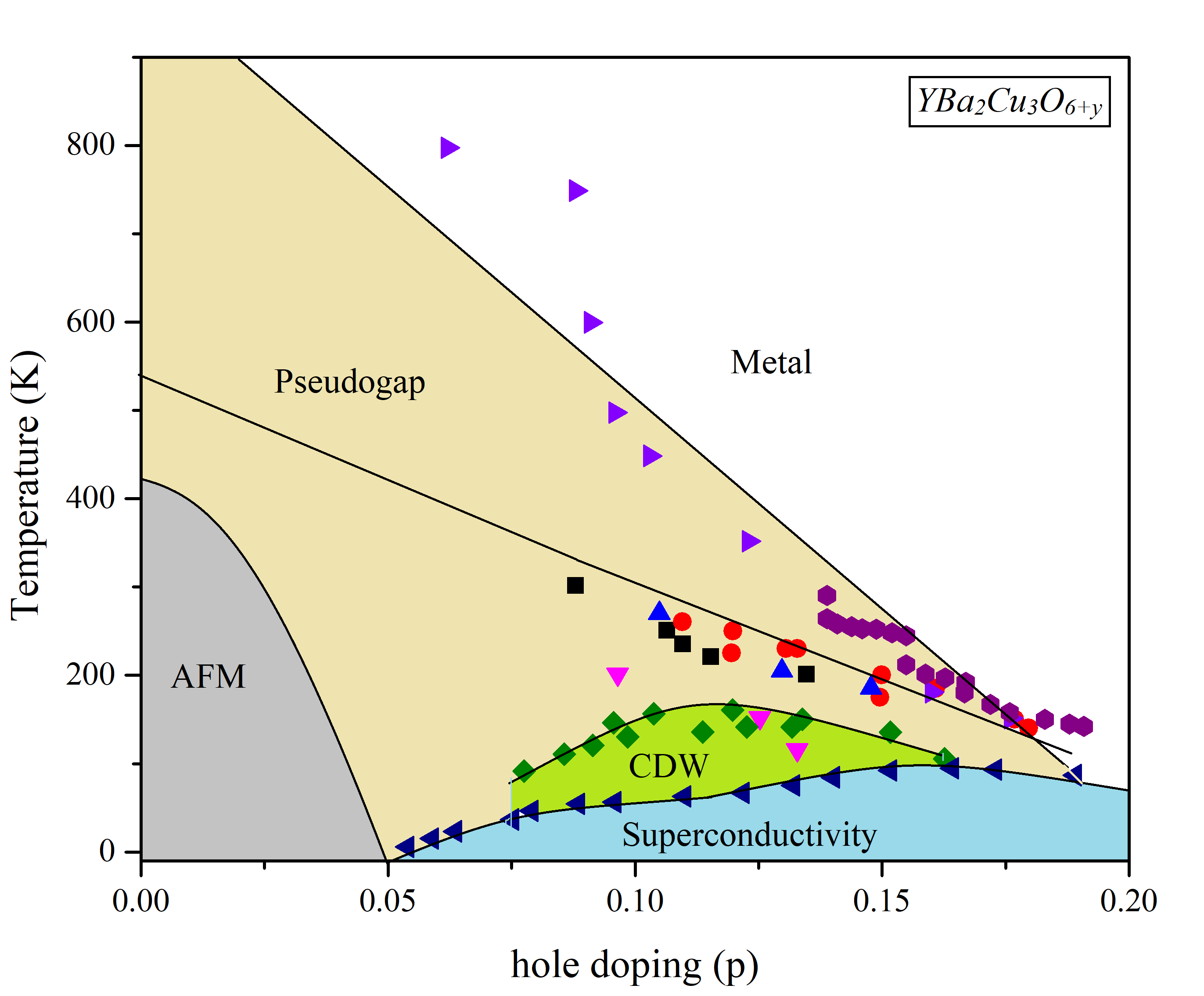}
    \caption{\justifying Phase diagram for $YBa_2Cu_3O_{6+\delta}$. The purple triangles represents the extracted $\Delta(p)$ from GTTA model. The violet hexagons represent the temperatures at which the second derivative of resistivity with respect to temperature is zero \cite{sacksteder2020quantized} i.e. the white line in Fig. 3. of \cite{ando2004condensed}. The other data points are re-plotted from Fig. 3 of \cite{sato2017thermodynamic}.}
  \label{YBCO-phase}
\end{figure}

 $\Delta(p)$, which corresponds to the PG crossover of the material, decreases roughly linearly as the hole doping is increased. Using $\Delta(p)$ extracted by GTTA model, an updated phase diagram for $YBa_2Cu_3O_{6+\delta}$ is drawn and shown in Fig. \ref{YBCO-phase}. $\Delta(p)$ obtained by GTTA model are marked in purple rectangles. This pseudogap crossover is in good agreement with the temperature at which the second temperature derivative of resistivity is zero \cite{sacksteder2020quantized}. It is extracted from the resistivity measurements performed by Ando et al \cite{ando2004condensed}. They performed the resistivity curvature mapping of the in-plane resistivity data for several cuprates \cite{ando2004condensed}. Such experiments on $YBa_2Cu_3O_{6+\delta}$ system have not been reported for temperatures beyond 300K. This may be due thermal history memory and a large step change in resistivity observed at higher temperatures \cite{sacksteder2020quantized}. However, PG crossover calculated from GTTA model is in line with the PG crossover obtained via resistivity curvature mapping \cite{ando2004condensed}. The data The short range charge density wave (CDW) and superconducting region in the phase diagram (Fig. \ref{YBCO-phase}) are reproduced from Fig. 3. of \cite{sato2017thermodynamic}. \

\begin{figure}[h!]
    \begin{subfigure}[b]{0.5\columnwidth}
    \includegraphics[width=\textwidth]{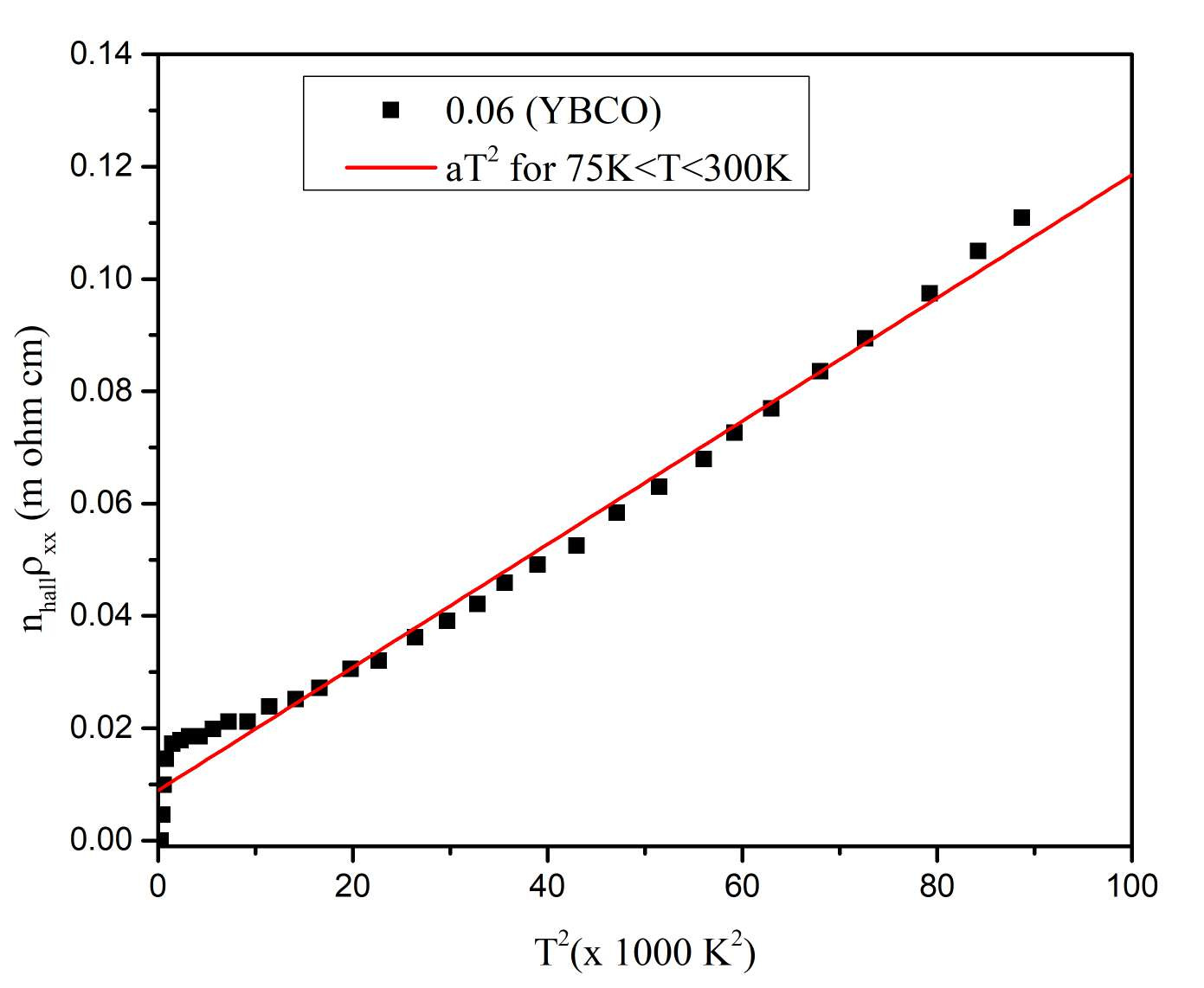}
    \caption{$p=0.06$}
  \end{subfigure}%
  \hfill
  \begin{subfigure}[b]{0.5\columnwidth}
    \includegraphics[width=\textwidth]{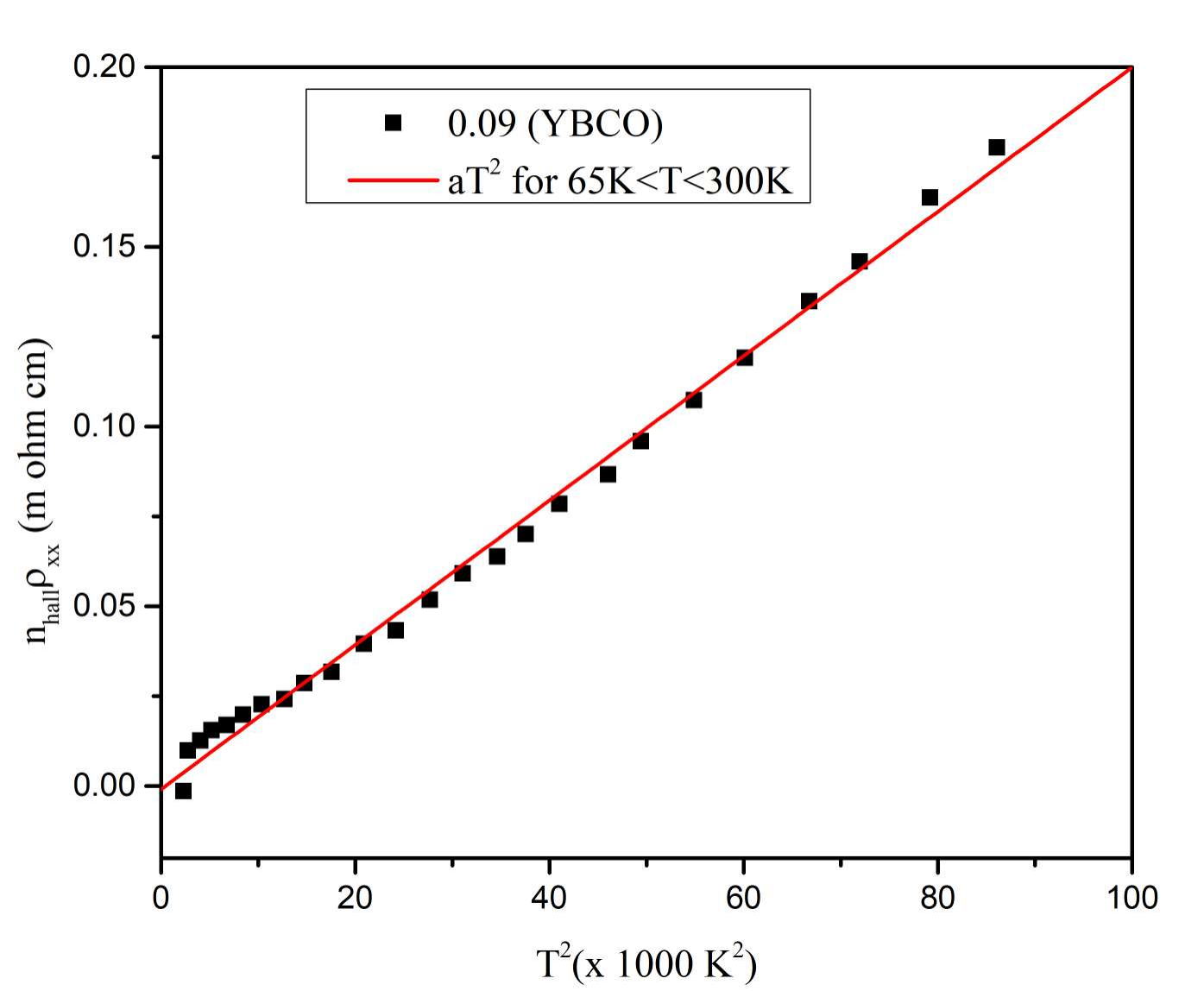}
    \caption{$p=0.09$}
  \end{subfigure}%

\medskip
\begin{subfigure}[b]{0.5\columnwidth}
    \includegraphics[width=\textwidth]{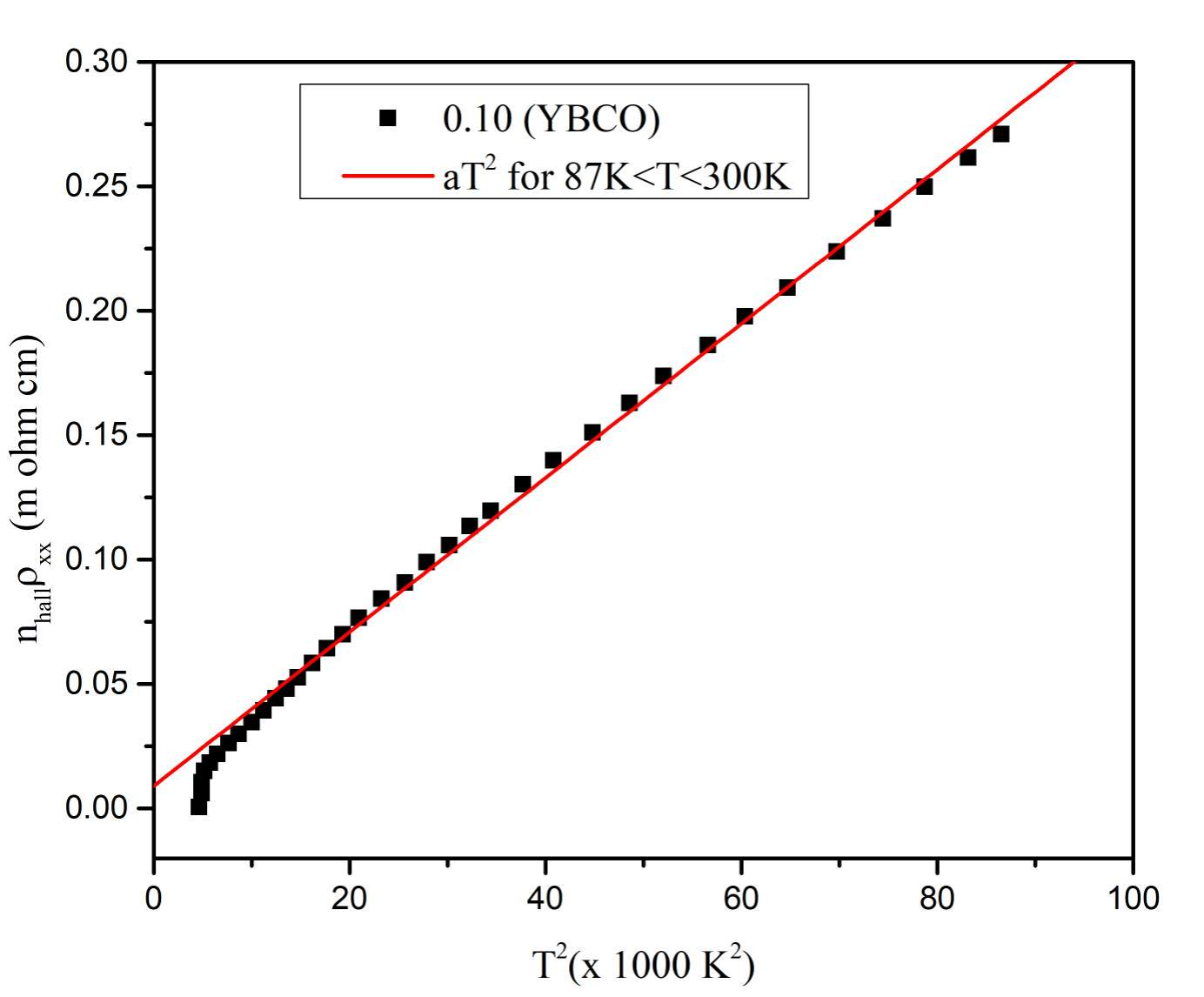}
    \caption{$p=0.10$}
  \end{subfigure}%
 \hfill
\begin{subfigure}[b]{0.5\columnwidth}
    \includegraphics[width=\textwidth]{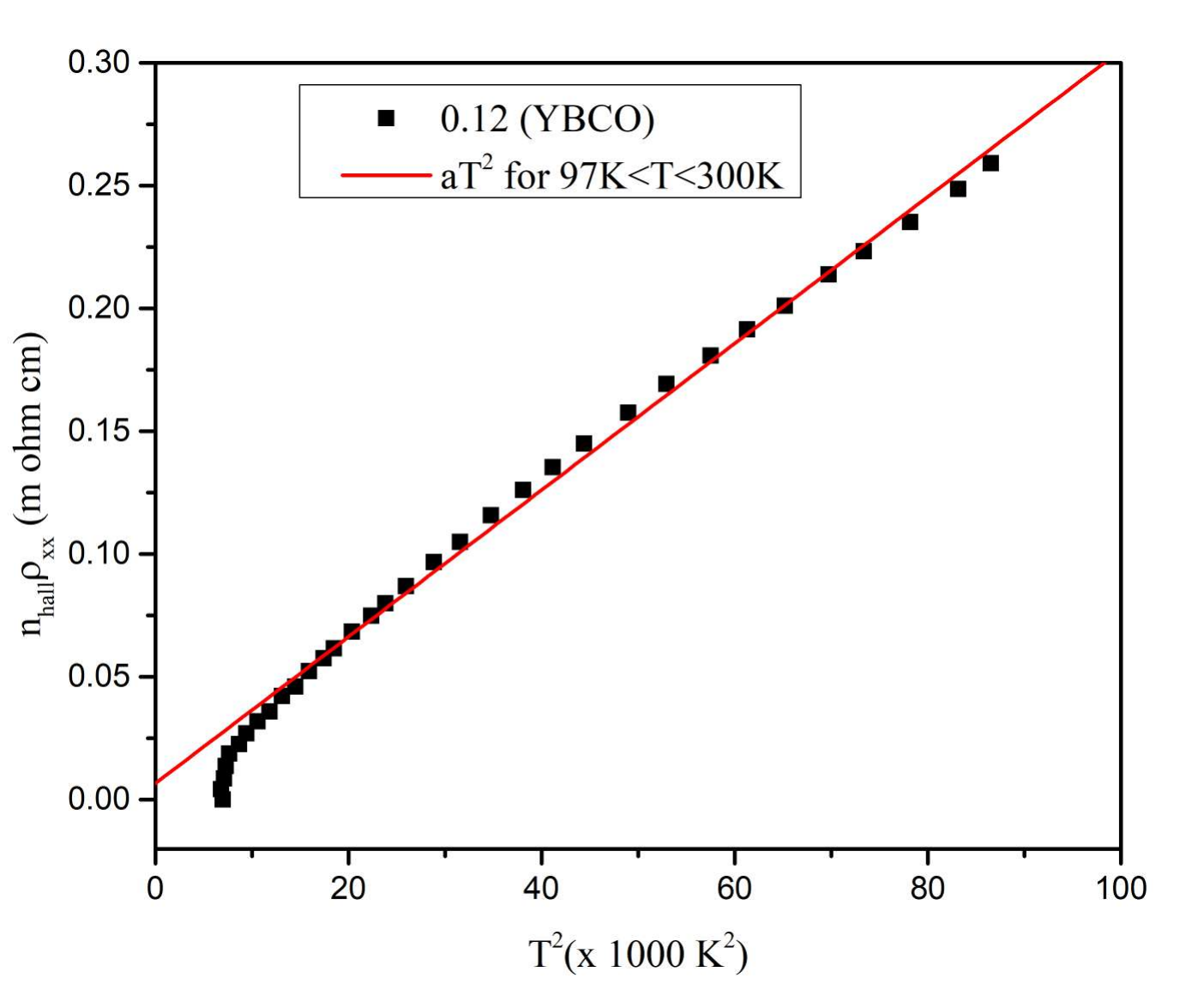}
    \caption{$p=0.12$}
  \end{subfigure}%

\medskip
\begin{subfigure}[b]{0.5\columnwidth}
    \includegraphics[width=\textwidth]{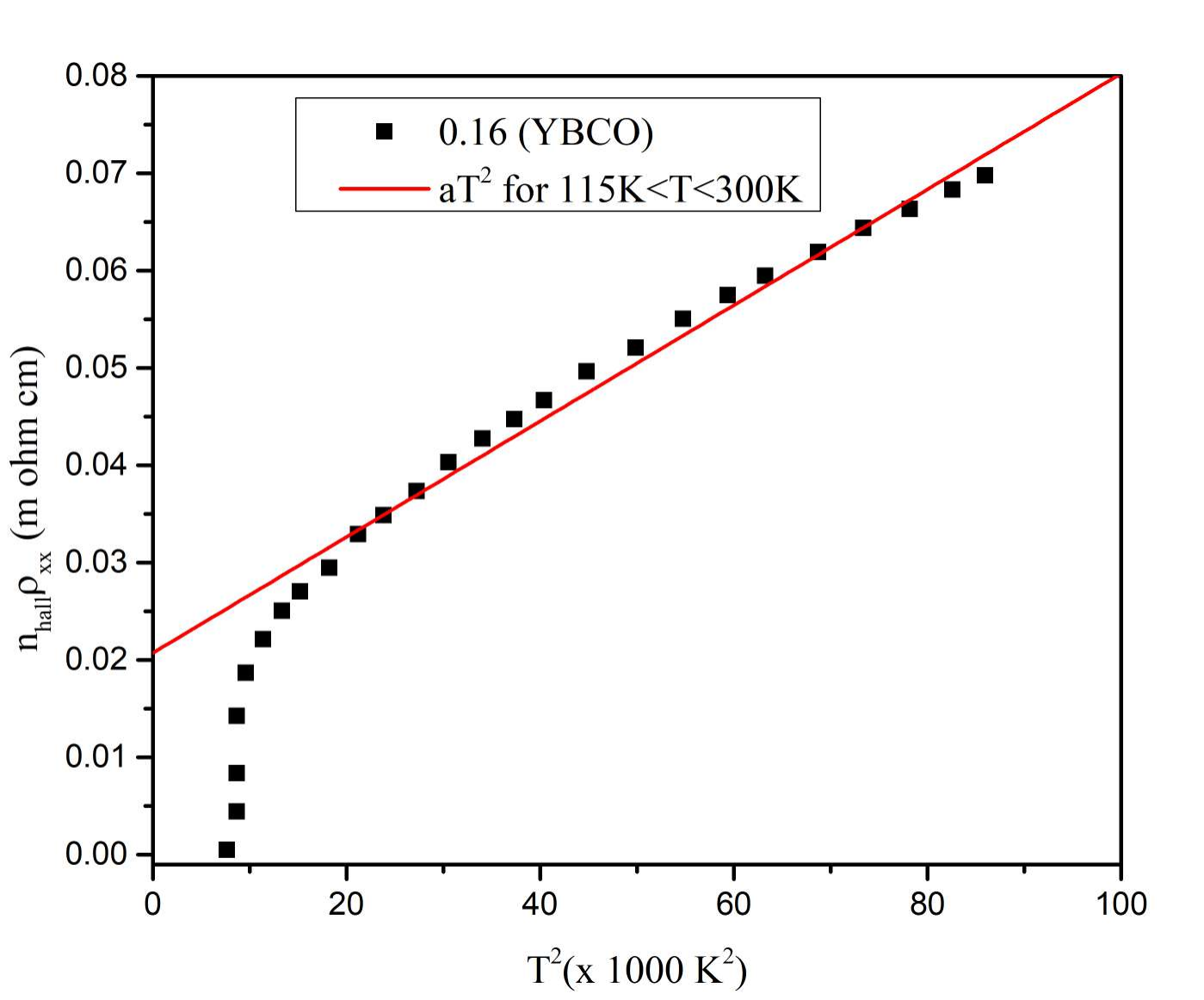}
    \caption{$p=0.16$}
    \end{subfigure}%
    \centering
\begin{subfigure}[b]{0.5\columnwidth}
    \includegraphics[width=\textwidth]{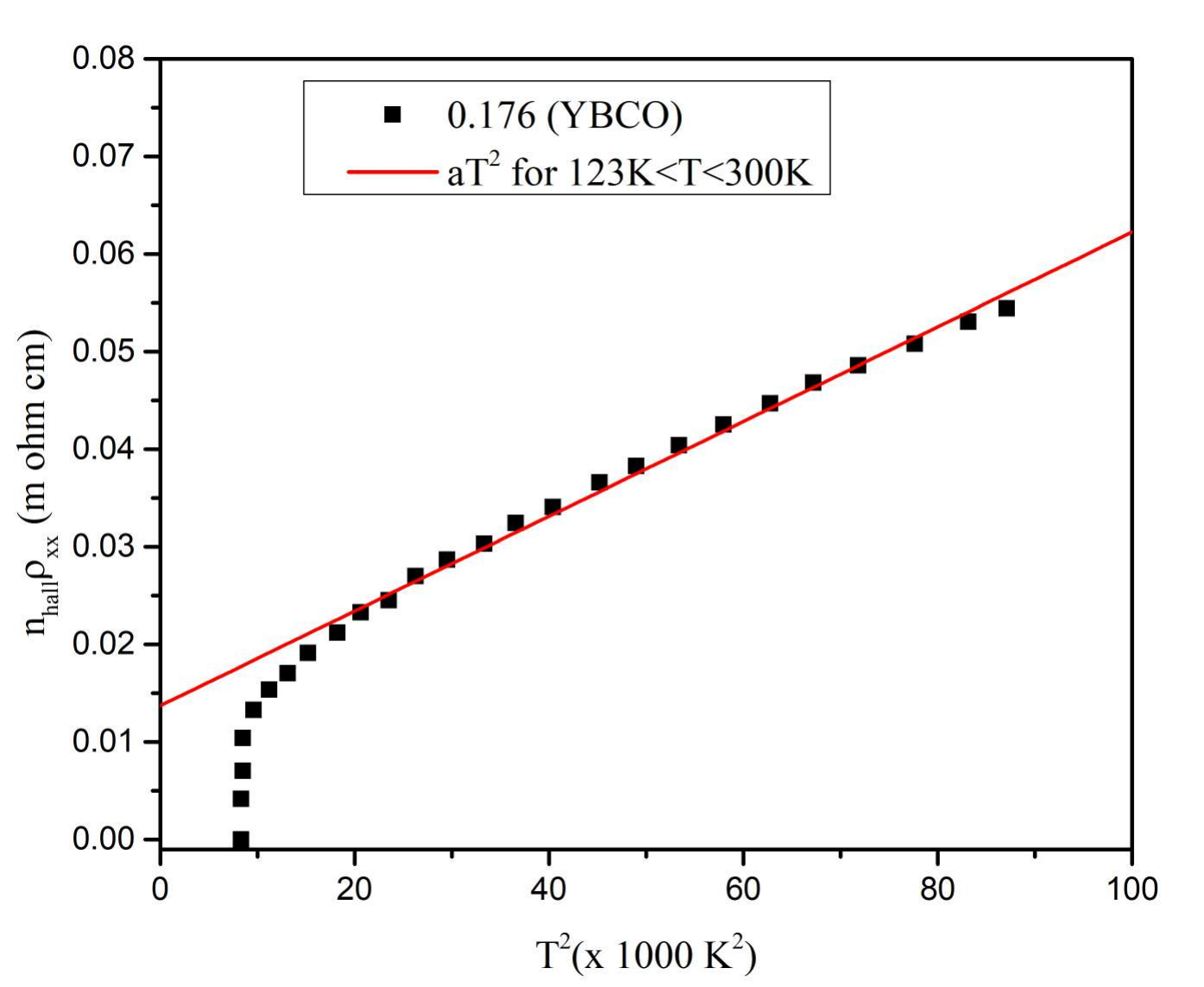}
    \caption{$p=0.176$}
    \end{subfigure}%
    \centering
  \caption{\justifying The $\rho(\delta,T)n_{Hall}(\delta,T)$ (cot$\theta_H$) vs $T^2$ of $YBa_2Cu_3O_{6+\delta}$. $n_{Hall}$ is calculated using the GTTA model by extracting $R_H$ of $YBa_2Cu_3O_{6+\delta}$ from Fig. 3 of \cite{segawa2004intrinsic}. $\rho(\delta,T)$ is extracted from Fig 3(a,b). of \cite{PhysRevLett.93.267001}. A reasonably good $T^2$ dependence of $\rho(\delta,T)n_{Hall}(\delta,T)$ is observed for hole doping concentrations $p=0.10$ and $p=0.12$.}
\label{YBCO-cot}
\end{figure}

The in-plane resistivity data for $YBa_2Cu_3O_{6+\delta}$ is extracted from Fig. 3(a,b) of \cite{PhysRevLett.93.267001}. To understand the temperature dependence of cot$\theta_H$ using GTTA model, the temperature evolution of $\rho(\delta,T)n_{Hall}(\delta,T)$ for $YBa_2Cu_3O_{6+\delta}$ is shown in Fig. \ref{YBCO-cot}. It is observed that $\rho(\delta,T)n_{Hall}(\delta,T)$ deviates considerably from $T^2$ at low temperatures. At $p=0.06$, a moderately good fit is seen for $75K\leq T\leq300K$. The $T^2$ fitting is improved as the hole doping concentration increases from $p=0.09$ to $p=0.12$, however the lower bound of the temperature after which $\rho(\delta,T)n_{Hall}(\delta,T)$ obeys the $T^2$ law also increases. At $p=0.12$, $\rho(\delta,T)n_{Hall}(\delta,T)$ shows $T^2$ dependency from $\sim97K$. Fig. \ref{YBCO-cot}(e) shows the temperature variation of $\rho(\delta,T)n_{Hall}(\delta,T)$ at the optimal doping. It is clear that $\rho(\delta,T)n_{Hall}(\delta,T)$ does not show a perfect $T^2$ behaviour. However, the fitting improves for the overdoped material for a smaller temperature range as observed in Fig. \ref{YBCO-cot}(f).

\section{\label{sec:level1}Case of $TlBa_{1+x}La_{1-x}CuO_{5}$}
 Taking forward the theme of the unifying principle (as magnetic correlations become stronger (longer ranged in space and longer lived in time), ``tying down" of electrons happens and that leads to an effective reduction in carrier density as temperature is reduced) we move forward to study another cuprate $TlBa_{1+x}La_{1-x}CuO_{5}$.
 
 Manako and Kubo in 1994 studied the transport properties of $TlBa_{1+x}La_{1-x}CuO_{5}$ \cite{PhysRevB.50.6402}. This material shows transition from insulator to superconductor and metal as $x$ increases \cite{manako1989superconductivity}. At $x=0$, $R_H$ is almost independent of temperature and a semiconductor like behaviour is seen from its resistivity \cite{PhysRevB.50.6402}. The Hall coefficients $R_H(T)$ for $x=0.1, 0.2, 0.3, 0.4$ and $0.5$ for this cuprate are reported in Fig. 6. of \cite{PhysRevB.50.6402}. $R_H(T)$ decreases with $x$ i.e. effective carrier density increases with $x$. The GTTA analysis of this material is performed by using $R_H(T)$ data points in Fig. 6. of \cite{PhysRevB.50.6402}. $n_0(x)$, $n_1$ and $\Delta(x)$ are extracted using equation (1) of Appendix A. \ 
 \begin{figure}[h!]
 \centering
  \subfloat[$n_0(x)$]{\includegraphics[width=0.8\linewidth]{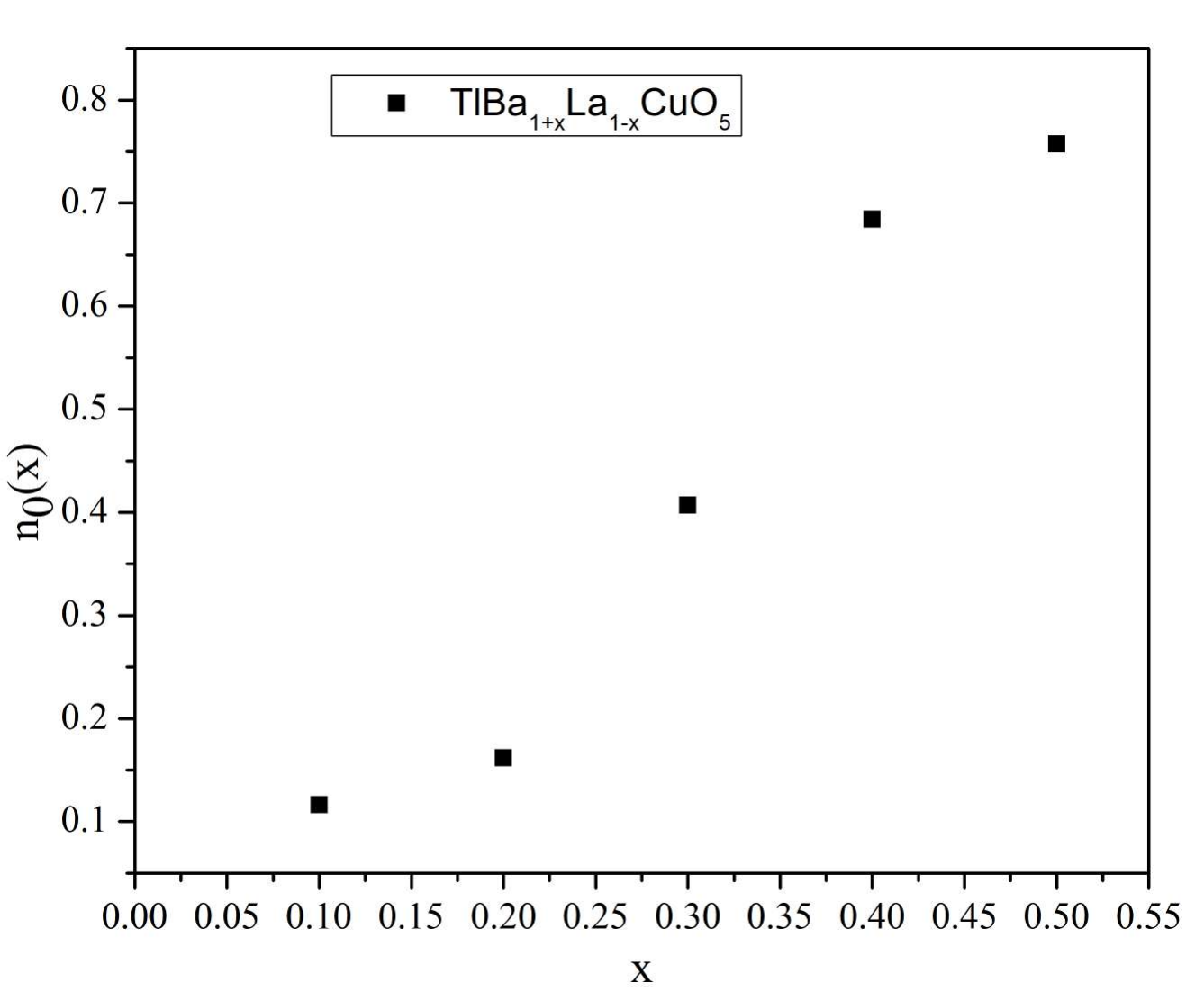}}\\
  \subfloat[$\Delta(x)$]{\includegraphics[width=0.8\linewidth]{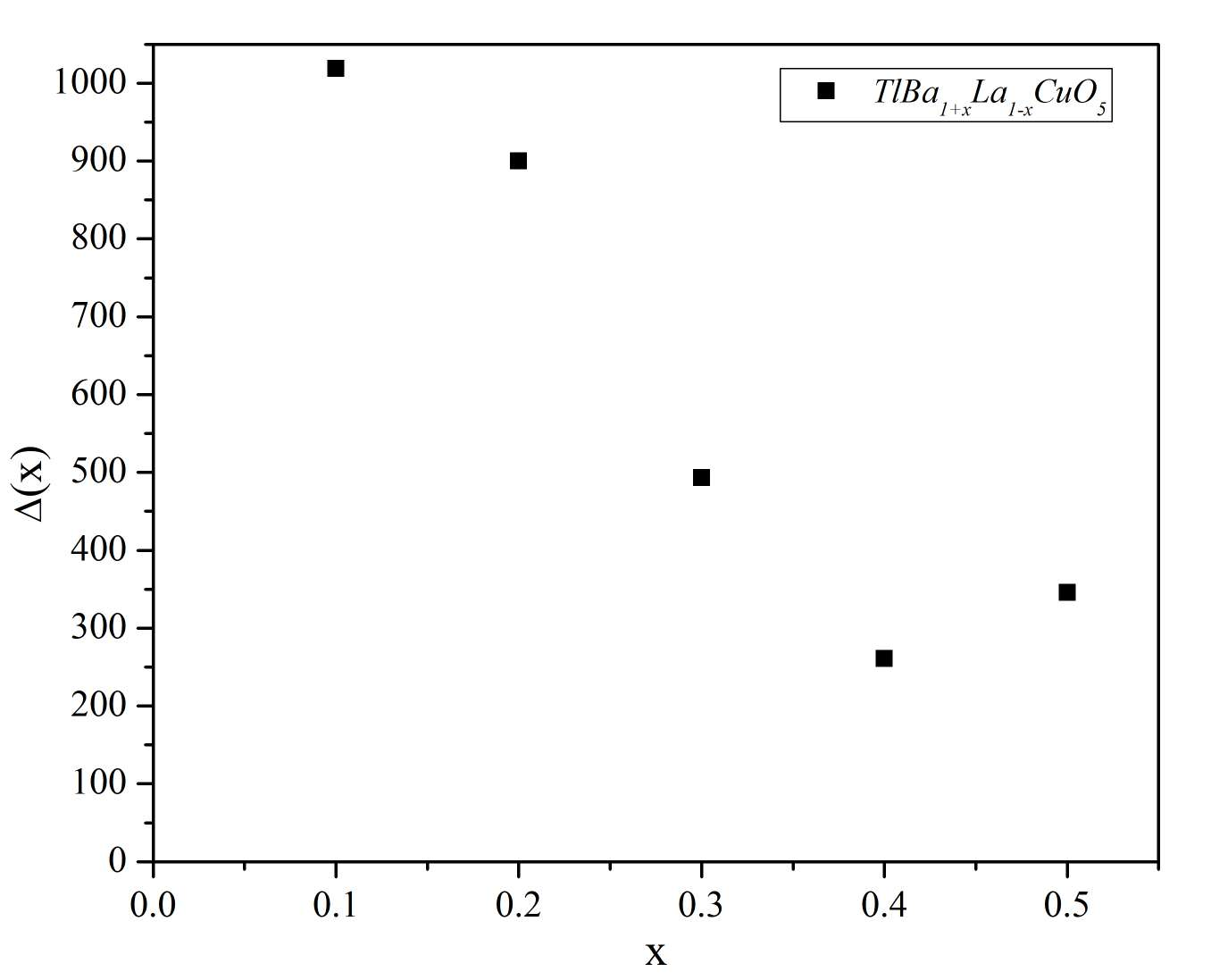}}
  \caption{\justifying The doping dependence of (a) $n_0(x)$ and (b) $\Delta(x)$ of $TlBa_{1+x}La_{1-x}CuO_{5}$ obtained using the GTTA model from the $R_H(T)$ extracted from Fig. 6 of \cite{PhysRevB.50.6402}.}
\label{TBLCO-GTTA}
\end{figure}

\begin{figure}[h!]
 \centering
  \includegraphics[width=0.8\linewidth]{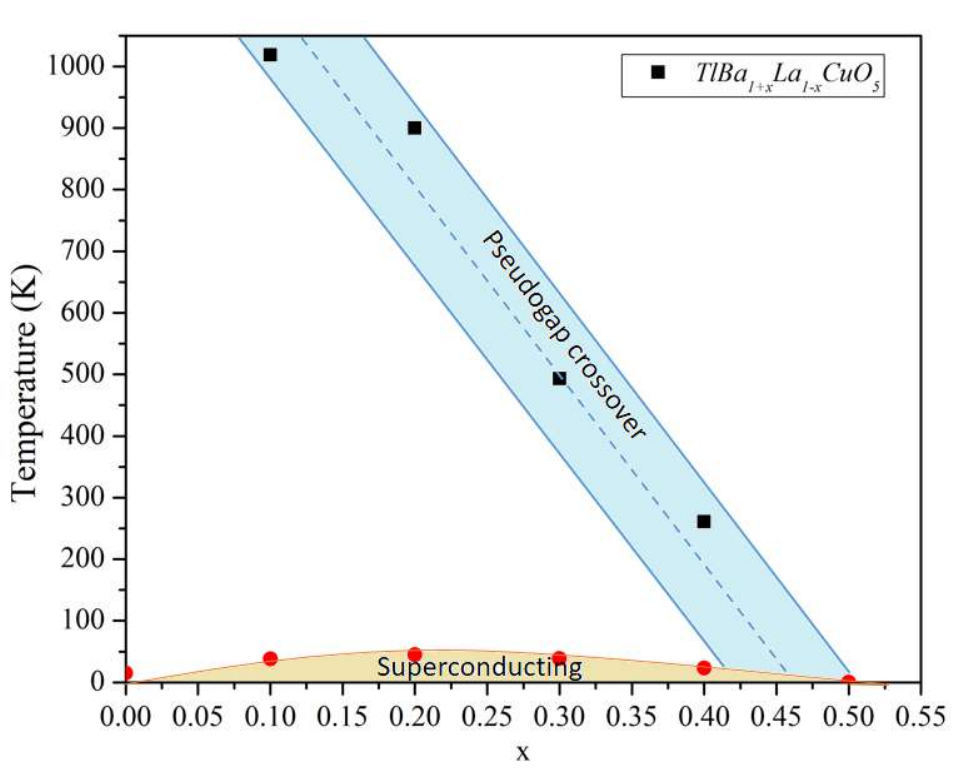}
  \caption{\justifying Updated phase diagram of $TlBa_{1+x}La_{1-x}CuO_{5}$. The $\Delta(x)$, obtained using the GTTA model, are marked in solid black squares. The PG crossover is shaded light blue. The superconducting dome is drawn from the $T_C$ values extracted from Fig. 3 of \cite{PhysRevB.50.6402}.}
\label{TBLCO-phase}
\end{figure}

 $n_1$ for this material is also roughly constant ($\sim1.0$) and falls for $x=0.4$. Fig. \ref{TBLCO-GTTA} shows the temperature dependence of $n_0(x)$ and $\Delta(x)$. $n_0(x)$, increases with doping concentrations quite rapidly from $x=0.1$ to $x=0.4$. Although, $n_0(x)$ is greater at $x=0.5$ than $x=0.4$, $n_0(x)$ seems to saturate at larger dopings. This could be hinting at crossover of $n_H\sim x$ to $n_H\sim 1+x$. As seen in Fig. \ref{TBLCO-GTTA}(b), the $\Delta(x)$ decreases roughly linearly till $x=0.4$. This indicates that the pseudogap crossover exists till $x=0.4$ after which the material is metallic. An updated phase diagram for $TlBa_{1+x}La_{1-x}CuO_{5}$ is drawn and displayed in Fig. \ref{TBLCO-phase}, wherein an effective PG crossover region is marked from the extracted $\Delta(x)$ from equation (1) of appendix A. The superconducting dome is drawn from the $T_c$ values extracted from Fig. 3 of \cite{PhysRevB.50.6402}. We suspect that at $x\simeq0.4$ there is a quantum critical point and the PG crossover should end there.\

\begin{figure}[h!]
    \begin{subfigure}[b]{0.5\columnwidth}
    \includegraphics[width=\textwidth]{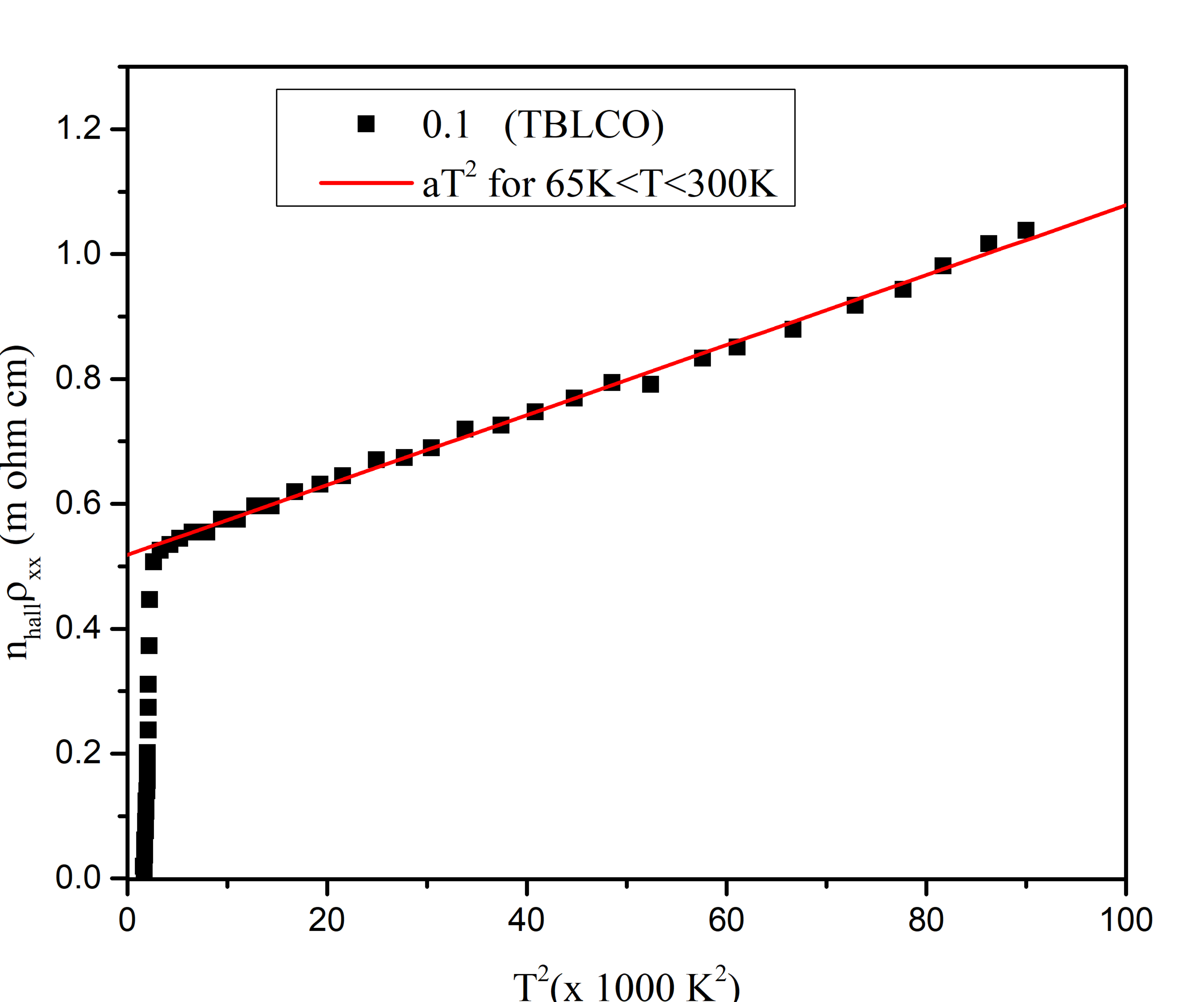}
    \caption{$x=0.1$}
  \end{subfigure}%
  \hfill
  \begin{subfigure}[b]{0.5\columnwidth}
    \includegraphics[width=\textwidth]{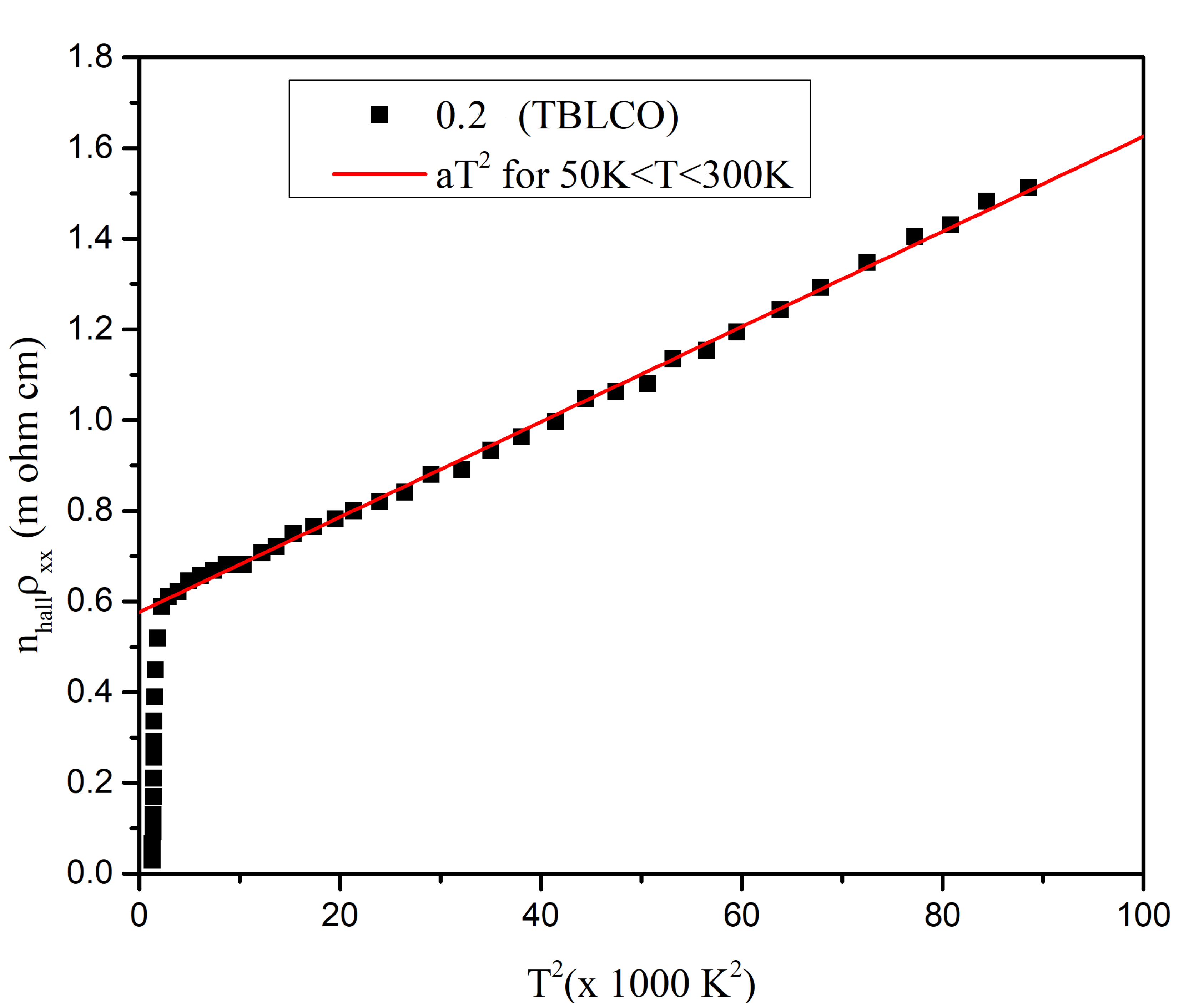}
    \caption{$x=0.2$}
  \end{subfigure}%

\medskip
\begin{subfigure}[b]{0.5\columnwidth}
    \includegraphics[width=\textwidth]{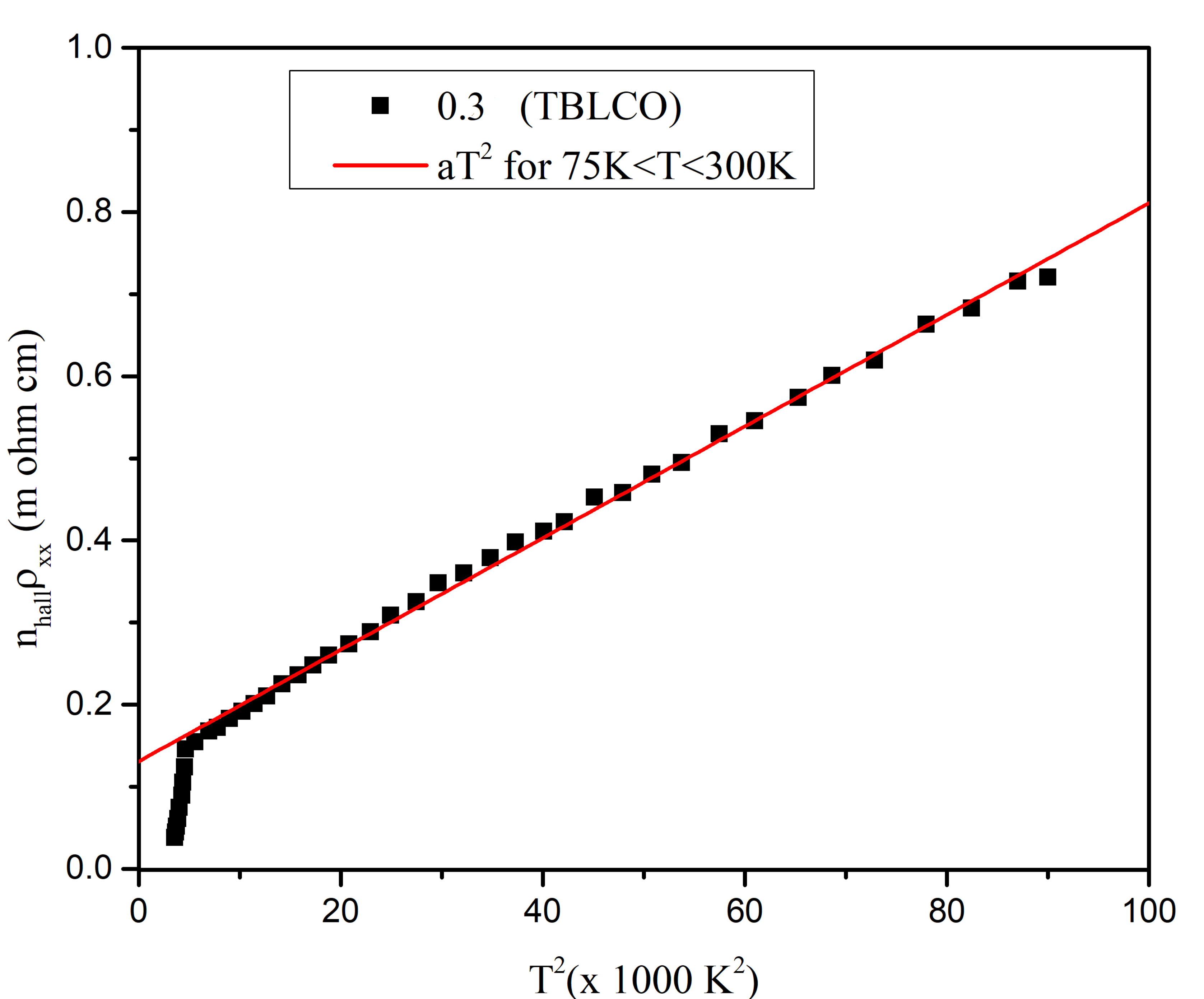}
    \caption{$x=0.3$}
  \end{subfigure}%
 \hfill
\begin{subfigure}[b]{0.5\columnwidth}
    \includegraphics[width=\textwidth]{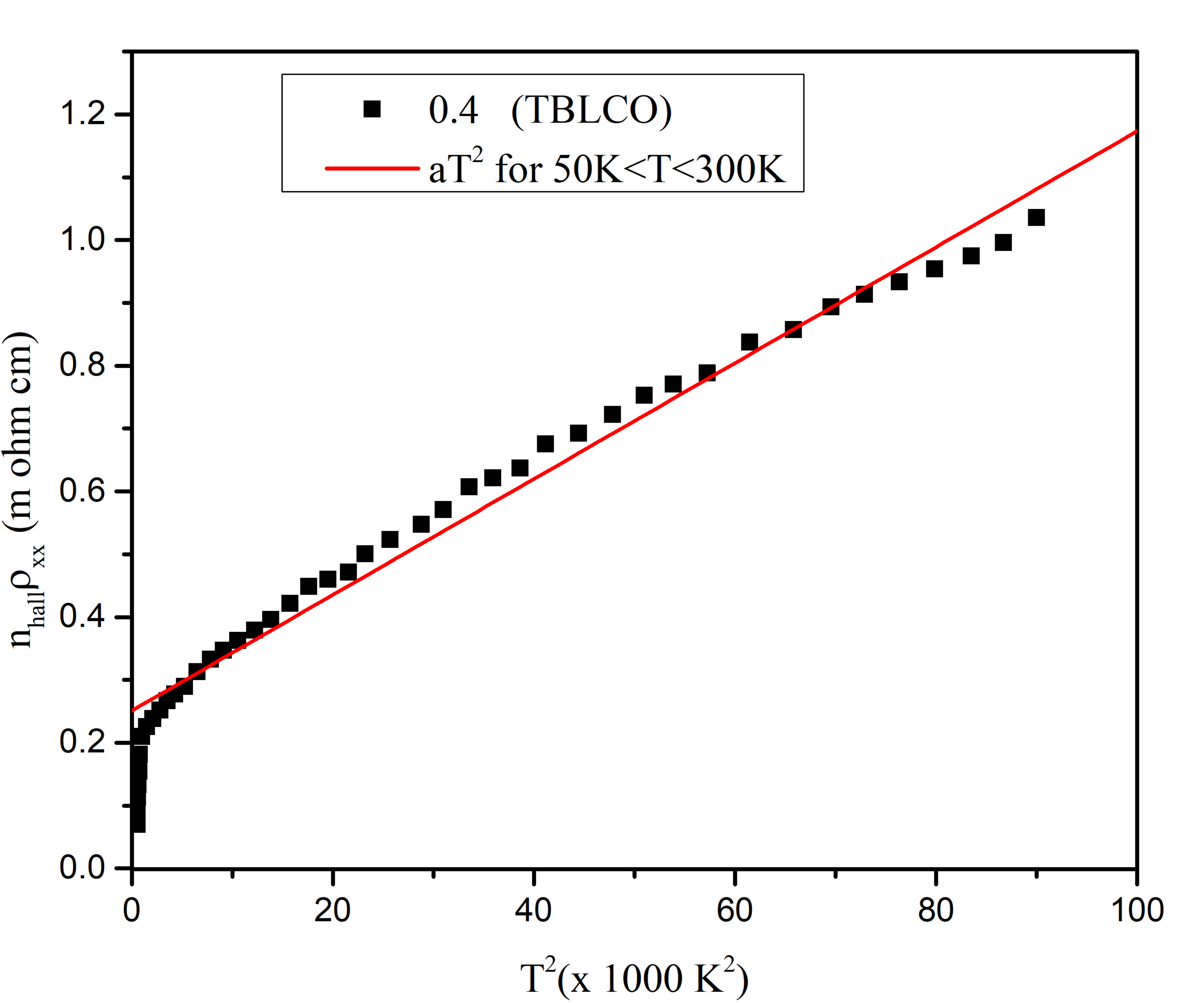}
    \caption{$x=0.4$}
  \end{subfigure}%
  \hfill
\begin{subfigure}[b]{0.5\columnwidth}
    \includegraphics[width=\textwidth]{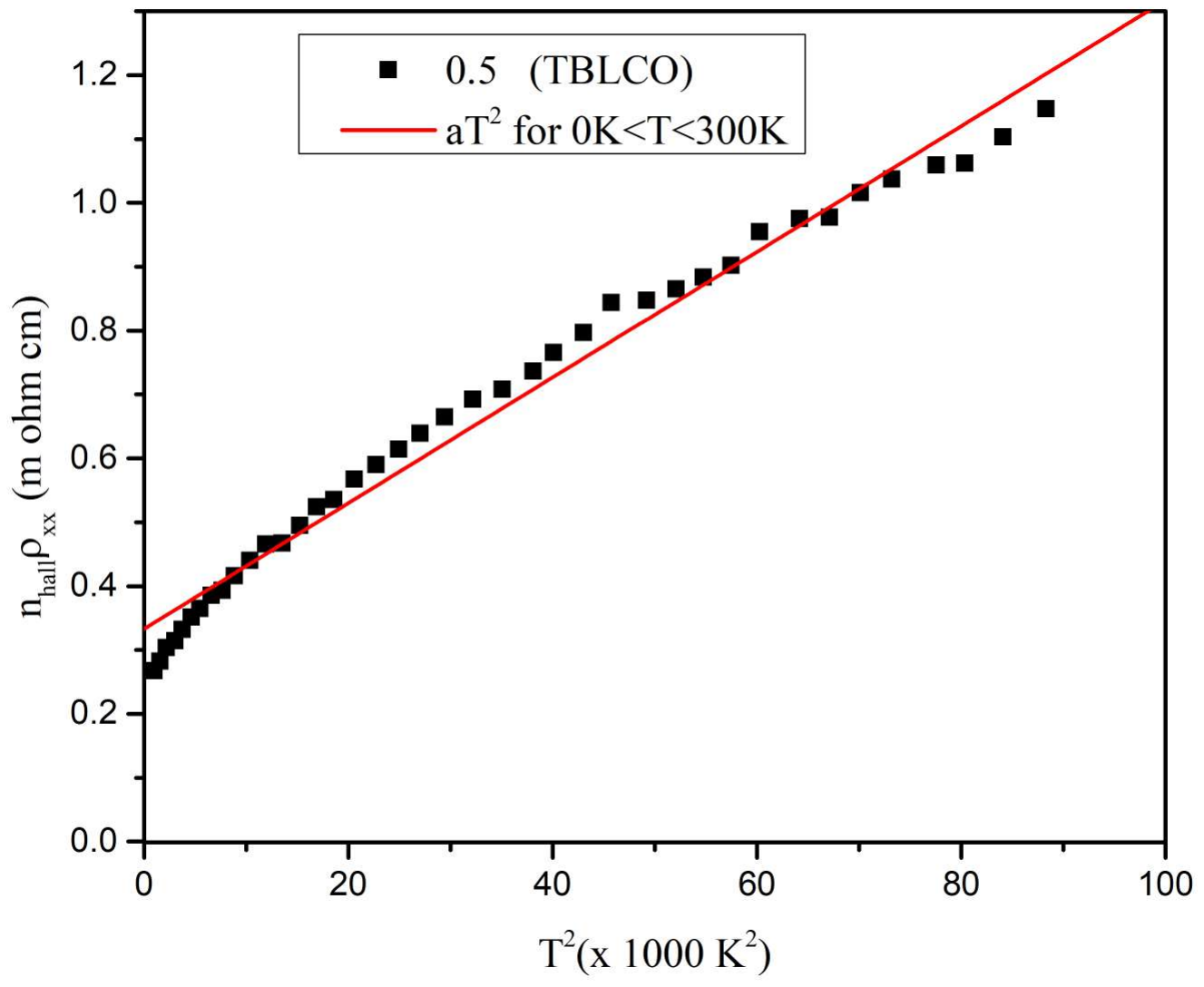}
    \caption{$x=0.5$}
    \end{subfigure}%
    \centering
  \caption{\justifying The $\rho(x,T)n_{Hall}(x,T)$ (cot$\theta_H$) vs $T^2$ of $TlBa_{1+x}La_{1-x}CuO_{5}$. $\rho(x,T)$ is extracted from Fig. 5 of \cite{PhysRevB.50.6402}. $n_{Hall}$ is the effective charge carriers obtained using GTTA model to the $R_H$ data reported in Fig. 6 of \cite{PhysRevB.50.6402}. The $\rho(x,T)n_{Hall}(x,T)$ obeys the $T^2$ law for $x$ ranging from 0.1 to 0.3 after which it deviates.}
  \label{TBLCO-cot}
\end{figure}

The authors of \cite{PhysRevB.50.6402} have also studied the temperature dependence of in-plane resistivity data, which is roughly T-linear (reported in Fig. 5 of \cite{PhysRevB.50.6402}). However, $\rho(x,T)$ does not seem to follow a pattern with respect to $x$ unlike $R_H$. The authors reason this with the possibility of grain boundary effects \cite{PhysRevB.50.6402}. The inverse Hall mobility obeys $T^2$ law (Fig. 8 of \cite{PhysRevB.50.6402}) for samples with $x\leq0.3$ after which it deviates, which according to the authors of \cite{PhysRevB.50.6402} is due to disorder of $CuO_2$ planes. We analyze the temperature dependence of cot$\theta_H$ using the the GTTA model. Fig. \ref{TBLCO-cot} displays the $\rho(x,T)n_{Hall}(x,T)$ vs $T^2$ plots of $TlBa_{1+x}La_{1-x}CuO_{5}$ for the mentioned doping concentrations. As seen in Fig. \ref{TBLCO-cot}(a), an excellent $T^2$ fit is observed for $65K\leq T\leq300K$ at $x=0.1$. At $x=0.2$ the temperature range for which an excellent $T^2$ linear behaviour of $\rho(x,T)n_{Hall}(x,T)$ is $65K\leq T\leq300K$. A reasonably good fit is obtained for $x=0.3$. However, on further doping i.e. $x\geq0.4$, a deviation from the $T^2$ law is observed which can be clearly noticed in Fig. \ref{TBLCO-cot}(d,e) This is consistent with the behaviour of the inverse Hall mobility seen in the experimental results reported in \cite{PhysRevB.50.6402}. This is also consistent with the behaviour of cot$\theta_H$ observed in $La_{2-x}Sr_xCuO_4$ beyond $x\simeq0.15$.\

\section{\label{sec:level1}Case of $Bi_2Sr_{2-x}La_xCuO_{6}$}

$Bi_2Sr_{2-x}La_xCuO_{6}$ is another superconducting cuprate which exhibits anomalous transport properties in its normal state. Pure $Bi_2Sr_{2}CuO_{6}$ is an overdoped system and substitution of La atoms reduces the number of holes in the system. In 1999, Ando et al studied the in plane resistivity $\rho(T)$ and Hall coefficient $R_H$ of this system at various La concentrations \cite{ando1999nonuniversal}.  Ono et al have also studied the metal to insulator transition of $Bi_2Sr_{2-x}La_xCuO_{6}$ \cite{ono2000metal}. They report that the transition takes place at $p\sim \frac{1}{8}$ (where $p$ is obtained by comparing the normalized Hall coefficient ($R_H/V$) of $Bi_2Sr_{2-x}La_xCuO_{6}$ with that of $La_{2-x}Sr_xCuO4$). In 2003, Balakirev et al studied the low temperature Hall coefficients for this material \cite{balakirev2003signature}.

\begin{figure}[h!]
 \centering
  \subfloat[$n_0(p)$]{\includegraphics[width=0.8\linewidth]{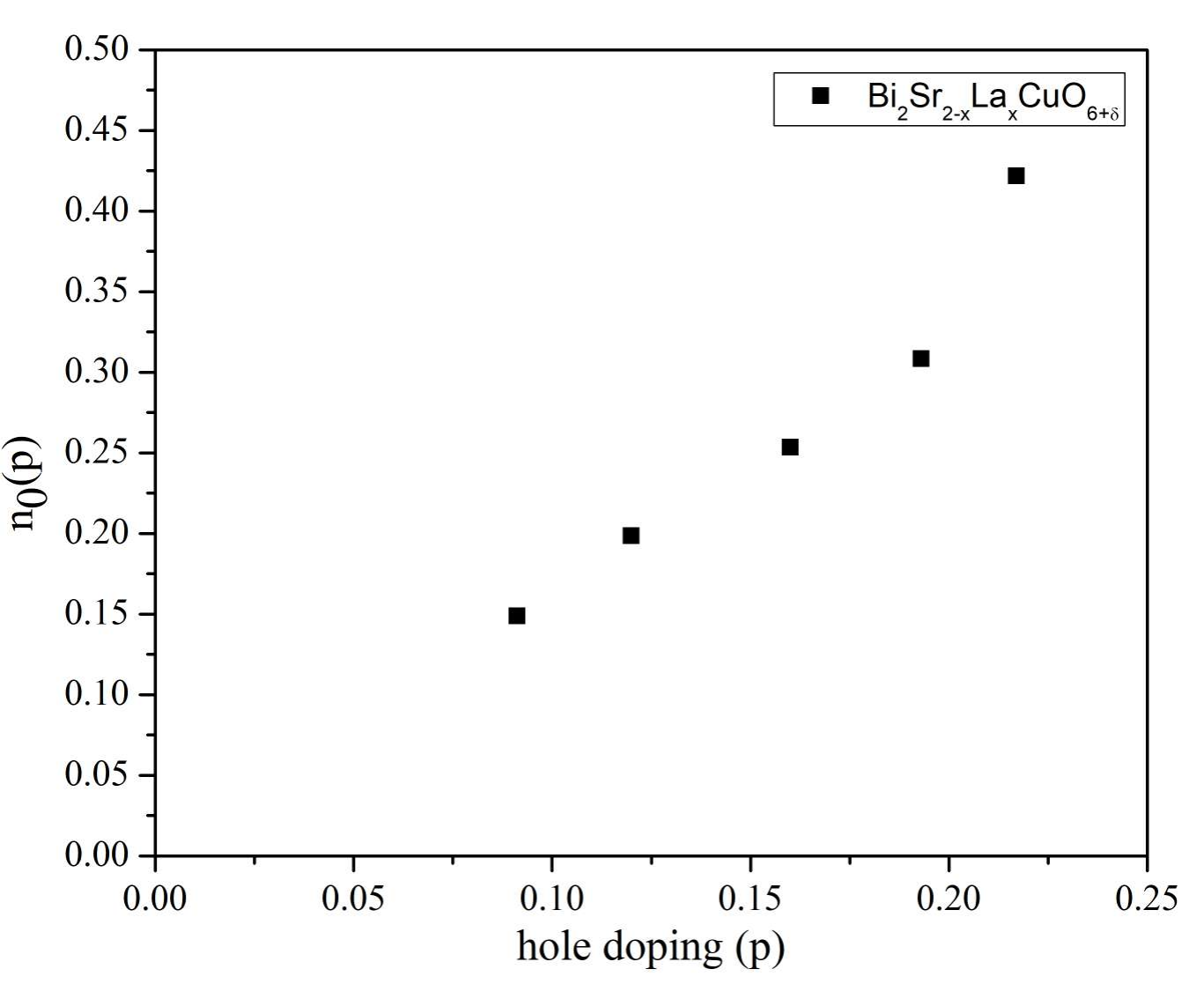}}\\
  \subfloat[$\Delta(p)$]{\includegraphics[width=0.8\linewidth]{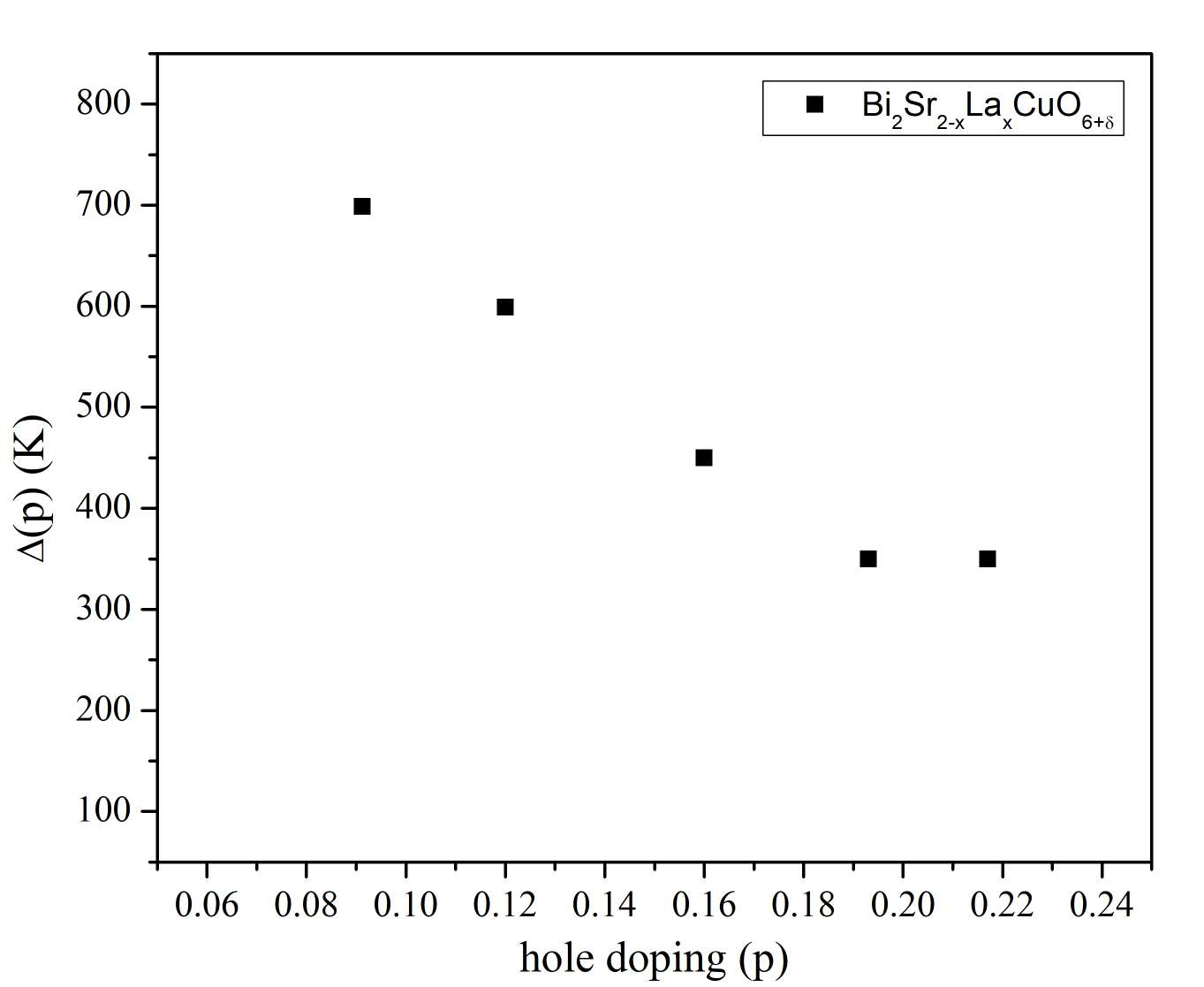}}
  \caption{\justifying The doping dependence of (a) $n_0(p)$ and (b) $\Delta(p)$ of $Bi_2Sr_{2-x}La_xCuO_{6}$ obtained using the GTTA model from the $R_H(T)$ extracted from Fig. 3(a) of \cite{ando1999nonuniversal}.}
\label{BSLCO-GTTA}
\end{figure}
\
In this work, we study the temperature dependent $R_H$ of $Bi_2Sr_{2-x}La_xCuO_{6}$ reported in \cite{ando1999nonuniversal} using GTTA model. The $R_H(T)$ values are extracted from Fig 3(a) of \cite{ando1999nonuniversal} and $n_0(x)$, $n_1$ and $\Delta(x)$ are calculated using equation (1) of Appendix A. The number of hole doped per Cu ($p$) is calculated from the law; $\frac{T_c}{T_{cmax}}=1-82.6(p-0.16)^2$. $n_0(p)$ and $\Delta(p)$ as a function of $p$ is shown in Fig. \ref{BSLCO-GTTA}. As observed in other cuprates, $n_0(p)$ increases rapidly as the hole doping concentration increases i.e. as La doping decreases. A roughly linear decrease in $\Delta(p)$ with increase in $p$ is observed. $\Delta(p)$ seems to saturate after $p\sim0.19$.\

\begin{figure}[h!]
 \centering
  \includegraphics[width=1.0\linewidth]{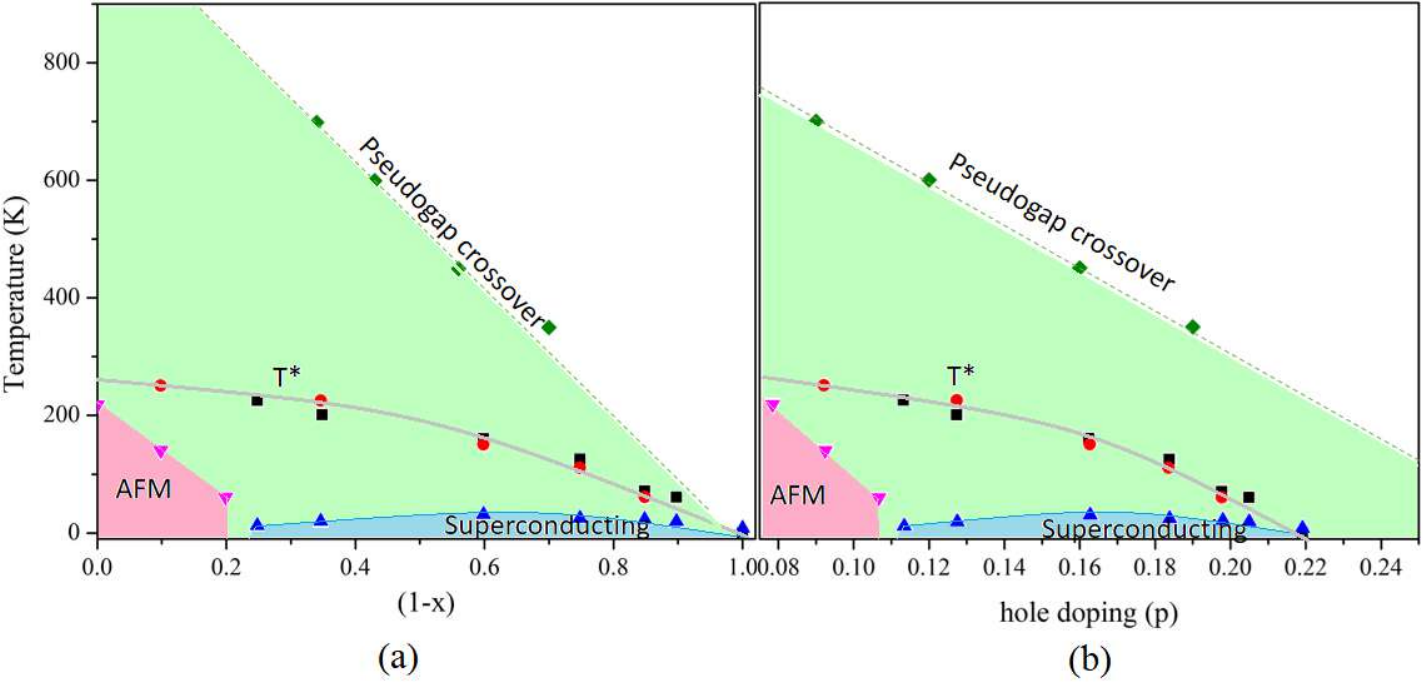}
  \caption{\justifying Updated phase diagram of $Bi_2Sr_{2-x}La_xCuO_{6}$ with respect to (a) La concentration $1-x$ and (b) hole doping $p$ calculated using the universal law. The $\Delta(x)$, obtained using the GTTA model, are marked in solid green squares. The PG crossover is shaded green. The red solid circles \cite{zheng2005critical}and solid black squares marks \cite{PhysRevB.50.6402} the characteristic temperature ($T^*$) measured from NMR experiments. The superconducting dome is drawn from the $T_C$ values extracted from Fig. 5 of \cite{PhysRevB.50.6402}.}
\label{BSLCO-phase}
\end{figure}

As it has been mentioned in section IV and V, this $\Delta(p)$ obtained from GTTA model can be correlated to a PG crossover, which in case of $La_{2-x}Sr_xCuO4$ and $YBa_2Cu_3O_{6+\delta}$ agrees very well with the experimental results. We also draw an updated phase diagram of $Bi_2Sr_{2-x}La_xCuO_{6}$ displayed in Fig. \ref{BSLCO-phase}. In case of $Bi_2Sr_{2-x}La_xCuO_{6}$, inspite of knowing the La concentration ($x$) the exact hole concentration cannot be calculated. The ``universal relation" that we have used for calculating $p$ which works very well for $YBa_2Cu_3O_{6+\delta}$, may not hold in the $Bi_2Sr_{2-x}La_xCuO_{6}$ \cite{ando2000carrier}. Therefore, we plot the phase diagram of $Bi_2Sr_{2-x}La_xCuO_{6}$ with respect to the La doping $x$ and the effective hole doping concentration $p$ (Fig. \ref{BSLCO-phase}). 

\begin{figure}[h!]
    \begin{subfigure}[b]{0.5\columnwidth}
    \includegraphics[width=\textwidth]{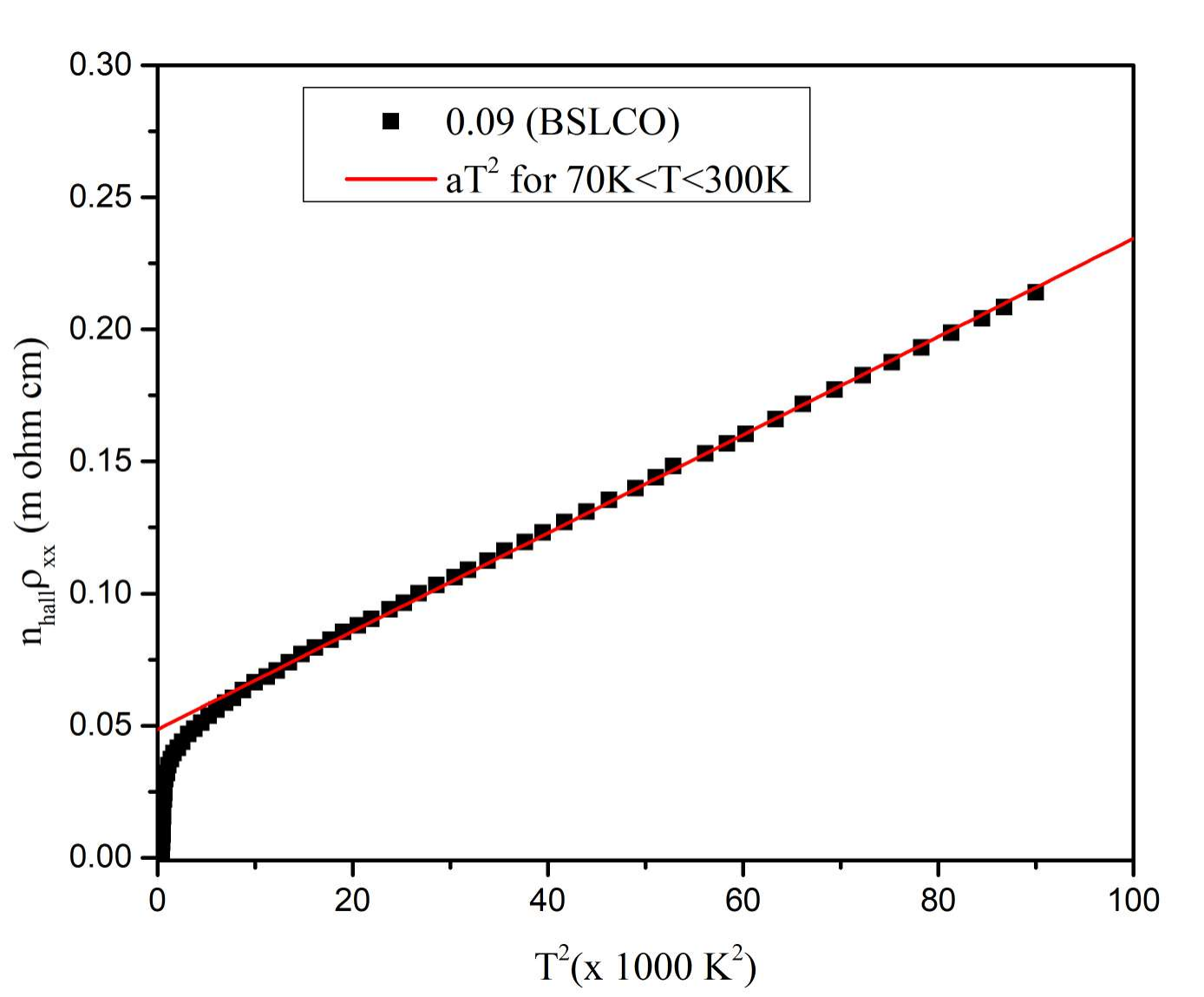}
    \caption{$p=0.09$}
  \end{subfigure}%
  \hfill
  \begin{subfigure}[b]{0.5\columnwidth}
    \includegraphics[width=\textwidth]{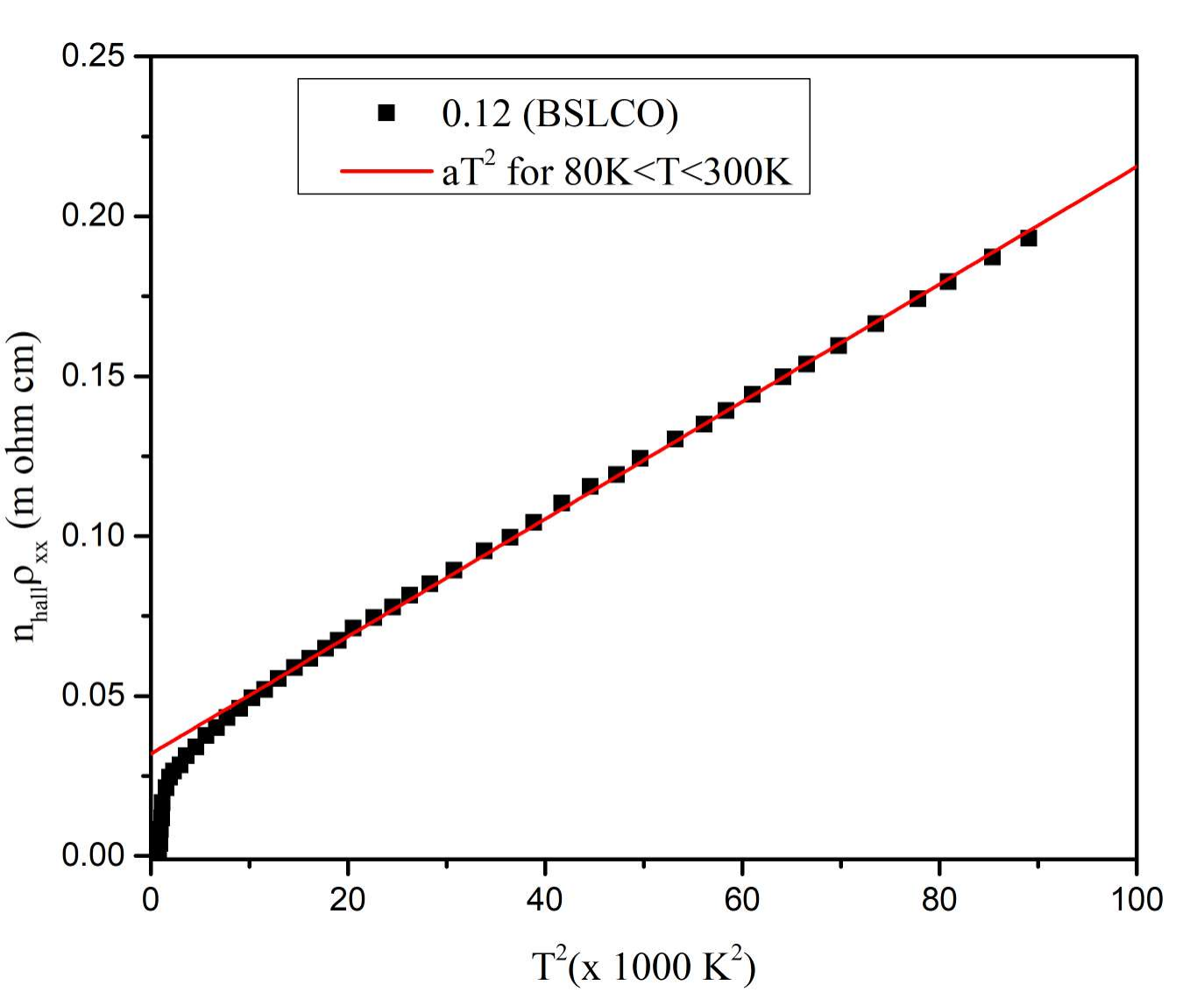}
    \caption{$p=0.12$}
  \end{subfigure}%

\medskip
\begin{subfigure}[b]{0.5\columnwidth}
    \includegraphics[width=\textwidth]{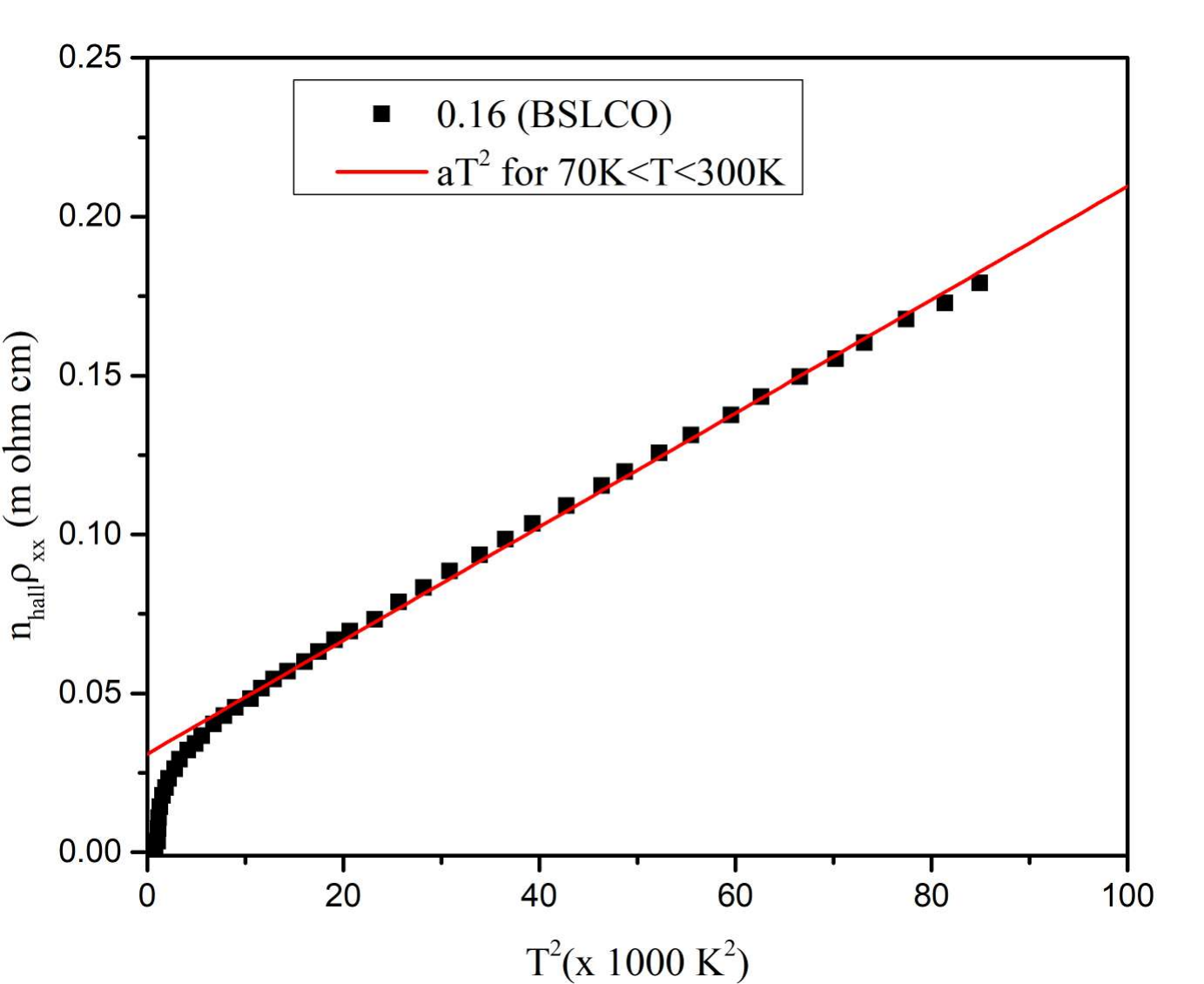}
    \caption{$p=0.16$}
  \end{subfigure}%
 \hfill
\begin{subfigure}[b]{0.5\columnwidth}
    \includegraphics[width=\textwidth]{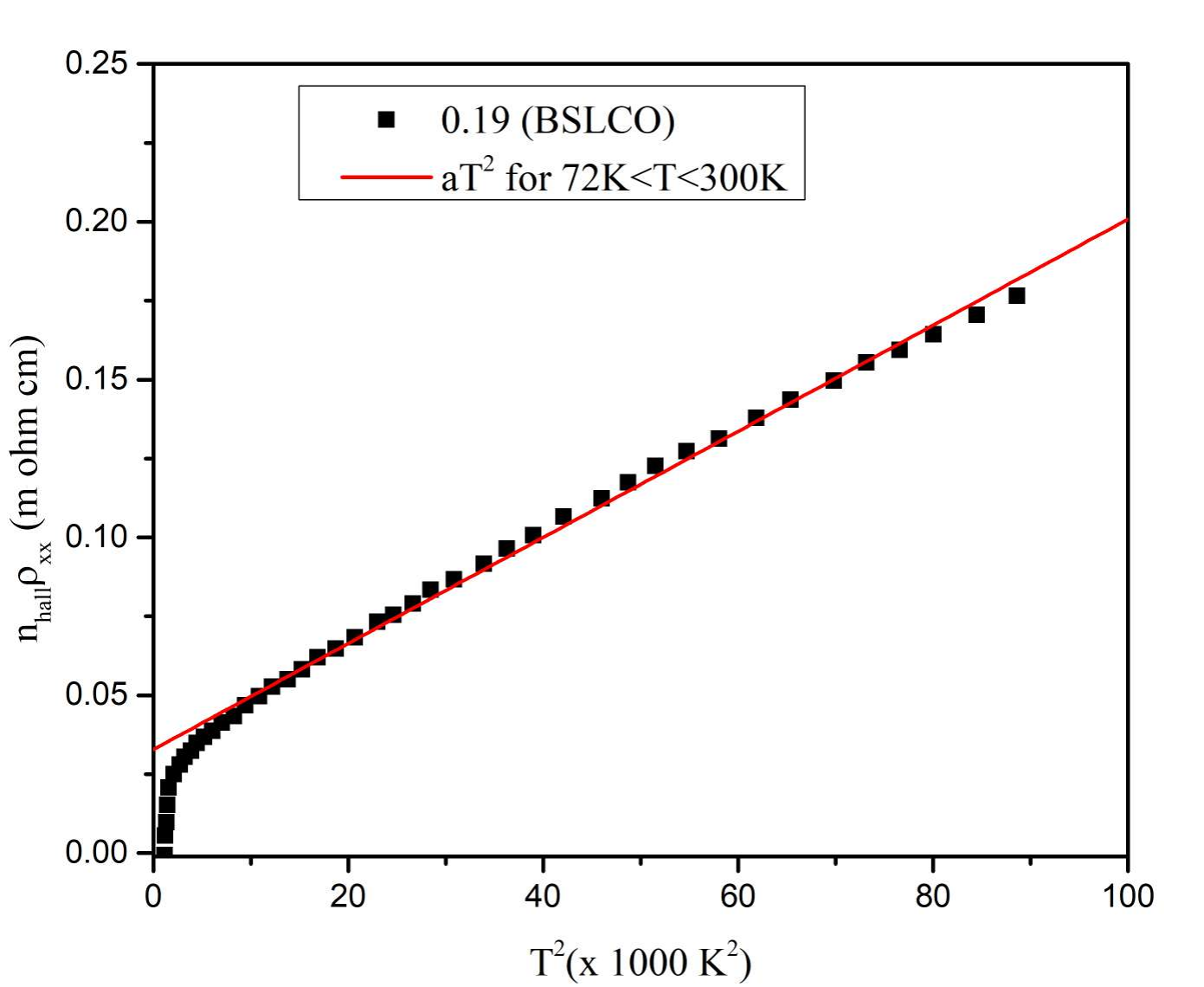}
    \caption{$p=0.19$}
  \end{subfigure}%
  \hfill
\begin{subfigure}[b]{0.5\columnwidth}
    \includegraphics[width=\textwidth]{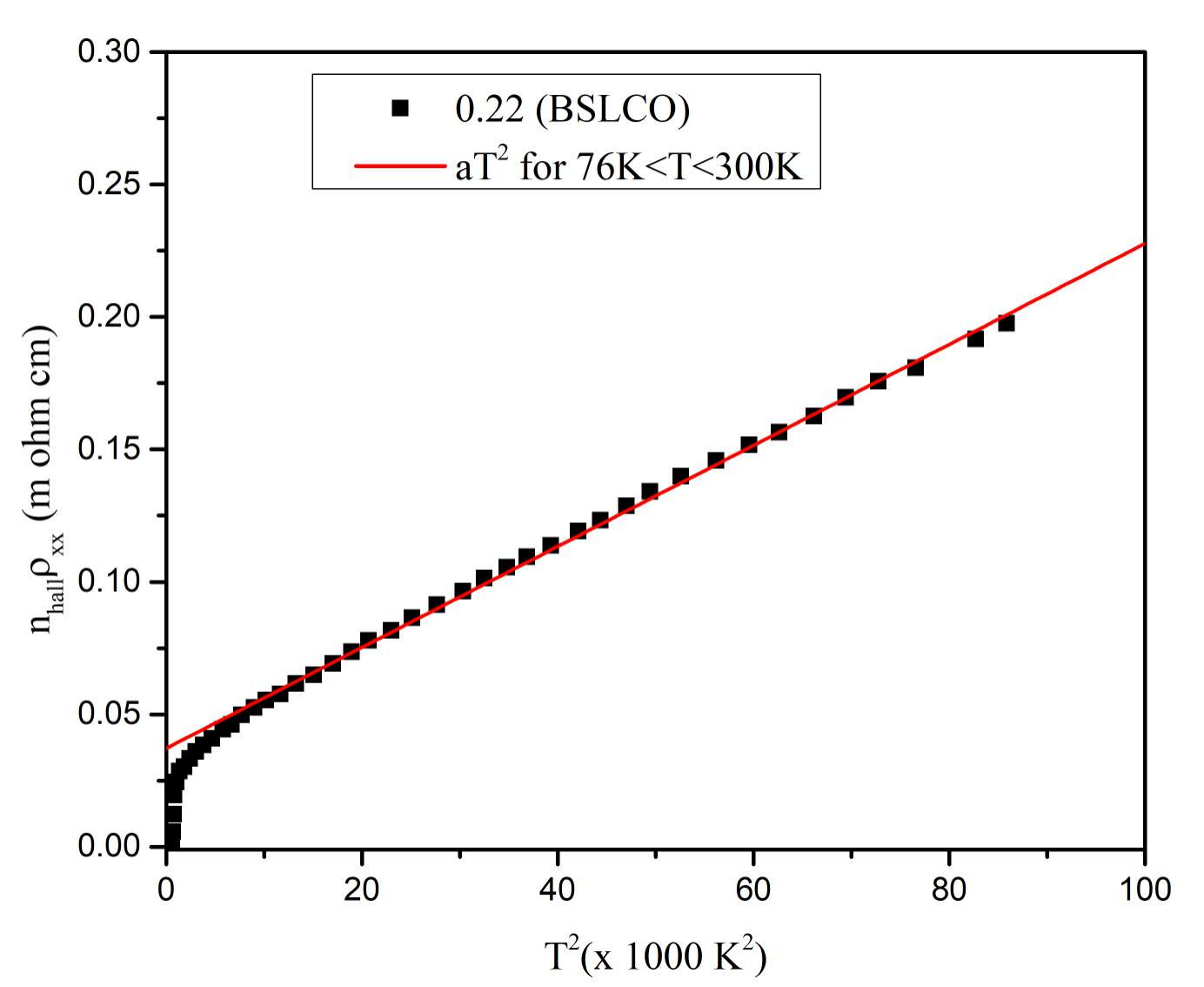}
    \caption{$p=0.22$}
    \end{subfigure}%
    \centering
  \caption{\justifying The $\rho(x,T)n_{Hall}(x,T)$ (cot$\theta_H$) vs $T^2$ of $Bi_2Sr_{2-x}La_xCuO_{6}$. The $R_H(T)$ values are extracted from Fig 3(a) of \cite{ando1999nonuniversal} to calculate $n_{Hall}(x,T)$ using GTTA model. The $\rho(x,T)$ data is extracted from Fig 1 of \cite{ando1999nonuniversal}. $\rho(x,T)n_{Hall}(x,T)$ is found to obey the $T^2$ for all the mentioned doping concentrations.}
  \label{BSLCO-cot}
\end{figure}

The authors of \cite{ando1999nonuniversal} also reported the Hall angle cot$\theta_H$ of this system for different La concentrations. The Hall angle of $Bi_2Sr_{2-x}La_xCuO_6$ crystals also obey the $T^\alpha$ power law, where $\alpha$ decreases with increase in carrier concentrations \cite{ando1999nonuniversal}. The value of $\alpha$ was 2 for underdoped samples and it decreased as the doping increased. Thus the temperature dependence of $\rho(x,T)n_{Hall}(x,T)$ is also analyzed for $Bi_2Sr_{2-x}La_xCuO_6$, where $n_{Hall}$ is obtained by using the GTTA model. The $\rho(x,T)$ data is extracted from Fig 1 of \cite{ando1999nonuniversal}. The resistivity $\rho(x,T)$ shows a perfect T-linear behaviour at the optimal doping ($x=0.44$). \

Fig. \ref{BSLCO-cot} shows the temperature dependence of $\rho(x,T)n_{Hall}(x,T)$. It is observed that $\rho(x,T)n_{Hall}(x,T)$ exhibits $T^2$ behaviour for all the doping concentrations studied. 

\

\section{\label{sec:level1}Summary of results and discussion}

\begin{itemize}
  \item In this work we put forward the idea of a ``unifying principle" which governs the Hall coefficient and Hall angle behaviour of systems near magnetic instability and we validate it for $Cr_{1-x}V_x$, $V_{2-y}O_3$ and some high-$T_c$ superconducting cuprate systems (studied in this work). The unifying principle essentially states that when the temperature of materials which is near a magnetic instability is reduced a gradual ``tying down" of electrons take place as the magnetic correlations become longer-ranged and longer-lived. This leads to a ``loss" in carrier number with decrease in temperature, implying that the carrier density is not constant rather it is a temperature dependent function. This fact is reflected in the temperature dependent Hall coefficients $R_H$ observed for such materials. 
  \item $Cr_{1-x}V_x$ and $V_{2-y}O_3$ are 3D antiferromagnetic materials which have temperature dependent $R_H$ \cite{yeh2002quantum,rosenbaum1998temperature}. The antiferromagnetic fluctuations observed even beyond the antiferromagnetic order are responsible for ``tying down" of electrons as $R_H$ (carrier density $n$) is temperature dependent. Presence of an effective pseudogap due to these antiferromagnetic fluctuations in $Cr_{1-x}V_x$ was suggested by the authors of \cite{yeh2002quantum} and in this work we quantitatively demarcate it.
  \item The GTTA model (a phenomenological model) is a mathematical expression of the unifying principle stated and it is an ad-hoc expression(not obtained from a microscopic calculation). Better mathematical expressions based on microscopic model must be found. GTTA model explains the temperature dependence of $R_H$ in the form of equation (1) (Appendix A) where, the term $\Delta(x)$ can be correlated to an effective PG of thermal activation nature. It is already proven to efficiently model $R_H$ for $La_{2-x}Sr_xCuO_4$ and $Bi_2Sr_{2-x}La_xCuO_{6+\delta}$ \cite{gor2006interplay,gor2008mobility}. For $La_{2-x}Sr_xCuO_4$, $\Delta(x)$ were in agreement with previous known PG signatures. GTTA model also reproduces the $T^2$ behaviour for $La_{2-x}Sr_xCuO_4$ \cite{gor2008mobility}. In addition, for the case of $YBa_2Cu_3O_{6+\delta}$, $\Delta(x)$ matched very well with the PG crossover obtained from the study of its resistivity \cite{sacksteder2020quantized}.
  \item The Hall coefficients of $Cr_{1-x}V_x$ and $V_{2-y}O_3$ are studied using the GTTA model (discussed in Appendix A) and,$n_0(x)$ and $\Delta(x)$ values are extracted. Using the $\Delta(x)$ (effective PG crossover), updated phase diagrams of $Cr_{1-x}V_x$ and $V_{2-y}O_3$ (Fig. \ref{CrV-phase} and Fig. \ref{VO-phase}) are drawn and PG crossovers are depicted.
  \item In case of $Cr_{1-x}V_x$, the PG crossover increases rapidly till the critical doping ($x_c=0.035$) after which it increases roughly linearly. The doping dependence of $n_0(x)$ for $Cr_{1-x}V_x$ is in good agreement with doping dependence of carrier concentration at 0K \cite{yeh2002quantum}. In case of $V_{2-y}O_3$ having a insulator to metal crossover, the PG crossover decreases roughly linearly with increase in V-deficiency. PG crossover demarcated in this work should be checked by experiments. Some experiments have been proposed in section I to this end. 
  \item The high-$T_c$ superconducting cuprates also have a strongly temperature dependent $R_H$ which again can be understood from the temperature dependence of carrier density $n(T)$ (unifying principle). In this work, the temperature dependence of $R_H$ is also rationalized using the GTTA model for some cuprates and a quantitative value of their PG crossover is calculated. The PG crossover ($\Delta(x)$) for $La_{2-x}Sr_xCuO_4$ matched with $\Delta(x)$ obtained by Gor'kov and Teitel'baum which was in agreement with the ARPES experiments \cite{gor2006interplay}. In case of $YBa_2Cu_3O_{6+\delta}$, the $\Delta(x)$ falls in line with the the temperature at which the second temperature derivative of resistivity is zero. An updated phase diagram for $YBa_2Cu_3O_{6+\delta}$ (Fig. \ref{YBCO-phase}) is drawn incorporating the GTTA results \cite{sacksteder2020quantized}. The PG crossover of $TlBa_{1+x}La_{1-x}CuO_{5}$ and $Bi_2Sr_{2-x}La_xCuO_{6}$ is also predicted using GTTA model. 
  \item The discrepancy in the temperature dependence of in-plane resistivity $\rho(T)$ (T-linear behaviour) and the Hall angle $cot\theta_H$ ($T^2$ behaviour) which is observed in $Cr_{1-x}V_x$, $V_{2-y}O_3$ and high-$T_c$ superconducting cuprates is also addressed in this work. Anderson had addressed this issue by using the idea of ``two-relaxation time". We show that using the ``unifying principle" (temperature dependent $n(T)$) the Hall effect in these materials can be understood without incorporating two different relaxation times. There is only one relaxation rate which scales as $T^2$. The resistivity is T-linear because its temperature dependence is not only governed by relaxation rate but also by the temperature dependent carrier density $n(T)$. Similar ideas are also advocated in \cite{barivsic2022high}.
  \item We have analyzed the temperature dependence of $\rho(x,T)n_{Hall}(x,T)$ for $Cr_{1-x}V_x$, $V_{2-y}O_3$ and some high-$T_c$ superconducting cuprates ($n_{Hall}(x,T)$ is the carrier density calculated using the GTTA model). It scales as $T^2$ similar to $cot\theta_H$. Deviations from the $T^2$ law is observed for the overdoped region for most of the systems. The $\rho(x,T)n_{Hall}(x,T)$ ($\propto$$cot\theta_H$) qualitatively matches exceptionally well with the experimental $cot\theta_H$. In fact, in the case of $La_{2-x}Sr_xCuO_4$, we quantitatively obtained its structural phase boundary from $\rho(x,T)n_{Hall}(x,T)$ which is in close agreement with the experimental phase boundary. Thus, an updated phase diagram with the structural phase boundary is also reported (Fig. \ref{LSCO-phase}). 
  \item In conclusion, the temperature $R_H(T)$ and the anomalous transport properties (in-plane resistivity $\rho(T)$ and Hall angle $cot\theta_H$) is governed by a unifying principle. For all the magnetic systems studied in this article, this unifying principle is justified. However, in-order to check its general validity it must be tested for a greater variety of systems. 
\end{itemize}

\appendix
\section{Appendix A: Brief overview of the Gor'kov Teitel'baum Thermal activation model}
As discussed in Section 1, in the paramagnetic phase of magnetic materials, as the temperature is reduced the carrier density decreases due to the progressive "tying down" of electrons. Considering this carrier density $n_{Hall}$ to be analogous to the Fermi surface area, the reduction in $n_{Hall}$ corresponds to loss in Fermi surface. Thus the antiferromagnetic fluctuations in such materials beyond $T_N$ may form an effective pseudogap in the electronic spectrum. We agree with the authors of \cite{yeh2002quantum} and quantify the PG crossover. The temperature dependence of $R_H$ and the effective pseudogap of magnetic materials is studied quantitatively using the Gor'kov Teitel'baum Thermal Activation model.\ 

In 2006, Gor'kov and Teitel'baum developed a phenomenological model which described the temperature and doping dependence of the Hall coefficients $R_H(T)$ for $La_{-x}Sr_xCuO_4$ \cite{gor2006interplay}. They rationalized the experimental Hall effect data by introducing doping and temperature dependent carrier density $n_{Hall}(x,T)$. The relation between the Hall coefficients and effective density is $R_H=\frac{1}{n_{Hall}(x,T)e}$. According to Gor'kov and Teitel'baum:
\begin{equation}
    n_{Hall}(x,T)=n_{0}(x)+n_{1}(x)e^{[-\Delta(x)/T]}
    \label{GTTA}
\end{equation}
The total carrier density $n_{Hall}(x,T)$ is divided into a temperature independent term $n_0(x)$ and a temperature dependent term $n_{1}(x)e^{[-\Delta(x)/T]}$. $n_0(x)$ depends only on the amount of external hope doping for $x<0.07$ (in case of $La_{-x}Sr_xCuO_4$), on further doping $n_0(x)$ deviates considerably from the linear behaviour. The temperature dependent term is a thermal excitation or activation term. Equation (1) is the defining equation of the Gor'kov Teitel'baum Thermal Activation (GTTA) model. The model was successfully applied to $La_{1-x}Sr_xCuO_4$ and it reproduced the Hall effect data reasonably well \cite{gor2006interplay}. Extracted parameters $n_0(x)$, $n_1(x)$ and $\Delta(x)$ behaves in the following way: For $x<0.19$, $n_1(x)$ is roughly constant ($\simeq2.8$). An abrupt decrease in $n_1(x)$ is observed for $x>0.19$. This signifies an important physical change in the system as discussed in \cite{singh2024review}. The $\Delta(x)$ is interpreted as the thermal activation energy of electrons. It is the effective gap between the states in the nodal and anti-nodal region of the Brillouin zone. A linear decrease in $\Delta(x)$ with respect to $x$ is observed for $x<0.20$ \cite{gor2006interplay}. The $\Delta(x)$ values extracted from equation (1) agreed exceptionally well with that reported in ARPES measurements of $La_{1-x}Sr_xCuO_4$ \cite{yoshida2003metallic,yoshida2006systematic}.

\section{Appendix B: On the issue of "two-relaxation times"}
In spite of the extensive research done in the past 30 years on the high-temperature superconducting cuprates, their normal state transport properties are not well understood. The in-plane DC resistivity and the Hall coefficients of these materials show temperature dependence quite different from standard metals \cite{gurvitch1987resistivity,ginsberg1998physical}. Cuprates are found to have T-linear in-plane resistivity at optimal doping wheres the Hall coefficients have a complex temperature dependence (very roughly $R_H\propto\frac{1}{T}$). Chien et al in 1991 suggested that the peculiar transport properties of cuprates can be described in terms of the Hall angle cot$\theta_H$ which is the ratio of in-plane resistivity to the Hall resistivity. cot$\theta_H$ shows $T^2$ behaviour \cite{chien1991effect}. This indicates discrepancy in the scattering rates observed from the resistivity and Hall angle cotangent.\

In 1991, Anderson explained the unique transport properties of cuprates by using two different relaxation times for carriers perpendicular ($\tau$) and parallel ($\tau_H$) to the Fermi surface \cite{PhysRevLett.67.2092}. The relaxation time $\tau$ when carriers are accelerated by electric field varies as $\frac{1}{T}$ whereas the relaxation time ($\tau_H$) of carriers when accelerated by magnetic field varies as $\frac{1}{T^2}$. Using the equation of motion of a charged particle in presence of magnetic and electric field with two distinct relaxation times leads to quadratic temperature dependence for $cot\theta_H$ and inverse temperature dependence for the Hall coefficient \cite{singh2024review}.\

Nevertheless, there are various other explanations reported in literature which does not require two different relation rates for carriers accelerated by electric field and magnetic field. Anderson in his approach assumed the effective carrier density $n$ to be constant with respect to temperature $T$. If $n$ depends on $T$, the transport properties could be explained without assuming two different longitudinal and transverse relaxation rates, because then the temperature dependence on the scattering rate $\tau(T)$ would come from both resistivity $\rho(T)$ and carrier density $n(T)$. \ 

Several reports on cuprates indicates variation of the carrier concentration with temperature. Early on, Hirsch and Marsigilio modelled the normal state properties of high-$T_c$ oxides by assuming it to be a Fermi liquid and $n$ changes with temperature \cite{HIRSCH1992355}. In 1992, Kubo and Manako investigated the transport properties of Tl-doped cuprates which suggested that $n$ varied with temperature \cite{KUBO1992378}. A single band model was proposed by Xing and Lui to explain the linear temperature dependency of $\rho$ and $n$ \cite{xing1991single}. Alexander and Mott used the bipolaron theory to understand the doping and temperature dependence of in-plane and c-axis resistivity \cite{alexandrov1996coherent}. Levin and Quader developed a model to study $\rho(T,x)$ for underdoped cuprates. They assumed that the non-degenerate and degenerate carriers have distinct relaxation rates suggesting that the carrier density is temperature dependent \cite{levin2000plane}. In 2008, the anomalies observed in the transport properties were reasoned by introducing $n(T)\propto T$ because of the formation of bound states of electron and hole like quasiparticles \cite{luo2008alternative}. The GTTA model, which is a phenomenological model, discussed in Appendix A also accounts for the temperature dependence of carrier density $n$. In this work, GTTA model is used to rationalize the temperature dependence of $n$ of materials near magnetic instability.

\bibliography{two-relaxation-times}
\end{document}